\documentclass[final, 12pt, a4paper]{elsarticle}
\journal{Physics Reports}

\usepackage{lineno}	

\usepackage{amssymb}	
\usepackage{amsmath}	
\usepackage{trfsigns}	
\usepackage{nicefrac}	
\usepackage{xcolor}
\usepackage[urlcolor=blue, colorlinks=true]{hyperref}      
\usepackage{booktabs}	

\biboptions{sort&compress} 

\begin{document}
\begin{frontmatter}

\title{Chaos in time delay systems, an educational review}

\author[label1]{Hendrik Wernecke\corref{corauthor}}
\ead{wernecke@th.physik.uni-frankfurt.de}	
\author[label2]{Bulcs\'{u} S\'{a}ndor}
\ead{bulcsu.sandor@phys.ubbcluj.ro}		
\author[label1]{Claudius Gros}
\ead{gros@th.physik.uni-frankfurt.de}		

\address[label1]{Institute for Theoretical Physics, Goethe University, Frankfurt/Main, Germany}
\address[label2]{Department of Physics, Babe\c{s}-Bolyai University, Cluj-Napoca, Romania}
\cortext[corauthor]{Corresponding author.}

\begin{abstract}
The time needed to exchange information in the physical world
induces a delay term when the respective system is modeled by differential
equations. Time delays are hence ubiquitous, being furthermore
likely to induce instabilities and with it various kinds of chaotic 
phases. Which are then the possible types of time delays, induced 
chaotic states, and methods suitable to characterize the resulting 
dynamics? This review presents an overview of the field that includes 
an in-depth discussion of the most important results, of the standard 
numerical approaches and of several novel tests for identifying chaos. 
Special emphasis is placed on a structured representation that
is straightforward to follow. Several educational examples are 
included in addition as entry points to the rapidly developing field 
of time delay systems.
\end{abstract}


\end{frontmatter}


\newpage
\tableofcontents
\newpage

 

\section{Introduction}
\label{sec:intro}


The field of dynamical systems characterized by retarded 
interactions and time delays is rapidly developing. New 
concepts have been emerging in the last years together 
with an increasing palette of applications and tools to 
analyze field data. Against this backdrop we present here
a review focusing in particular on recent developments and
readability. Aiming to make the review accessible also to newcomers 
in the field we supplement selected concepts with basic
educational examples.

\subsection{Time delays in theory and nature}

Dynamical systems with time delays are present in many fields \cite{just2010delayed},
including engineering, mathematics, biology, ecology and physics.
Especially well studied are optoelectronic circuits and laser coupled 
systems \cite{erneux2017introduction, soriano2013complex},
which may be considered to be model systems for delayed interactions.
A range of novel phenomena have emerged in the past two decades 
from both extensive theoretical modeling efforts and experimental 
studies. Examples are the implementation of echo-state networks via
the time sequencing of a single non-linear optical element with
time delayed feedback \cite{larger2012photonic}, the optoelectronic
realization of multi-stable delay systems, i.\,e.\ of systems 
with coexisting attractors \cite{houlihan2004experimental},
noise-induced resonances in delayed feedback systems \cite{masoller2002noise},
neuronal oscillations in feedforward delay networks 
\cite{payeur2015oscillatorylike}, and the discovery of
anticipating chaotic synchronization in autonomous
\cite{voss2000anticipating,masoller2001anticipation} and driven systems
\cite{ciszak2003anticipating}.
Delayed feedback is employed moreover for the control of chaotic 
\cite{pyragas1992continuous,scholl2008handbook} and of noise-induced 
dynamics \cite{janson2004delayed}. It has been furthermore shown 
that multistability can arise from delay coupling
\cite{foss1996multistability,foss2000multistability}.

Systems with constant time delays have been especially
well studied, in part due to the precise timing capabilities 
of optoelectronics systems and lasers. Recent work addresses 
also non-constant time delays, which are known to be core
to the dynamics of biological systems \cite{macdonald1978lecture},
such as for the brain \cite{stepan2009delay,rahman2015dynamics,deco2009key},
but which can be relevant also for photonic systems \cite{martinez2015dynamical}.

Turning and milling processes have become alternative prototype systems
for the study of the impact of time delays \cite{insperger2005state,bachrathy2011state},
in particular in relation to the question of how to control nonlinear 
delay systems \cite{hovel2010control}. The vibrations of the tool 
cutting a rotating workpiece during milling can be modeled incorporating
constant time delays \cite{stone2004stability}, time-varying 
delays \cite{otto2013application}, or a retardation depending on 
the state of the workpiece \cite{insperger2005state}, viz of the
dynamical system, with the latter allowing for an efficient
suppression of vibrations \cite{otto2013application}.

For comparatively simple mechanical systems, such as the 
stick-balancing task
\cite{milton2009time,sieber2004complex,campbell2008friction}, the 
influence of different types of delay have been studied extensively.
The analysis of more complex systems, like climate models, for which
the interaction of the atmosphere and the ocean may be characterized 
by distinct types of time-varying and/or state-dependent time delays,
is in contrast substantially more demanding \cite{keane2017climate}.

Besides a variety of new systems and time delay induced phenomena,
novel methods and classification schemes for time delay dynamics 
have been proposed. Examples are partially predictable chaotic 
motion, as it can be found in delayed and classical dynamics systems 
\cite{wernecke2017test}, and a type of laminar chaos inherent to certain
delay systems \cite{muller2018laminar}, with the latter being
closely related to a specific classification of time-varying delays
in terms of conservative and dissipative delays \cite{otto2017universal}.
A novel spatio-temporal representation of delay systems allows
furthermore for an interpretation in analogy to one-dimensional 
spatially extended systems \cite{arecchi1992two}, and as such for 
an intuitive understanding of delayed dynamics 
\cite{yanchuk2017spatio,masoller1997spatiotemporal}.

\subsection{Outline}

For the groundwork we present in Sect.~\ref{sect_state_histories} 
a formal definition of time delay systems, and of the respective 
configuration and phase spaces, which will  be followed in 
Sect.~\ref{sec:lyapunov} by a discussion of the distinct ways 
local and global Lyapunov exponents may be defined for delay 
systems. The introduction then concludes with an educational
analysis of the stability of fixed points in delay systems, for 
which several approaches to evaluate Lyapunov spectra are compared.
Sect.~\ref{sec:types} and \ref{sec:tests} are then devoted 
respectively to comprehensive overviews of the most important 
types of time delay systems and of the dynamics, with the
numerical methods being treated in Sect.~\ref{sec:numerics}.

\subsection{States and state histories}
\label{sect_state_histories}

A comprehensive class of delay differential equations are of the form
\begin{align}
\dot{x}(t)&=F\big(x(t), x(t-\tau)\big)\,,
\label{eq:dde}
\end{align}
where $\tau$ is the delay and $x$ a state in configuration 
space. To simplify the discussion, most definitions and examples 
presented throughout this review are given, as for (\ref{eq:dde}),
for systems characterized by a single scalar variable~$x$ and a
single constant time delay~$\tau$.

The trajectories of a delay differential equation (DDE) such as 
(\ref{eq:dde}) are uniquely defined by their associated initial 
functions $\varphi(t)$ on an initial time interval,
\begin{equation}
x(t) =\varphi(t), \qquad\quad
t\in[t_\text{o}-\tau,t_\text{o}]\,.
\label{eq:initialfunction}
\end{equation}
Delay differential equations (DDE) are, as a consequence, formally 
infinite dimensional. A state in phase space is hence not uniquely
determined by $x=x(t)$, but by the state history
\begin{align}
\mathbf{X}(t)&=\{ x(t^\prime)\}\,,
\qquad\quad t^\prime \in[t-\tau, t]\,.
\label{eq:statehistory} 
\end{align}
In analogy we define the directed distance vector
$\mathbf{d}(t)$ between two state histories 
as
\begin{equation}
\begin{aligned}
 \mathbf{d}(t)&=\mathbf{X}_1(t)-\mathbf{X}_\text{o}(t)
\\ &=\{ x_1(t^\prime)-x_\text{o}(t^\prime)\}\,,
\qquad\quad t^\prime \in[t-\tau, t]\,,
\end{aligned}
\label{eq:disthist} 
\end{equation}
with the respective norm $d=d(t)$ being
\begin{align}
 d(t)&=
\left(
\frac{1}{\tau}\int_{t-\tau}^{t}\!\!\mathrm{d}t^\prime\;
\lvert x_1(t^\prime)-x_\text{o}(t^\prime)\rvert^\gamma
\right)^{1/\gamma}\,, 
\label{eq:distance}
\end{align}
where $\gamma=2$ for a Euclidean metric. For
$\gamma=1$ one has the Manhattan norm, which
corresponds to the average distance between
two trajectories, when averaging over a time
interval $\tau$. We will work here with
a Euclidean space of state histories.

\subsection{Lyapunov exponents}
\label{sec:lyapunov}

The classical definition of Lyapunov exponents,
as established for ordinary dynamical systems,
can be generalized to time delay systems. We 
distinguish here between local Lyapunov exponents~$\Lambda_j\in\mathbb{C}$
\cite{abarbanel1992local,abarbanel1991variation},
which are complex numbers,
and real-valued global Lyapunov exponents~$\lambda_j\in\mathbb{R}$
\cite{eckmann1985ergodic,abarbanel1992local},
among which the largest one, the maximal (global)
Lyapunov exponent $\lambda_\text{max}$ is of particular
interest.
Futher, note that the finite-time Lyapunov exponents
(cf.~Sect.~\ref{sec:benettin}) are different from
local Lyapunov exponents.

\subsubsection{Local Lyapunov exponents}

For the local stability of a DDE~(\ref{eq:dde}) one considers the time evolution
of a small perturbation $\delta$ (see also Sect.~\ref{sec:fixedpoint})
\begin{equation}
 \dot{\delta}(t) = J_\text{o}\,\delta(t)+J_\tau\,\delta(t-\tau)\,.
 \label{eq:stabilityjacobian}
\end{equation}
Here we have denoted with $J_\text{o}$ the instantaneous Jacobian and with 
$J_\tau$ the delayed Jacobian, which are defined by
the partial derivatives of the flow $F$ with respect to the instantaneous and
the delayed state, respectively \cite{lakshmanan2011delay}:
\begin{equation}
 J_\text{o}=
\frac{\partial F\big(x(t),x(t-\tau)\big)}{\partial x(t)}, \quad\qquad
J_\tau=
\frac{\partial F\big(x(t),x(t-\tau)\big)}{\partial x(t-\tau)}\,.
 \label{eq:jacobians}
\end{equation}
Here, the Jacobians $J_\text{o},J_\tau$ are scalar quantities, as we only
consider scalar systems~(\ref{eq:dde}), i.\,e.\ $x\in\mathbb{R}$.
For DDE in $N$ dimensions, the Jacobians are
$N\times N$ matrices.
Note that both Jacobians depend on the actual state $x(t)$ and 
on the delayed state $x(t-\tau)$ of the system.

In the case of ordinary differential equations (ODE), i.\,e.\ 
without delay $\tau=0$, $J_\tau=0$,
the $N$, generally complex eigenvalues 
of the Jacobian $J_\text{o}$ are termed local Lyapunov exponents.
For a one-dimensional ODE,
the instantaneous Jacobian $J_\text{o}\in\mathbb{R}$ coincides
with the only local Lyapunov exponent.
One may study local Lyapunov exponents anywhere in phase space, 
even though they are typically used to classify fixed points 
as foci, saddles and nodes \cite{gros2015complex}.

In order to generalize the concept of local Lyapunov exponents
$\Lambda_j$ for finite delays $\tau>0$, one may approximate
any DDE by a finite-dimensional Euler map (see~Sect.~\ref{sec:eulermapjacobian}).
Then the local Lyapunov exponents $\Lambda_j$ of the DDE
can be estimated at every point in the phase space of the delayed system
from the eigenvalues of the map's Jacobian matrix.
As a special case one may, on the other hand, directly evaluate
the local Lyapunov exponents for the delayed system
at a fixed point of DDE~(\ref{eq:dde})
via a characteristic equation (see~Sect.~\ref{sec:fixedpoint})
\cite{lakshmanan2011delay,abarbanel1992local}.
Note that we do not use the term local Lyapunov exponents
to refer to finite-time Lyapunov exponents
(cf.~Sect.~\ref{sec:global_maximal_Lyapunov_exponents}).

\begin{figure}[t]\centering
\includegraphics[width=0.5\textwidth]{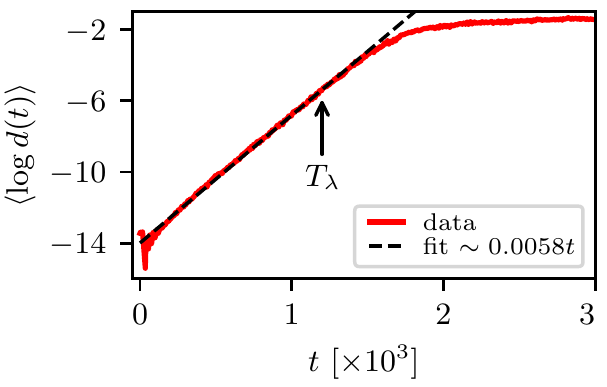}
\caption[Extracting the maximal Lyapunov exponents from the distance of trajectories]
{\label{fig:lyapfromdist}
The logarithmic distance $\log(d)$ between two trajectories
of the Mackey-Glass system (cf.~Sect.~\ref{sec:constdelay}),
for $\tau=17.20$ and averaged over $100$ pairs with initial 
distance $\delta=10^{-6}$.
The slope of the linear fit (dashed line) retrieves the maximal 
Lyapunov exponent $\lambda_\text{max}=0.0058$ as defined by
Eq.~(\ref{eq:maxlyapexp}).
The Lyapunov prediction time~$T_\lambda=1071$,
as defined in Sect.~\ref{sec:lyapunovpredictiontime}, is marked to indicate 
the average time it takes until the distance between a pair of 
trajectories has reached $d=10^{-2}$.
Note that $\log10^{-6}\approx-13.8$ and $\log10^{-2}\approx-4.6$.
}
\end{figure}

\subsubsection{Global and maximal Lyapunov exponents}
\label{sec:global_maximal_Lyapunov_exponents}

An initially small distance $\delta$ between two 
trajectories, as defined by (\ref{eq:distance}), may be 
assumed to evolve exponentially,
\begin{align} 
d(t)&=\delta\;\e^{\lambda_\text{max}t},
\quad\qquad
 \lambda_\text{max}=\lim\limits_{t\to\infty}\lim\limits_{\delta\to0}
\frac{1}{t}\log\left(
\frac{d(t)}{\delta} \right)\,,
\label{eq:maxlyapexp} 
\end{align}
which defines the largest Lyapunov exponent $\lambda_\text{max}$.
Note that the limit of an infinitesimal small initial distance and 
an infinitely long divergence is subject to the constraint that
overall distances are finite for bounded dynamical systems.

The formal definition (\ref{eq:maxlyapexp}) has been extended 
to the more general concept of finite-time Lyapunov exponents
\cite{lapeyre2002characterization,pikovsky2016lyapunov}, and finite-size
Lyapunov exponents \cite{karolyi2010finite}. Lyapunov exponents 
may be extracted directly from data series \cite{abarbanel1992local},
as illustrated in Fig.~\ref{fig:lyapfromdist}, where the initial slope 
of the logarithmic distance $\log(d)$ is used to approximate 
the maximal Lyapunov exponent $\lambda_\text{max}$.

The largest global Lyapunov exponent $\lambda_\text{max}=\lambda_1$ 
captures the rate of divergence in the direction of the fastest 
divergence of trajectories. Further Lyapunov exponents $\lambda_j$,
with $j>1$, describe then the remaining directions. In a system with 
an infinite number of dimensions there is potentially an infinite
number of distinct Lyapunov exponents $\lambda_j$, with the entirety 
being called the Lyapunov spectrum. It is common to order the
exponents by size,
\begin{equation}
\lambda_\text{max}=\lambda_1\geq\lambda_2\geq\ldots\,.
\label{1_order_lambda_j}
\end{equation}
Computationally the Lyapunov spectrum is computed in general 
resorting to Benettin's method
\cite{benettin1980lyapunov,skokos2010lyapunov,wolf1985determining,farmer1982chaotic},
which will be detailed out in Sect.~\ref{sec:benettin}.

\subsubsection{Global Lyapunov exponents for maps}
\label{sec_Lyapunov_maps}

As an alternative to the numerical treatment one may extract 
the Lyapunov spectrum (\ref{1_order_lambda_j}) from the Euler 
map, which we will define in Sect.~\ref{sec:eulermap}.
For this approach one needs to know the time evolution operator 
$M(t)$ explicitly, a precondition holding for the Euler
map and in general for discrete maps, for which $M(t)$ is 
given by a suitable product of the map's Jacobian matrix
\cite{sandor2018world}.

We consider the distance vector $\mathbf{d}_j(t)$ between the 
state histories of two trajectories, $x_\text{o}(t)$ and $x_j(t)$,
where $x_\text{o}(t)$ is a reference orbit. Neglecting mathematical 
subtleties \cite{ott2002chaos,eckmann1985ergodic},
one may assume that the time evolution of $\mathbf{d}_j(t)$ 
is governed by the time evolution operator,
\begin{align}
 \mathbf{d}_j(t)&=M(t)\;\boldsymbol{\delta}_j\,,
\end{align}
where $\boldsymbol{\delta}_j=\mathbf{d}_j(0)$ 
is the vector corresponding to the initial distance 
$\delta$, which we take to be small.
The norm of the distance vector can then 
be expressed with 
\begin{align}
\lVert \mathbf{d}_j(t) \rVert &=
\sqrt{\boldsymbol{\delta}_j^\intercal M^\intercal(t)
M(t)\boldsymbol{\delta}_j}=
\sqrt{\boldsymbol{\delta}_j^\intercal
U(t)\boldsymbol{\delta}_j}
\label{eq:distnorm} 
\end{align}
as a function of the matrix $U(t)=M^\intercal(t) M(t)$, 
where $M^\intercal$ and $\boldsymbol{\delta}_j^\intercal$
are the transpose of $M$, which is a matrix
\cite{shimada1979numerical,sandor2018world}, and 
respectively of $\boldsymbol{\delta}_j$.
With $U$ being real and symmetric, its eigenvalues 
$\alpha_j(t)$ and the corresponding eigenvectors 
$\mathbf{e}_j$ are also real. One has
furthermore that $\alpha_j(t)\ge0$ holds, as
\begin{equation}
\lVert M \mathbf{e}_j \rVert^2
= \mathbf{e}_j^\intercal M^\intercal M \mathbf{e}_j 
= \mathbf{e}_j^\intercal U \mathbf{e}_j 
= \alpha_j \lVert \mathbf{e}_j\rVert^2\,.
\label{1_alpha_positive}
\end{equation}
Choosing the $j$th eigenvector $\mathbf{e}_j$ of $U$ to be 
aligned with the initial distance $\boldsymbol{\delta}_j$ 
one then obtains 
\begin{align}
 \lVert\mathbf{d}_j(t)\rVert&=\sqrt{\alpha_j(t)}\;\lVert
\boldsymbol{\delta}_j \rVert
\end{align}
for the evolution of the distance $\lVert\mathbf{d}_j(t)\rVert$
between two state histories. Using (\ref{eq:maxlyapexp}),
we may then express the $j$th global Lyapunov exponent 
$\lambda_j$ in terms of the $j$th eigenvalue $\alpha_j(t)$ of 
$U(t)$:
\begin{align}
\lambda_j&=\lim\limits_{t\to\infty}\frac{\log\alpha_j(t)}{2t}\,.
\label{eq:deflyap}
\end{align}
This expression is useful when extracting Lyapunov exponents
from the Euler map (cf.\ Sect.~\ref{sec1_euler_map}), as
we will detail out in Sect.~\ref{sec:eulermapjacobian}.
Eq.~(\ref{eq:deflyap}) shows in particular that the spectrum 
of Lyapunov exponents is well defined.

\subsection{Educational example: Stability of a fixed point}
\label{sect_educational_example}

In order to discuss several notions related to the stability
of a fixed point we consider with
\begin{align}
\dot{x}(t)&=-x(t-\tau)
\label{eq:exampledde}
\end{align}
the simplest time delay system \cite{sieber2011characteristic}.
The evolution of small perturbations around $x^*$ are determined 
by the local Lyapunov exponent, which depends in turn on the 
delay time $\tau$. The stability of the fixed point
in terms of the Lyapunov exponent can be evaluated by the standard 
analytic ansatz, as discussed in the following 
Sect.~\ref{sect_analytic_ansatz}, and via the Euler map 
(cf.\ Sect.~\ref{sec1_euler_map}). Numerical methods for the 
evaluation of both the maximal Lyapunov exponent and of the 
Lyapunov spectrum, such as the Benettin method \cite{benettin1980lyapunov},
will be treated later in Sect.~\ref{sect_Lyapunov_exponents}.
Here we will use Benettin's approach for benchmarking.

\begin{figure*}[t]\centering
\includegraphics[width=1.\textwidth]{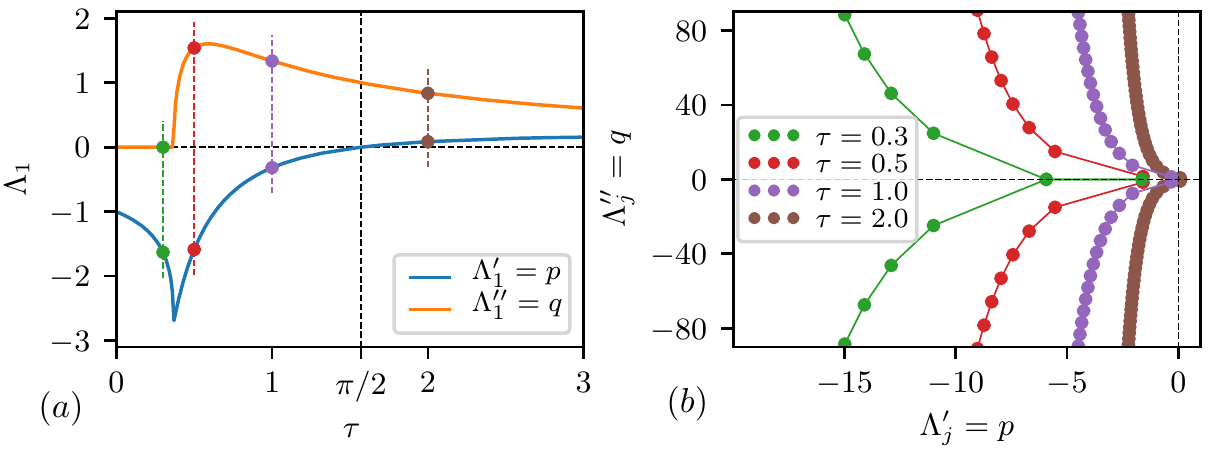}
\caption[Spectrum of local Lyapunov exponents from an analytic ansatz]
{\label{fig:exansatz}
The spectrum of local Lyapunov exponents
$\Lambda_j=\Lambda_j^{\prime}+\imath \Lambda_\text{j}^{\prime\prime}$ 
for the DDE~(\ref{eq:exampledde}) at the fixed point $x^*=0$
in terms of the roots of (\ref{eq:l2cond}).
Note that the spectrum is countably infinite.
($a$) Real and imaginary part (shown is one of the two branches) 
of the largest exponents $\Lambda_1$ as a function of the delay $\tau$.
The real part changes sign at $\tau=\pi/2$ (vertical dashed line,
cf.~Eq.~(\ref{eq:char_p_0})).
Bullets and vertical lines indicate the maximal local Lyapunov exponent
of the spectra shown in the right-hand panel (colors matching).
($b$) The imaginary part $\Lambda_\text{j}^{\prime\prime}$ 
as a function of the real part $\Lambda_j^{\prime}$.
Lines are guides to the eye.
}
\end{figure*}

\subsubsection{Analytic ansatz for local Lyapunov exponents}
\label{sect_analytic_ansatz}

Close to the fixed point $x^*=0$ the dynamics of (\ref{eq:exampledde}) 
can be approximated by the exponential ansatz $x(t)\propto\exp(\Lambda t)$ 
for the complex local Lyapunov exponents
$\Lambda_j=\Lambda_j^{\prime}+\imath \Lambda_j^{\prime\prime}\to\Lambda=p+\imath q$,
where we drop the index and denote with $p=\Lambda_j^{\prime}$
the real part and with $q=\Lambda_j^{\prime\prime}$ the imaginary 
part (cf.\ Sect.~\ref{sec:fixedpoint}).
The characteristic equation is consequently \cite{just2000eigenvalue}
\begin{align}
\Lambda &= -\e^{-\Lambda\tau},
\qquad\quad
\Lambda=p+\imath q\,,
\label{eq:chareq}
\end{align}
which can be separated into a real and an imaginary part:
\begin{equation}
p = -\e^{-p\tau}\cos(q\tau),
\qquad\quad
q =  \e^{-p\tau}\sin(q\tau)\,.
\label{eq:char_eq_p_q}
\end{equation}
This equation has, as a graphical inspection shows, an infinite
number of solutions, which we may order with respect to the
real part: $p_1\ge p_2\ge \dots$. The fixed point is stable 
when $p_1<0$, viz when $\cos(q_1\tau)>0$. The transition occurs, 
as shown in Fig.~\ref{fig:exansatz}, for
\begin{equation}
p_1=0, \quad\quad
q_1=1, \quad\quad
\tau=\pi/2, \quad\quad
q_1\tau=\pi/2\,, 
\label{eq:char_p_0}
\end{equation}
viz when the time delay $\tau$ starts to be out-of-phase 
with the period $2\pi/q$ of the Lyapunov oscillation.
Eliminating $p$ from (\ref{eq:char_eq_p_q}) one obtains the
transcendental equation
\begin{align}
q &= \e^{q\tau/\tan(q\tau)}\sin(q\tau),
\qquad\quad
p = - q/\tan(q\tau)\,,
\label{eq:l2cond}
\end{align}
for the imaginary part $q$ of the local Lyapunov exponent.
Note that (\ref{eq:l2cond}) has a countable but infinite
number of roots, the local Lyapunov spectrum, which can be found 
numerically, e.\,g., via bisection.

For any solution of Eq.~(\ref{eq:l2cond}) with non-vanishing imaginary
part $q\neq0$ there exists a complex conjugate solution -- a necessary
condition when $x=x(t)$ is real.
Thus, the Lyapunov spectrum is symmetric with respect to
the sign of the imaginary part, viz when interchanging
$q\leftrightarrow (-q)$.
In Fig.~\ref{fig:exansatz} the numerical solution of 
Eq.~(\ref{eq:l2cond}) for different values of the delay 
time $\tau$ are given.

All roots have negative real parts, $p<0$, when the delay 
is small, viz when $\tau<\pi/2$. The fixed point $x^*=0$ is then
attracting. Above the transition $\tau=\pi/2$ at least one
Lyapunov exponent is positive, with the number of positive
exponents increasing with increasing delay $\tau$. The
fixed point is then repelling.

\begin{figure*}[t]\centering
\includegraphics[width=1.\textwidth]{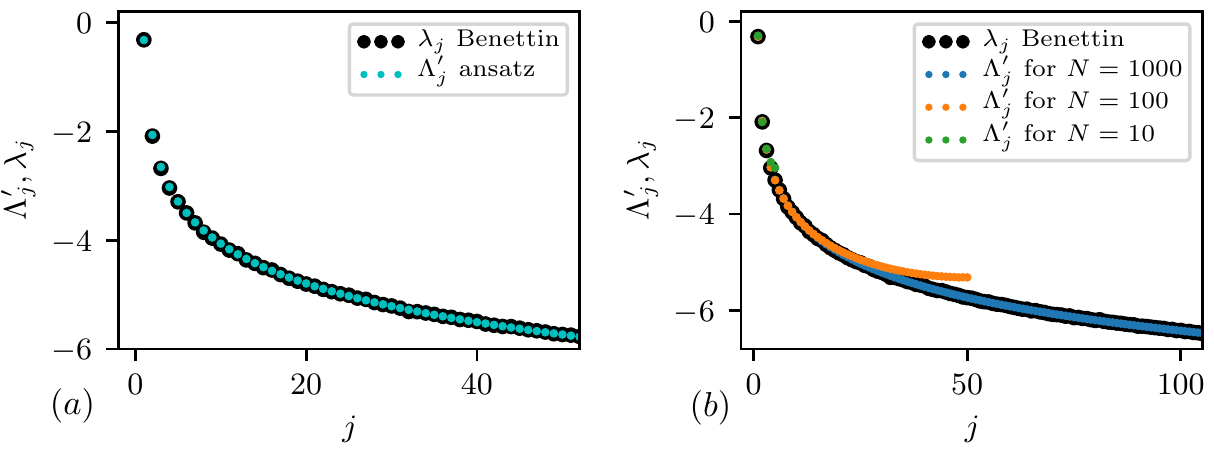}
\caption[Comparison of different methods for computing Lyapunov exponents]
{\label{fig:exspectra}
A comparison of methods determining the stability of the fixed point
$x^*=0$ of the DDE~(\ref{eq:exampledde}) for $\tau=1$
(showing only values for non-negative imaginary parts
$\Lambda_j^{\prime\prime},\lambda_j^{\prime\prime}\geq0$).
($a$) The real part $\Lambda_j^\prime$ 
of the exponents solving (\ref{eq:chareq}) are in agreement with the Lyapunov exponents 
$\lambda_j$ computed with Benettin's method (black bullets, cf.~Sect.~\ref{sec:benettin}).
($b$) The real part $\Lambda^\prime_j$ of the (local) Lyapunov exponents,
as estimated from the eigenvalues of the $N\times N$ Jacobian (\ref{eq:exjac})
of the Euler map (colored dots), 
in comparison with the Lyapunov exponents computed with Benettin's 
method (black bullets).
The agreement improves rapidly with increasing resolution $N$ of the Euler map.
}
\end{figure*}

In Fig.~\ref{fig:exspectra} the real part 
$\Lambda_j^{\prime}$ of the roots of (\ref{eq:l2cond}) 
is shown in comparison with the Lyapunov exponents 
$\lambda_j$ obtained numerically using Benettin's
approach (cf.\ Sect.~\ref{sec:lyapunov}). One finds
point per point agreement.

When the delay vanishes $\tau\to0$ the DDE~(\ref{eq:exampledde})
turns into an ordinary differential equation (ODE) and the dimensionality of
the system reduces from an infinite number of dimension to one dimension.
In consequence the spectrum of local Lyapunov exponents~$\Lambda_j$ collapses
onto a single exponent $\Lambda_1=\partial \dot{x}/\partial x=-1$,
which approaches its value from below when decreasing the delay.
The real parts of the rest of the spectrum diverges with the second largest
local Lyapunov exponent $\lim_{\tau\to0}\Lambda_2=-\infty$
leading to a compactification of dimensions.

\subsubsection{Euler map}
\label{sec1_euler_map}

One may discretize time, such that the delay interval $\tau$ 
is subdivided into $N-1$ segments of length $\varDelta t$, 
as described in Sect.~\ref{sec:eulermapjacobian}. A
DDE is such transformed to a discrete map, the Euler map.

For Eq.~(\ref{eq:exampledde}) the $N\times N$ Jacobian matrix 
of the Euler map is given by
\begin{align}
 J&=\left(
  \begin{array}{cccccc}
   -\varDelta t & 0            & \multicolumn{2}{c}{\cdots}   & 0 & 1\\
   -\varDelta t & -\varDelta t & 0 & \cdots & 0 & 1\\
    \vdots      &              & \ddots &\ddots&\vdots& \vdots \\
    \vdots      &              & \ddots &\ddots&0& \vdots \\
   -\varDelta t & \multicolumn{2}{c}{\cdots}   &  & -\varDelta t & 1\\
   -\varDelta t & \multicolumn{3}{c}{\cdots}  & -\varDelta t & 1-\varDelta t
  \end{array}
 \right)\,,\label{eq:exjac}
\end{align}
where the steps size $\varDelta t=\tau/(N-1)$ depends 
on the resolution $N$. From the $N$, in general complex 
eigenvalues $\sigma_j$ of (\ref{eq:exjac}), one can estimate 
the real parts $\Lambda^\prime_j$ of the $N$ largest (local)
Lyapunov exponents of (\ref{eq:exampledde}).
For this purpose one uses the relation (cf.~Sect.~\ref{sec:eulermapjacobian})
\begin{equation}
\lVert \sigma_j\rVert^2 = (\sigma_j^{\prime})^2+
(\sigma_j^{\prime\prime})^2 \to \e^{2\Lambda^\prime_j\tau}
\quad \text{ for }N\to\infty\,
\label{eulerMap_e_lambda}
\end{equation}
for the modulus of complex numbers, 
which follows from (\ref{eq:deflyap}).
From the relation of the complex eigenvalue $\sigma_j$ and
the complex local Lyapunov exponent~$\Lambda_j$,
\begin{equation}
 \sigma_j^\prime+\sigma_j^{\prime\prime}\to
  \exp\Big(\tau\big( \Lambda_j^\prime +
  \imath \Lambda^{\prime\prime}_j
 \big)\Big)\,,
\end{equation}
one can also extract the imaginary part $\Lambda_j^{\prime\prime}$
modulo $2\pi/\tau$ (cf.~Sect.~\ref{sec:eulermapjacobian}).

Fig.~\ref{fig:exspectra} shows the results for $\tau=1$ and 
a series of $N$, in comparison to the Lyapunov exponent obtained
with the Benettin method (cf.\ Sect.~\ref{sec:benettin}). The largest
Lyapunov exponents are approximated well even for a limited 
resolution $N\sim10$.


\begin{figure*}[t]\centering
  \includegraphics[width=\textwidth]{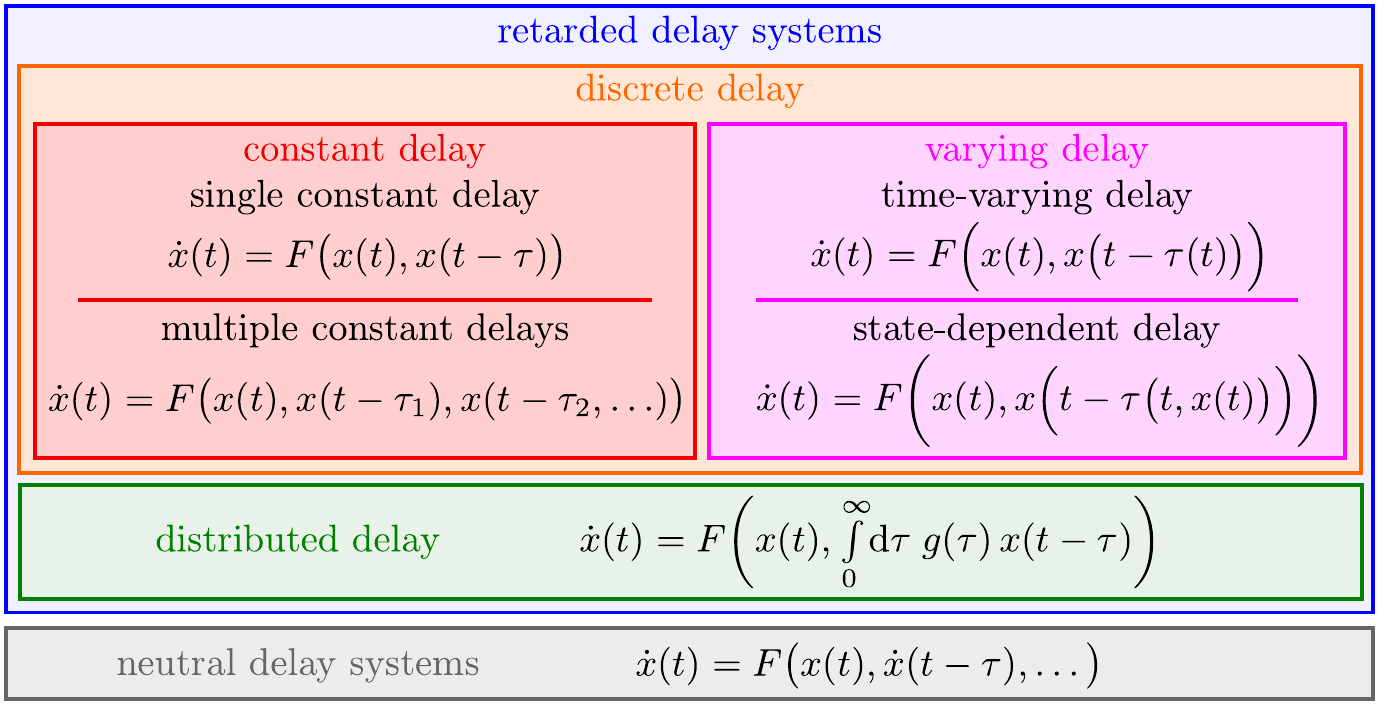}
  \caption[Overview and categorization of different time delays]
  {\label{fig:overview}
    Overview and categorization of the different time delays discussed in 
Sect.~\ref{sec:types}.
  }
\end{figure*}

\section{Types of time delay systems}
\label{sec:types}

A large class of delay differential equations (DDE) take the form
of a continuous-time dynamical system of the type
\begin{align}
\dot{x}(t) &= F\big( x(t), \alpha\big)\,,
\label{eq:ddegeneral}
\end{align}
where $x(t)\in\mathbb{R}$ denotes the state of the system parameterized 
by the time $t$. The flow $F\,:\,\mathbb{R}\times\mathbb{R}\to\mathbb{R}$ 
depends on the current state $x(t)$ and on a delay function $\alpha\in\mathbb{R}$,
which we will specify later on for the distinct types of time delays 
For simplicity the DDE~(\ref{eq:ddegeneral}) is chosen to be scalar and
the flow to be autonomous, with the latter implying that $F$ is not an 
explicit function of time. A summary of the most important types
time delays is presented in Fig.~\ref{fig:overview}.

\subsection{Single constant time delay}
\label{sec:constdelay}

The simplest but non-trivial delay function $\alpha=x(t-\tau)$ 
incorporates a single constant time delay $\tau>0$ \cite{gros2015complex}.
The corresponding DDE depends then on a single past state:
\begin{align}
\dot{x}(t) &= F\big(x(t), x(t-\tau)\big)\,. 
\label{eq:ddesingleconst}
\end{align}
An example for this type of DDE has been discussed previously, 
see Eq.~(\ref{eq:exampledde}). 

Single constant time delays are experimentally realized in optical 
laser systems \cite{williams2013synchronization}, where they
can be used to generate chaotic communication \cite{vanwiggeren1999chaotic},
that is communication channels suitable for private communication
\cite{mensour1998synchronization}. Examples of theoretical 
investigations using this type of DDE include 
the modeling of traffic dynamics by car-following models 
\cite{jiang2001full} and word recognition with time delayed 
neural networks \cite{lang1990time}. 

A possible reference system for a DDE is the limit of
vanishing time delay, viz the case $\alpha\to0$ 
in (\ref{eq:ddegeneral}). Systems with stable 
instantaneous evolution will become unstable, as
illustrated in Fig.~\ref{fig:exansatz}, when the 
length $\tau$ of the time delay becomes larger than the 
time scale of the instantaneous dynamics \cite{gros2015complex}.
This observation has led to the suggestions that
modern democracies may be generically unstable 
\cite{gros2017entrenched}. The instability would
result in this context from the growing mismatch 
between the ongoing acceleration of the instantaneous 
political dynamics, as defined by the time scale of 
opinion swings, and a delayed feedback that is entrenched 
in the election cycle.

A reference example for a DDE with a delay induced instability 
is the Mackey-Glass system \cite{mackey1977oscillation}:
\begin{align}
\dot{x}(t)&=\frac{a\,x(t-\tau)}{1+\big(x(t-\tau)\big)^c}-b\,x(t)~.
\label{eq:mackeyglass}
\end{align}
The typical choice for the parameters, $a=0.2$, $b=0.1$ and $c=10$,
ensures that the trivial fixed point $x=0$ is unstable for
all time delays \cite{farmer1982chaotic} and that the non-trivial
fixed point $(a/b-1)^{1/c}=1$ is stable for small time delays.
Increasing the time delay $\tau$ one observes first periodic oscillations 
and then a transition to chaos \cite{farmer1982chaotic}. Originally 
designed to describe the production of blood cells, the Mackey-Glass 
is now considered a standard example of deterministic chaos 
\cite{tel2006chaotic}, for which it is widely used for bench marking results
\cite{marsden1993interdisciplinary,wolf1985determining,grassberger1983measuring}.
The Mackey-Glass system will serve in this review as a reference system 
for the discussion of chaos, as presented in Sect.~\ref{sec:tests}.

\begin{figure*}[t]\centering
\includegraphics[width=0.9\textwidth]{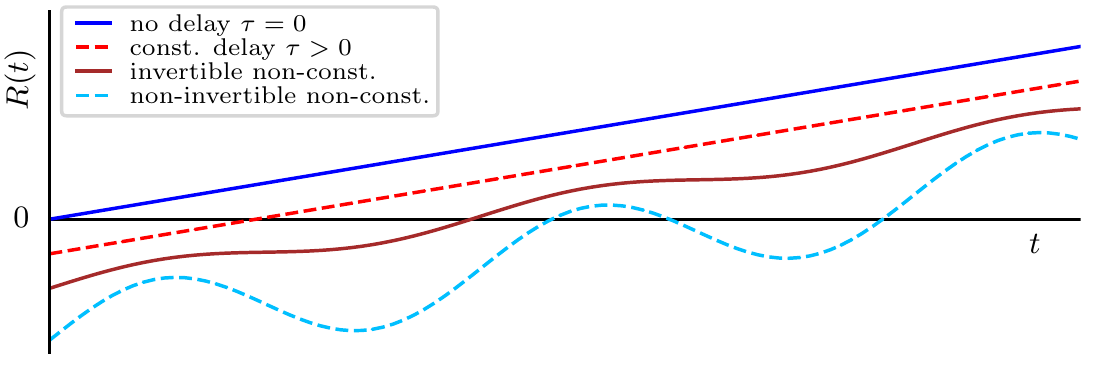}
\caption[Access functions corresponding to different time delays]
{\label{fig:accessFunction}
Illustration of selected continuous access functions 
$R(t)=t-\tau(t)$. An access function is non-invertible if $\dot{R}<0$.
}
\end{figure*}

\subsection{Multiple constant time delays}
\label{sec:multipleconst}

For systems with multiple constant time delays $\tau_1,\tau_2,\ldots>0$ the
delay function $\alpha=\alpha(x(t-\tau_1),x(t-\tau_2),\ldots)$ depends on several
corresponding past states.
Multiple constant time delays are used to study, e.\,g., synchronization properties
in heterogeneous networks \cite{otto2018synchronization,masoller2005random}.
Experimentally systems with multiple constant time delays are realized in coupled
optoelectronic oscillators \cite{williams2013synchronization}, where the
combination of different time delays is used to create states of full or
partial synchronization. In a modified Stuart-Landau model 
\cite{stuart1960non} two distinct time delays induce instabilities that 
exhibit spatio-temporal pattern formation and turbulence 
\cite{yanchuk2014pattern}.
In time delay systems with state-switching the dynamics becomes 
more robust to noise, when two distinct time delays are incorporated
\cite{kyrychko2018enhancing}.

The destabilization of a stationary state in systems with multiple constant
delays can happen via different types of bifurcations
\cite{shayer2000stability}. It has been shown \cite{ahlborn2004stabilizing}, 
on the other hand, that multiple time delay feedback may suppress chaotic 
dynamics in Chua's circuit \cite{chua1993universal}. We note that
chaos can be suppressed quite in general by stabilizing fixed points
or by inhibiting noise modulations~\cite{jaurigue2016suppression}.
In addition we mention that an increase of the time delay leads
to an improvement in the performance in act-and-wait feedback systems
\cite{insperger2006act}.

\subsection{Time-varying delay}
\label{sec:timevardelay}

For time-dependent non-constant time delays $\tau=\tau(t)$ 
the delay differential equation reads
\begin{align}
\dot{x}(t)&=F\Big(x(t),x\big(R(t)\big)\Big)\,,
\qquad\quad
R(t)=t-\tau(t)\,,
\end{align}
where we have defined with $R(t)$ the access function
(or access map \cite{otto2017universal}). Discontinuous
or non-invertible access functions are generically not 
considered.
Periodically varying delays
\cite{ghosh2007synchronization,muller2018laminar}, like
a sinusoidal variation
\begin{align}
 \tau(t)&=\tau_\text{o}+A\sin(\omega t) \label{eq:tddtperiodic}
\end{align} 
with mean $\tau_\text{o}$ and amplitude $A$, become
non-invertible whenever $|A\omega|>1$. An example
is shown in Fig.~\ref{fig:accessFunction}. Periodic
time delays may be used to stabilize systems that are 
strongly chaotic in the limit of fixed time delay, viz 
when $A\to0$. In general, a periodically varying delay 
is incorporated to non-linear delayed feedback in order 
to study the effect on synchronization \cite{senthilkumar2007delay} 
or on chaotic behavior \cite{madruga2001effect}. Implemented 
in electronic circuits, periodically varying delay have been
shown to stabilize unstable orbits \cite{jungling2012experimental}.

\begin{figure*}[t]\centering
 \includegraphics[width=1.0\textwidth]{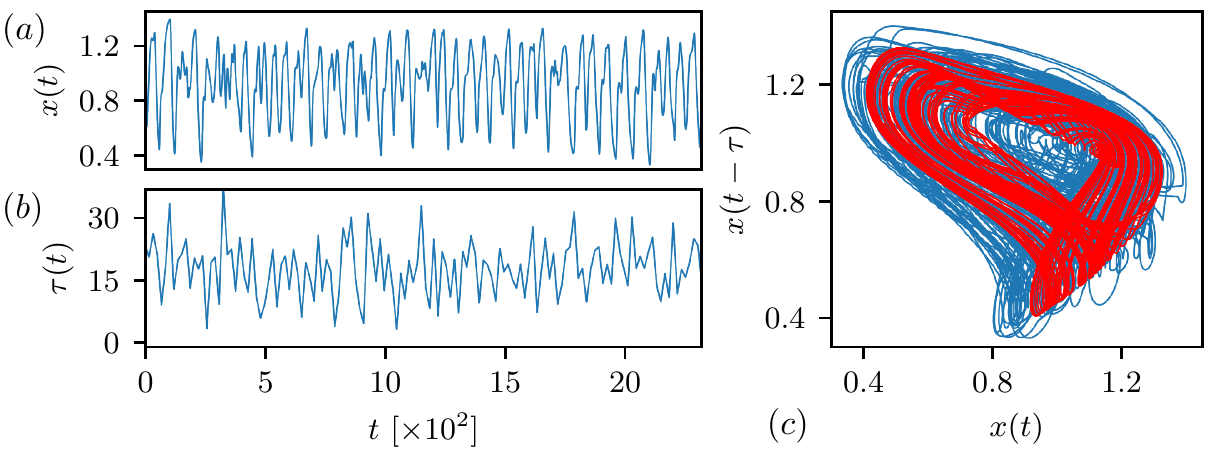}
\caption[Stochastically time-varying delay]
{\label{fig:stochasticvartd}
The Mackey-Glass system~(\ref{eq:mackeyglass})
with stochastically varying time delay (\ref{eq:tddtstochastic}).
($a$)~Chaotic time series $x(t)$ and ($b$) stochastic time delay $\tau(t)$
with mean $\mu_\tau=\tau_\text{o}=17.20$ and variance $\sigma_\tau=7$
over time.
($c$)~Attractors in the $x(t)$, $x(t-\tau)$ projection with varying delay
(blue, cf. panels ($a,b$)) and fixed time delay $\tau=\tau_\text{o}$ (red).
Figure replicated from~\cite{kye2004characteristics}.
}
\end{figure*}

The dynamics of stochastically varying 
time delays \cite{kye2004characteristics},
\begin{align}
 \tau(t)&=\tau_\text{o}+\int\limits_0^t\mathrm{d}t^\prime\,\xi(t^\prime)\,,
\label{eq:tddtstochastic} 
\end{align}
can be characterized on the other hand only by statistical
distributions. The stochastic process is described by the
random variable $\xi(t)$ generating
the noise distribution. Noise may prevent the collapse
of phase-space trajectories onto simple manifolds
\cite{kye2004characteristics,mensour1998synchronization}, as
observed regularly for systems with fixed time delays 
(cf.\ Sect.~\ref{sec:ppc} on partially predictable chaos).
This effect is illustrated in Fig.~\ref{fig:stochasticvartd} 
for the Mackey-Glass system~(\ref{eq:mackeyglass}). 
Stochastically time-varying delays are used in control schemes 
for communication networks \cite{nilsson1998stochastic} and for 
tuning fuzzy PID controllers \cite{pan2011tuning}. It has been 
shown moreover that the distribution of stochastically 
varying delay has an impact on the stability of the dynamics
\cite{yue2009delay}.

In systems with digital controllers the controlled signals are measured
at discrete times and with finite precision, a strategy called digital 
sampling~\cite{isermann2013digital}. Modern control systems belong 
mostly to this class of time delay systems.
With digital sampling the state of the system is detected 
with a certain sampling period, inducing a time-dependent delay between the
controlled system and the digital controller, which may in turn be expressed 
in terms of a discrete mapping~\cite{haller1996micro}.
For systems with differential control it has been shown that digital sampling
can exhibit micro-chaos \cite{haller1996micro}, which manifests 
itself as chaotic vibrations on comparably small length scales in the
controlled system. Micro-chaos can be permanent or appear 
transiently \cite{csernak2016multi}.

\subsection{State-dependent delay}
\label{sec:statedependent}

\begin{figure*}[t]\centering
\includegraphics[width=\textwidth]{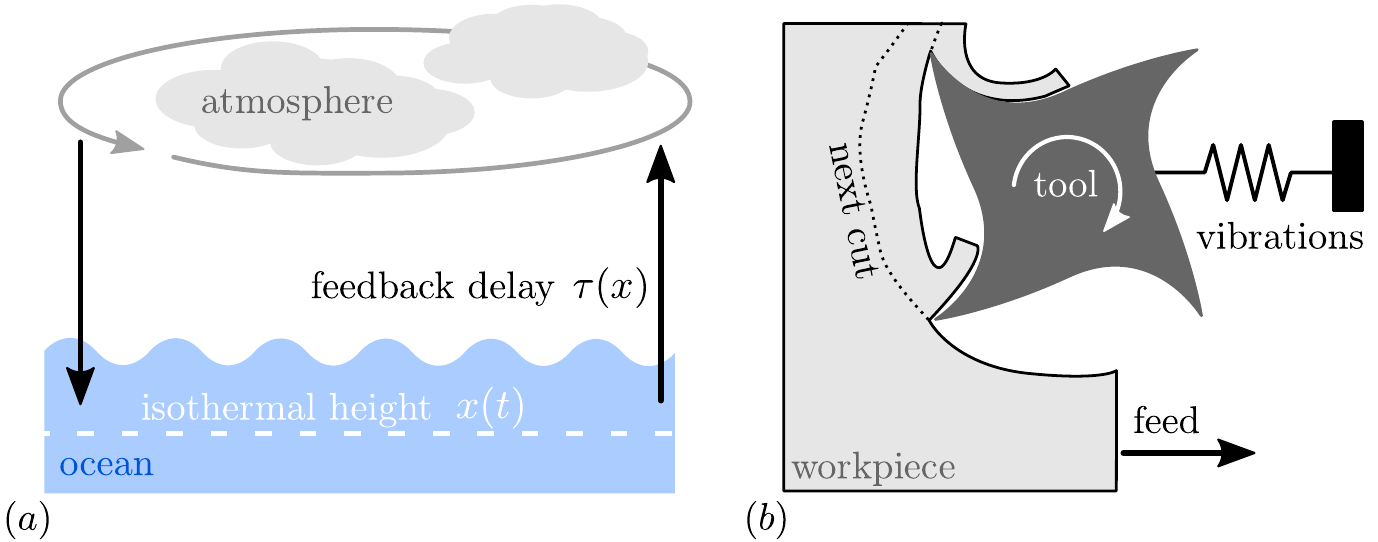}
\caption[Examples of systems with state-dependent time delay]
{\label{fig:statedepdelay}
Examples of systems with state-dependent time delay.  
($a$) Sketch of the interacting dynamics between the ocean
and the atmosphere. The delay in the feedback
mechanism depends on the internal state of the atmosphere or the ocean,
e.\,g. on the isothermal height $x(t)$ of the ocean.
Figure inspired by \cite{keane2017climate}.
($b$) Mechanical model for describing the cutting processes 
with a vibrating tool. The edge cut by the tool into the 
workpiece depends on the edge of the previous cut, which 
makes the delay state dependent.
Figure inspired by \cite{insperger2005state}.
}
\end{figure*}

The feedback mechanism generating time delays in physical systems
may depend on the state of the system itself \cite{hartung2006functional}.
A non-constant state-dependent time delay $\tau(t,x(t))$,
\begin{align}
\dot{x}(t)&=F\bigg(x(t),x\Big(t-\tau\big(t,x(t)\big)\Big)\bigg)
\end{align}
may then result. This type of time delay can be considered as 
an additional dimension to the dynamical system, adding further 
to the complexity.

In the DAO (Delayed Action Oscillator) paradigm of the ENSO 
(El Ni$\tilde{\mbox{n}}$o Southern Oscillation) climate model 
\cite{keane2017climate}, the delay induced by the mutual feedback 
mechanism of ocean and atmosphere depends on the physical state 
of either part of the system, see Fig.~\ref{fig:statedepdelay}.

Time delay systems with state-dependent delays are employed in control 
tasks, such as the balancing of an inverted pendulum with a PD controller
\cite{sieber2004complex}, or when modeling milling
processes \cite{insperger2005state} (cf.\ Fig.~\ref{fig:statedepdelay}).
Due to the vibrations of workpiece and tool, the chip thickness and shape of
each cut of the tool depends on the previous cut, which makes milling processes
with vibrations \cite{insperger2005state}, and turning processes
\cite{insperger2008criticality}, prototype systems for state-dependent delays.
Besides numerical simulations only few universal analytic results, 
such as a rigorous theory for linearizing state-dependent DDE
\cite{gyori2007exponential}, are known for state-dependent delays.

\begin{figure*}[t]\centering
\includegraphics[width=1\textwidth]{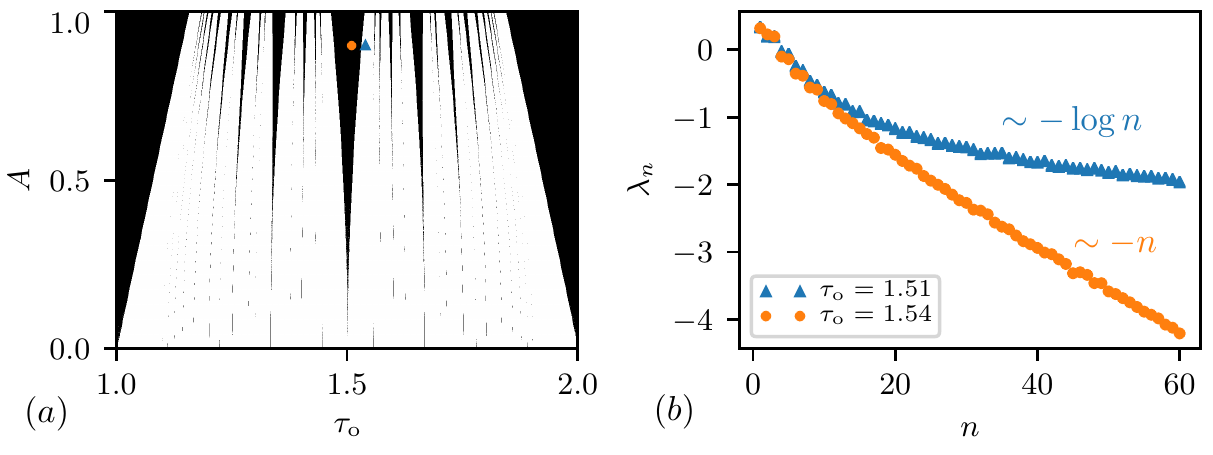}
\caption[Periodically varying time delay being conservative or dissipative]
{\label{fig:dissipativedelay}
($a$)~Parameter space of a periodically varying time delay 
(\ref{eq:tddtperiodic}) with amplitude $A$ and mean $\tau_\text{o}$. 
In the black regions the delay is dissipative, whereas it is
conservative in the white regions. The two regions in parameter space
are separated by fractal Arnold tongues.
($b$)~Lyapunov spectra $\lambda_n$ of an attractor in a time delay systems
with periodically varying time delay (\ref{eq:tddtperiodic})
(cf.~Sect.~\ref{sec:laminarchaos}, see \cite{muller2017dynamical,otto2017universal}).
Shown is the spectrum for a conservative (blue triangles, $\tau_\text{o}=1.51,A=0.9$)
and a dissipative (orange bullets, $\tau_\text{o}=1.54,A=0.9$) time delay.
For the first the asymptotic scaling of the spectrum is logarithmic
$\lambda_n\sim-\log n$, for the latter it is linear $\lambda_n\sim-n$.
Figures replicated from~\cite{muller2017dynamical}.
}
\end{figure*}

\subsection{Conservative vs.\ dissipative delay}
\label{sec:dissipativedelay}

It has been proposed that state-dependent time delays
may be classified to be either conservative or
dissipative \cite{muller2017dynamical,otto2017universal}. 
For invertible access maps $R(t)=t-\tau(x(t),t)$
(see Fig.~\ref{fig:accessFunction}), a
transformation $\Phi(t)=\varphi$ of the
time scale $t\to\varphi$ leads to a 
corresponding transformation of the access function
$R(t)\to \tilde{R}(\varphi)$:
\begin{align}
\tilde{R}(\varphi)&=\Phi\,R(t)\,\Phi^{-1},
\qquad\quad
t\to\varphi = \Phi(t)\,.
\end{align}
If the transformed access map $\tilde{R}(\varphi)=\varphi-\tau_\varphi$ 
is equivalent to the access map for a constant delay $\tau_\varphi>0$, then 
the delay is considered to be conservative 
\cite{otto2017universal}, otherwise it is said to be dissipative.
Conservative time delays are known under various names in different 
fields: Within engineering conservative delays are called variable 
transport delays
\cite{bresch2016implicit,otto2017transformations}, whereas they are
referred to as threshold delays in biological systems 
\cite{kuang1993delay,mahaffy1998hematopoietic}.

For periodically varying time delay (cf.~Eq.~(\ref{eq:tddtperiodic})),
the dissipative and conservative regions in parameter space are
fractionally divided by Arnold tongues (cf.~Fig.~\ref{fig:dissipativedelay}).
The mapping of time instances $t_n$ defined by the access function $R(t)$
\begin{align}
 t_{n+1}&=R(t_n)=t_n-\tau(t_n)\label{eq:accessmap}
\end{align}
is equivalent to a circle map \cite{muller2017dynamical},
when using sinusoidally varying time delays (\ref{eq:tddtperiodic}),
with dissipative time delays corresponding to chaotic behavior 
of the circle map (\ref{eq:accessmap}).

Conservative systems, which are equivalent to systems with a constant
time delay \cite{otto2017transformations}, tend to be less complex 
than dissipative systems, for which a new type of chaotic motion,
laminar chaos \cite{muller2018laminar}, has been found. See
Sect.~\ref{sec:laminarchaos}. The two classes differ furthermore
with respect to the scaling of the Lyapunov spectrum, which we will
define in Sect.~\ref{sec:lyapunov}. The well studied logarithmic 
scaling of the Lyapunov exponents $\lambda_n\sim-\log n$
for $n\to\infty$ holds for conservative delays
\cite{farmer1982chaotic}, as depicted in Fig.~\ref{fig:dissipativedelay}.
For dissipative delays a linear scaling $\lambda_n\sim-n$
is observed in contrast~\cite{otto2017universal}.

\subsection{Distribution of delays}
\label{sec:distributed}

\begin{figure*}[t]\centering
\includegraphics[width=\textwidth]{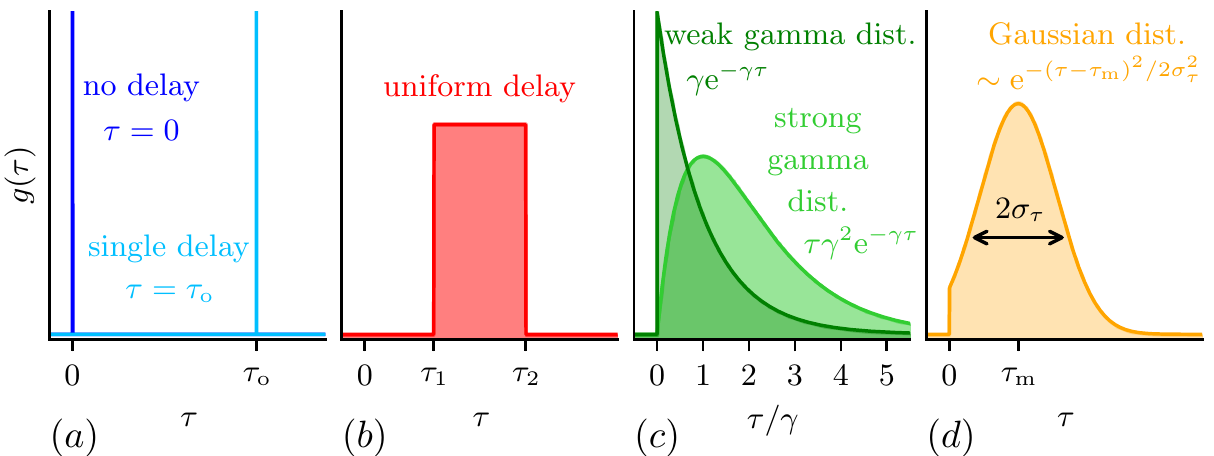}
\caption[Distributions of time delays]
{\label{fig:distrdelay}
  Different distributions $g(\tau)$ of delays $\tau\geq0$.  
  ($a$)~Dirac delta delay distributions are equivalent to single 
  constant delays.
  ($b$)~Uniform distribution of delay in the range $[\tau_1,\tau_2]$ with
  height $1/(\tau_2-\tau_1)$.  
  ($c$)~Two examples of gamma distributions
  characterized by the parameter $\gamma>0$: the weak gamma distribution,
  which decays exponentially, and the strong gamma distribution, as
  characterized by a pronounced contribution around $\tau=1/\gamma$.
  ($d$)~A Gaussian with mean $\tau_\text{m}$ and variance
  $\sigma_\tau^2$.
}
\end{figure*}

The discussion concerned hitherto discrete delays, 
that is systems for which the evolution of the current
state is influenced by distinct instances of the past. 
This is a valid approximation for, e.\,g., optical systems, 
for which there is only little variation of the delay.
However, biological \cite{macdonald1978lecture} and social
\cite{gros2017entrenched} systems may be described more
accurately by time delays that are drawn from a probability 
distribution $g(\tau)$,
\begin{align}
 g(\tau)\geq0 \,,\qquad \int\limits_0^\infty\mathrm{d}\tau\;g(\tau)&=1
\,,\qquad g(\tau\!<\!0)=0 \,,
\end{align}
of delays. The distribution vanishes for the sake of causality 
for negative delays. 

A distribution of time delays may enter in two ways. 
For the first possibility the dynamics as such is averaged
over the distribution of time delays:
\begin{equation}
\dot x(t) = \int\limits_0^\infty \mathrm{d}\tau \, g(\tau)
F\big(x(t), x(t-\tau)\big)\,.
\label{eq:ddedistr_F} 
\end{equation}
For a non-linear bare flow $F$ the delay differential equation is 
in this case not of the form given by Eq.~(\ref{eq:ddegeneral}).

A more common way to incorporate a distribution
of delays is to assume that the dynamics is
influenced solely by a weighted average $\alpha$ of
past states:
\begin{align}
\dot{x}(t) &= F\big(x(t),\alpha\big),
\qquad\quad
\alpha = \int\limits_0^\infty\!\mathrm{d}\tau\,g(\tau)\,x(t-\tau)\,.
\label{eq:ddedistr} 
\end{align}
The two approaches, (\ref{eq:ddedistr_F}) and
(\ref{eq:ddedistr}), coincide for linear dynamics.
The standard stability analysis of fixed points 
(cf.\ Sect.~\ref{sect_analytic_ansatz})
can be carried out also
for distributed time delays \cite{gros2017entrenched},
with the investigation being particularly straightforward
for a $g(\tau)$ which can be Laplace-transformed 
analytically~\cite{rahman2017aging}.

A selection of delay distributions is presented in 
Fig.~\ref{fig:distrdelay}. Dirac delta functions
correspond to fixed time delays and uniform time
delay distributions to the flat average over 
a past time interval $[t-\tau_1,t-\tau_2]$ 
(cf.~Fig.~\ref{fig:distrdelay}).
The latter has been employed for describing the
aging transition of a delay coupled network 
of oscillators \cite{rahman2017aging} and for the
delayed influences within advanced political systems.
\cite{gros2017entrenched}. Note, that standard mode 
decomposition (cf.~\cite{amann2007some}) may be 
used also for solving linear delayed dynamical systems with 
uniformly distributed delays \cite{rene2017mean}.

Distributions from the family of gamma distributions 
$g_{\gamma p}(\tau)\sim \tau^{p-1}\e^{-\gamma \tau}$ 
with parameters $\gamma$ and $p$ are typically chosen 
for their favorable analytic tractability \cite{hogg1995introduction}.
Two prominent examples, which have been employed to model 
biological systems \cite{rahman2015dynamics,busenberg1982use}, 
are depicted in Fig.~\ref{fig:distrdelay}.
For $p=1$, the weak limit, the gamma distribution 
$g_{\gamma 1}(\tau)$ corresponds to a pure exponential 
decay.
Systems with weakly gamma distributed delays can be reduced to
systems without delay (cf.~Sect.~\ref{sec:reducibledelay}).
The strong gamma distribution $g_{\gamma 2}$ for $p=2$
has in contrast a maximum around $\tau\sim1/\gamma$,
decaying thereafter exponentially. 

For a delay that varies 
randomly around a given mean, with mean $\tau_\text{m}$ 
and variance $\sigma_\tau^2$, a Gaussian distribution
$g_\text{G}(\tau)\sim\exp\big(-(\tau-\tau_\text{m})^2/2\sigma_\tau^2\big)$
is a suitable choice \cite{payeur2015oscillatorylike}
(cf.~Fig.~\ref{fig:distrdelay}). One may use
alternatively, in particular for the description of
neural systems, distributions of time delays that are
motivated by experiments \cite{payeur2015oscillatorylike}.

\subsection{Reducible time delay systems}
\label{sec:reducibledelay}

A large class of time delay differential systems can be 
characterized by a time delay function $\alpha$, viz they 
are of the type
\begin{align}
\dot{x}(t) & = F\big(x(t),\alpha\big)~.
\label{eq:dot_x_closed}
\end{align}
Examples are a single time delay, $\alpha = x(t-\tau)$,
time varying time delays, $\alpha = x(t-\tau(t))$, 
state dependent time delays, $\alpha = x(t-\tau(t,x(t)))$
and distributions of time delay, $\alpha = \int \mathrm{d}\tau g(\tau)x(t-\tau)$
as discussed respectively in Sect.~\ref{sec:constdelay},
\ref{sec:timevardelay}, \ref{sec:statedependent} and
\ref{sec:distributed}. 

We have seen in Sect.~\ref{sec:dissipativedelay}, that
it is sometime possible to find a transformation
between distinct types of delay functions $\alpha$,
which become then equivalent. Conservative 
time delays are in this framework equivalent to constant 
time delays.
For time delays that are distributed according to a distribution from the
family of gamma distributions \cite{smith2011distributed}
(cf.~Sect.~\ref{sec:distributed}, Fig.~\ref{fig:distrdelay}~($c$))
an even stronger reduction occurs, for which
the time evolution of the corresponding delay function $\alpha=\alpha(t)$
can be written in closed form
as \cite{otto2017transformations,worz1978global} 
\begin{align}
\dot{\alpha}(t) & = G\big(x(t),\alpha(t)\big)~.
\label{eq:dot_alpha_closed}
\end{align}
The equations of motion for the pair of variables
$\{x(t),\alpha(t)\}$ is manifestly closed in
terms of a system of ordinary differential equations (ODE),
when (\ref{eq:dot_alpha_closed}) holds together
with (\ref{eq:dot_x_closed}).

Systems for which (\ref{eq:dot_alpha_closed}) holds
are called reducible time delay systems
\cite{busenberg1982use,worz1978global}. As an example 
of a reducible system consider the linear DDE 
\cite{lenhart1985stability,cohen1979stable}
\begin{equation}
\dot{x}(t)=F\big(x(t),\alpha(t)\big),
\qquad\quad
\alpha(t)= \gamma\int\limits_0^\infty \mathrm{d}\tau\,\e^{-\gamma\tau}x(t-\tau)\,,
\end{equation}
where the delay function $\alpha$ is given by an exponentially distributed
average over past states. Taking the derivative of $\alpha$,
interchanging $\partial/\partial t$ with $-\partial/\partial \tau$
in the integral, and integrating in part, one obtains the closed form
\begin{align}
\dot{\alpha}(t)&=\gamma \big(x(t) -\alpha(t)\big)\,.
\end{align}
This reduction of a time delayed system to a system of coupled ODEs
is also called the linear chain trick
\cite{smith2011introduction,macdonald1978lecture}.
Note that the argument of $x=x(t)$ on the right-hand side
does not contain a time delay. Averaging over past states 
corresponds in this case to a dramatic dimensionality reduction, 
namely to the reduction of a formally infinite-dimensional delay
system to a 2-dimensional system of ordinary differential equations.
As a corollary we point out that there is no chaos in Mackey-Glass 
systems, see Eq.~(\ref{eq:mackeyglass}), with exponentially
distributed delay functions.

\subsection{Neutral delay systems}
\label{sec:neutral}

The delay systems discussed so far where functionally dependent
on past states. Systems of this type are called 
retarded delay systems. The delay may enter however 
also via a higher order derivative \cite{kuang1993delay}, 
f.\,i.\ via a first-order time derivative:
\begin{align}
 \dot{x}(t)&=F\big(x(t),\dot{x}(t-\tau_1),x(t-\tau_2)\big)\,.
\label{eq:ddeneutral}
\end{align}
The corresponding system is considered in this case to be 
neutral \cite{hale2013introduction, hovel2010control}.
Neutral delay differential equations (NDDE),
such as the neutral delay logistic equation
\cite{gopalsamy1988neutral}, occur in population dynamics
\cite{kuang1993delay}, where they describe, e.\,g., ecological 
systems with feedback mechanisms.

The analytic and numerical treatment of neutral delay systems is 
substantially distinct from that of retarded delay systems. 
Stability criteria \cite{wu2004new,he2004delay} and the concept 
of Lyapunov stability \cite{he2005augmented} needs to be adapted
in particular (cf.~Sect.~\ref{sec:lyapunov}). Leaving these
interesting questions apart, we will focus for the remainder
of this review on retarded delay systems.

\subsection{Networks with delay coupling}

\begin{figure*}[t]\centering
 \includegraphics[width=\textwidth]{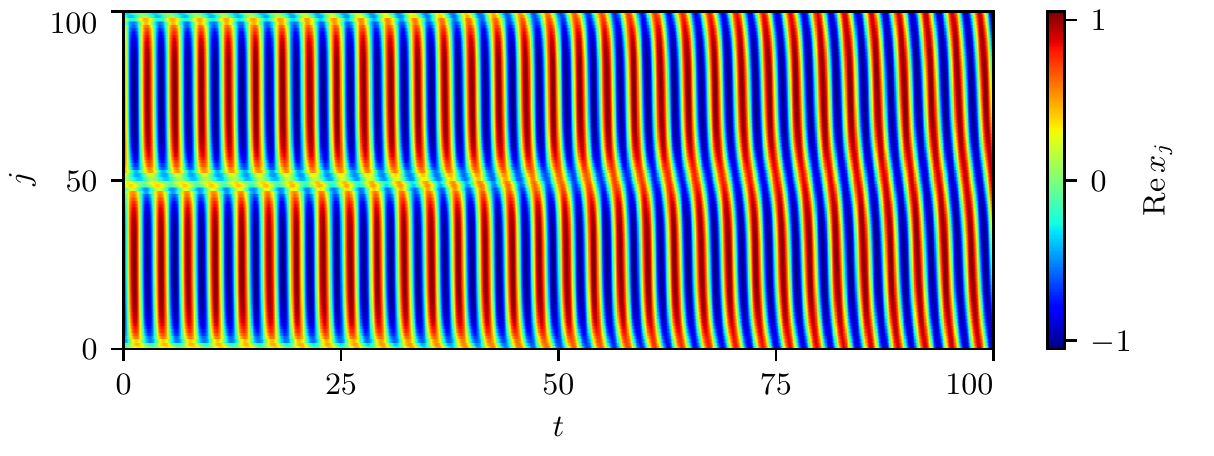}
 \caption[Chimera death in a delay-coupled ring of oscillators]
{\label{fig:chimeracontrol}
Real part $\operatorname{Re}x_j$ (indicated by color) of $100$ delay-coupled
Stuart-Landau oscillators on a ring indexed by $j$ over time $t$.
The coupling takes into account the ten nearest neighbors of an oscillator and
acts with a constant time delay $\tau=\pi$ which is equivalent to
the natural frequency $\omega=2$ of the oscillator.
Initially the oscillators show a transient chimera state,
i.\,e.\ a state in which a fraction of the oscillators is synchronized and 
another fraction is not synchronized.
The chimera disappears with increasing time and the systems ends up in a
state where all oscillators are synchronized with a phase-lag (chimera death).
Figure replicated from \cite[Fig.~3]{gjurchinovski2017control}.
}
\end{figure*}

Transmission delays are common in physical networks, 
where they may impact synchronization processes 
of functionally similar constituting units \cite{englert2010zero}.
Examples are optical systems \cite{soriano2013complex} and
gene expression networks \cite{tiana2013dynamics}.
In reaction-diffusion systems, such as the Gray-Scott model
\cite{kyrychko2009control,baba2002giant}, delays impact
the occurrence of self-organized spatio-temporal patterns.
For an overview of delay-coupled systems see \cite{flunkert2013dynamics}.
The synchronization of networks with delay coupling has been addressed
with a special focus on distributed delays \cite{kyrychko2014synchronization},
observing death and birth regions of amplitude synchronization \cite{atay2003distributed}.
Also, a general criterion for the synchronization of delay-coupled networks based
on the networks topology has been derived \cite{flunkert2010synchronizing}.

In neural networks with time delay couplings \cite{campbell2007time}, 
the synchronization of neurons may be studied with diffusive or 
with pulse-like delay couplings \cite{rossoni2005stability}. 
The delayed feedback of neural activity to the network has been
shown to be able to suppress noise induced dynamics and thus 
to stabilize brain activity \cite{masoller2008interplay}.
The type of delay, and its spatial distribution, have 
in general a pronounced influence on network activity
\cite{petkoski2016heterogeneity}.

From a more abstract perspective, the effect of time delay 
couplings on oscillatory systems has applications for 
control problems \cite{zakharova2013time}, as realizable 
in electric circuits \cite{reddy2000experimental}.
Chimera states are observed in this kind of delay-coupled
oscillatory networks \cite{gjurchinovski2017control,larger2013virtual},
that is states for which a finite fraction of the oscillators 
is synchronized, while the rest is fully desynchronized, 
i.\,e.\ chaotic (cf.~Fig.~\ref{fig:chimeracontrol}).
The interplay between the inherent dynamics of the network 
units and the delayed feedback can be used both
to stabilize partially synchronized states 
\cite{scholl2016synchronization}, and to control the 
lifetime of chimeras \cite{bohm2015amplitude}.
In optical systems delay coupling can give rise to two-dimensional chimeras
and soliton solutions \cite{brunner2018two}.

Another application of delay coupling is the realization of reservoir computing
networks \cite{maass2002real}, which are closely related to so-called echo state
networks \cite{jaeger2001echo}.
It has been shown that the time delayed feedback of a single optical unit allows
information processing in a reservoir like manner
\cite{appeltant2011information,larger2012photonic}.

An externally driven system is considered consistent
\cite{uchida2004consistency,oliver2015consistency,jungling2018consistency},
if the system produces the same output, when presented with a certain input,
independently of the initial internal state of the system.
The concept of consistency is therefore an important feature for information
processing networks. Further, the concept is closely related to synchronization
of chaotic units in a network \cite{jungling2018consistency}.
Consistency has been achieved with the help of time delay coupling for reservoir
computing networks \cite{nakayama2016laser} and other optical networks
\cite{oliver2015consistency,jungling2018consistency}.

\subsection{Long time delays}

Time delays are considered long if they act on a substantially 
longer time scale than the internal dynamics. This is the case, 
e.\,g.\ for coupled optical systems, when the optical feedback 
via fiber transmission is slower than the dynamics of the lasers
\cite{otto2012delay,soriano2013complex}. Time delays may hence
induce an additional time scale. In control theory 
\cite{albertos2009robust,camacho2007some}, long delays have a 
significant impact on stability regulation, with the consequence that 
the motion resulting from controlling the balance of an inverted 
pendulum differs qualitatively for short and long time 
delays \cite{milton2009time}.

Regarding the stability analysis of systems with long delays,
an equivalence between the dynamics in the vicinity of a 
fixed point and a generalized reaction diffusion process
has been worked out \cite{yanchuk2015spectrum}. In the
asymptotic limit, $\tau\to\infty$, the Lyapunov spectrum
may be rescaled by $1/\tau$, in terms of the real part,
with the resulting rescaled asymptotic spectrum being
continuous \cite{lichtner2011spectrum,yanchuk2015dynamical,pazo2010characteristic}. 
The stability of fixed points and limit cycles becomes in this 
sense independent of the exact value of the delay in the 
long-delay limit \cite{sieber2013stability,d2014stochastic}.



\section{Characterizing the dynamics of time delay systems}
\label{sec:tests}

We start with some preliminary remarks regarding the notation
used for the subsequent discussion of a range of approaches 
and measures that identify and describe regular and chaotic 
dynamics in time delay systems.

\subsection{Fixed points}
\label{sec:fixedpoint}

Fixed point attractors often constitute the starting point
when analyzing the dynamics of a time delay systems. The
entire state history collapses, with (\ref{eq:dde}) 
reducing to
\begin{align}
F(x^*,x^*)&=0,\qquad\quad
x^*= x(t)=x(t-\tau)\,.
\label{eq:fpcondition}
\end{align}
Linear DDE, like (\ref{eq:exampledde}), have the trivial
fixed point $x^*=0$, the Mackey-Glass system~(\ref{eq:mackeyglass})
the fixed point $x^*=1$ (for $a=2b$ and $c>0$).

For a standard stability analysis \cite{gros2015complex} one
considers a perturbation $\delta(t)$ to a given trajectory 
$x(t)$. For the DDE~(\ref{eq:dde}) one obtains
\begin{align}
 \frac{\mathrm{d}}{\mathrm{d}t}\Big(x(t)+\delta(t)\Big)&=
F\Big( x(t)+\delta(t), x(t-\tau)+\delta(t-\tau) \Big)\,,
\end{align}
which leads to Eq.~(\ref{eq:stabilityjacobian}) when 
expanding the flow $F$ into a first-order Taylor expansion
around the fixed point solution $x(t)\equiv x^*$.
Eq.~(\ref{eq:stabilityjacobian}) is itself a delay differential
equation. For a further treatment the state history of the 
perturbation $\delta(t)$ needs to be known on a time interval 
$[t-\tau,t]$, which is however normally not the case.

In the vicinity of a fixed point $x^*$ one can however assume
that the perturbation evolves exponentially,
$\delta(t)=\delta(0)\,\e^{\Lambda t}$, as characterized by 
the complex local Lyapunov exponent $\Lambda$ (cf.~Sect.~\ref{sec:lyapunov}).
The time evolution 
(\ref{eq:stabilityjacobian}) of the perturbation
reduces then to
\begin{equation}
\dot{\delta}(t)=J_\text{eff}\,\delta(t)\,,
\quad J_\text{eff}=J_\text{o}(x^*,x^*)+\e^{-\Lambda t}J_\tau(x^*,x^*)\,,
\label{eq:perturbationeffective}
\end{equation}
with an effective Jacobian $J_\text{eff}$ \cite{lakshmanan2011delay}.
Applying (\ref{eq:perturbationeffective}) to the exponential 
ansatz for the perturbation, one obtains the characteristic
equation
\begin{equation}
 \Lambda=J_\text{o}+\e^{-\Lambda t}J_\tau
 \label{eq:characteristicgeneral}\,,
\end{equation}
which is a transcendental equation solved by infinitely many 
local Lyapunov exponents
$\Lambda=\Lambda_j$.
A special case of (\ref{eq:characteristicgeneral})
is discussed in Sect.~\ref{sect_educational_example}.

The perturbation $\delta(t)$ lives in the $N\to\infty$
dimensional phase space of states histories. The
flow around a fixed point is governed therefore by
the local Lyapunov exponents $\Lambda_k$,
\begin{align}
x(t)=x^*+\delta(t),\quad\qquad
\delta(t)= \sum\limits_{k=1}^\infty c_k\,\e^{\Lambda_k t}\,,
\label{eq:anaansatz}
\end{align}
where $c_k$ and $\Lambda_k$ are complex (cf.\ Sect.~\ref{sect_analytic_ansatz}). 
For real states $x=x(t)$, as assumed here, the local Lyapunov exponents
come in complex conjugate pairs $\bar{\Lambda}_k=\Lambda_{k^\prime}$
whenever the imaginary part is non-zero.

Ordering the exponents $\Lambda_k$ with respect to the magnitude 
of the real part we have
\begin{equation}
\operatorname{Re}\Lambda_1\geq
\operatorname{Re}\Lambda_2\geq
\operatorname{Re}\Lambda_3\geq \ldots\,,
\label{2_Lambda_ordered}
\end{equation}
with the largest value $\operatorname{Re}\Lambda_1$
determining the stability of the fixed point. The
steady state solution $x=x^*$ is stable for
$\operatorname{Re}\Lambda_1<0$, and unstable otherwise.

\subsection{Types of chaotic motion}

Deterministic chaos \cite{tel2006chaotic, lorenz1963deterministic} can
be classified along a series of distinct criteria, which are not necessarily
mutually exclusive. This is in particular true for some
recent classifications schemes discussed in this section, which 
describe in part different features of chaotic motion.

\begin{figure*}[t]\centering
\includegraphics[width=0.70\textwidth]{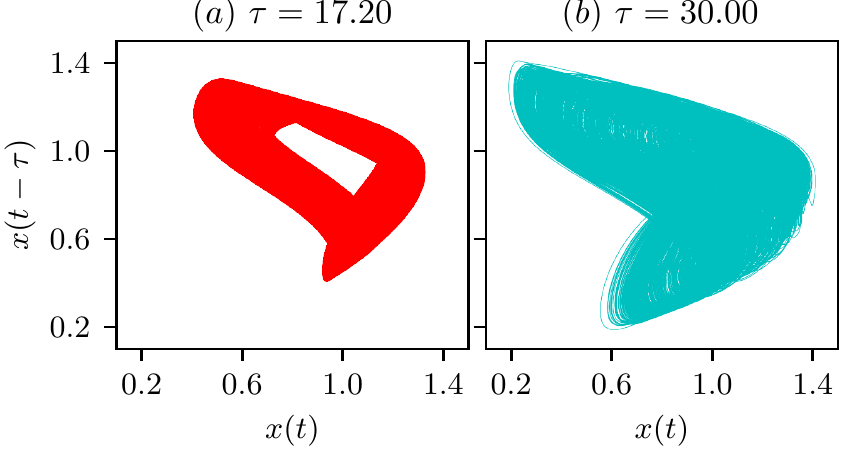}
\caption[Chaotic and hyperchaotic attractors of the Mackey-Glass systems]
{\label{fig:mgchaos}
Trajectories of chaotic attractors in the Mackey-Glass
system~(\ref{eq:mackeyglass}) in the stroboscopic projection 
$x(t)$-$x(t-\tau)$, for ($a$) $\tau=17.20$ and ($b$) $\tau=30.00$,
both sampled over $t\in[0,5\cdot10^4$].
}
\end{figure*}

\subsubsection{Delay induced chaos}
\label{sec:delayinduced}

Stable fixed points and limit cycles existing in the limit 
$\tau\to0$ are necessarily destabilized by a Hopf bifurcation
when increasing $\tau$ continuously \cite{gros2015complex,farmer1982chaotic}. 
The local Lyapunov exponents then first become complex.
Once the time delay~$\tau$ becomes large enough to be out of phase with the 
period $2\pi/\operatorname{Im}\Lambda_1$ of the oscillation,
a perturbation~$\delta(t)$ can increase in a self-reinforcing manner
(cf.~Fig.~\ref{fig:exansatz}).

This mechanism is well documented for the Mackey-Glass 
system (\ref{eq:mackeyglass}), for which a series of 
period doubling bifurcations leads to chaotic dynamics 
\cite{mackey1977oscillation,wei2007bifurcation}. 
The respective route to delay-induced chaos has been 
observed experimentally for a catalytic reaction 
\cite{khrustova1995delay}. We note, however, that
the limit cycle appearing beyond the first Hopf 
bifurcation may remain stable \cite{gros2017entrenched},
even tough it is non unexpected that chaos will eventually
show up, given that DDEs are formally infinite dimensional.

In Fig.~\ref{fig:mgchaos} we present the trajectories of 
two chaotic attractors of the Mackey-Glass system by a 
stroboscopic projection (cf.~Sect.~\ref{sec:poincare}). 
We will show in the next Section that these two 
attractors differ qualitatively in terms their 
cross-correlation functions.

\begin{figure*}[t]\centering
\includegraphics[width=1\textwidth]{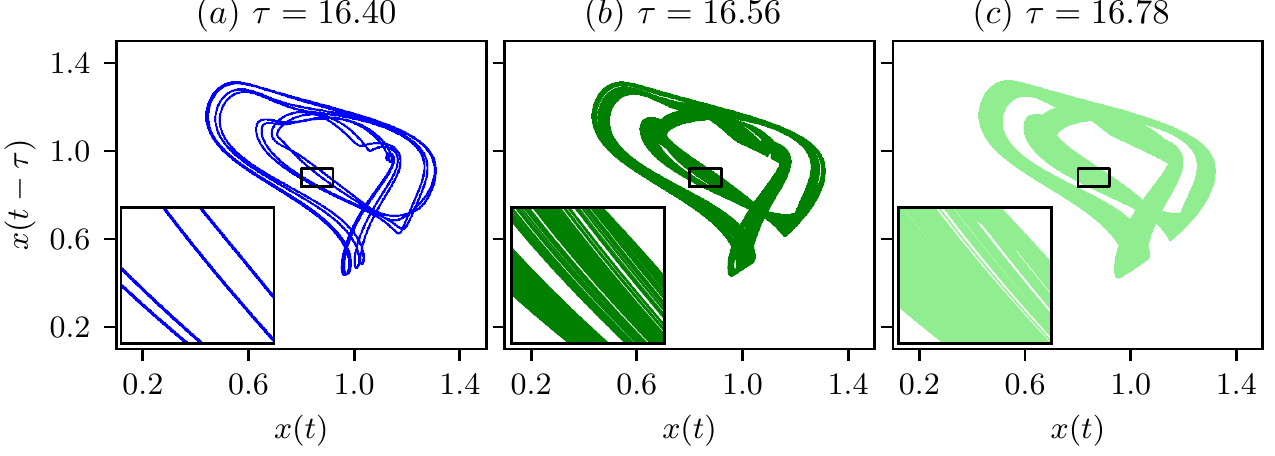}
\caption[Periodic and partially predictable chaotic (PPC) attractors of the 
Mackey-Glass system]
{\label{fig:mgppc}
Stroboscopic projection of attractors of the Mackey-Glass
system~(\ref{eq:mackeyglass}) for different values of the time delay $\tau$
and sampled over $t\in[0,5\cdot10^{4}]$; insets magnifying the indicated area.
($a$)~A limit cycle for $\tau=16.40$ with $8$ windings per period.
($b$)~A partially predictable chaotic (PPC) attractor for $\tau=16.56$ with
visible gaps in the fractal braid. ($c$)~A partially predictable chaotic (PPC)
attractor for $\tau=16.78$ with one extended fractal braid winding twice per
period.
}
\end{figure*}

\subsubsection{Partially predictable chaos}
\label{sec:ppc}

Chaotic attractors may fill a substantial part of the 
phase space, forming in this way a fractal structure 
(cf.~Fig.~\ref{fig:mgchaos}). On the other hand, 
one can observe chaotic attractors differing in shape 
overall only slightly from a periodic orbit
\cite{suzuki2016periodic,li2014dynamic,wernecke2018attractor}.
Such kind of attractors are also found for the Mackey-Glass
system~(\ref{eq:mackeyglass}), as presented in Fig.~\ref{fig:mgppc} 
in comparison with a regular limit cycle. The insets
magnifying the selected parts of the respective trajectories
show that the chaotic attractors consist of a fractal braids
with either a coarser structure, including gaps of
all sizes, or fine fractal filaments.

The difference between the chaotic attractors shown 
in Figs.~\ref{fig:mgchaos} and \ref{fig:mgppc} 
can be quantified by the cross correlation 
\begin{align}
 C(t)&=\Big\langle \big(x_\text{o}(t)-\mu\big)\big(x_1(t)-\mu\big) \Big\rangle/\sigma^2
\label{eq:correlation}
\end{align}
of a pair of trajectories $x_\text{o}(t)$ and $ x_1(t)$ in the 
vicinity of an attractor with mean $\mu$ and variance $\sigma^2$. 
Included in (\ref{eq:correlation}) is an average over respectively
initial conditions (for ordinary differential equations) and
initial functions (for delay systems), as indicated by
$\langle\cdot\rangle$. For delay systems one needs to average 
(\ref{eq:correlation}) in addition over a delay interval,
viz to add an integral $\int_{t-\tau}^t\mathrm{d}t^\prime(\ldots)$, as in the 
definition (\ref{eq:distance}) for the distance $d(t)$ 
between two state histories.

The cross-correlation is related via
\begin{align}
1-C(t)=d^2(t)/(2\sigma^2)
\label{eq:corrdist}
\end{align}
to the distance $d(t)$ between the two trajectories 
\cite{wernecke2017test}, where $d(t)$ is either
the instantaneous distance (for ordinary differential equations),
or the distance between state histories defined
by Eq.~(\ref{eq:distance}).

A pair of trajectories is initially maximally correlated, 
in the sense that $C(t\!=\!0)\to1$, when the initial 
distance of state histories $d(t\!=\!0)=\delta$ is 
small with respect to the extent $\sigma$ of the
attractor, viz when $\delta\ll\sigma$. This is 
clearly true independently of the type of the 
attractor under consideration. Inter-trajectory 
correlations are retained in the long-term limit 
$t\to\infty$ for regular motion, that is, e.\,g., for fixed points 
and limit cycles, but fully lost for chaotic attractors
\cite{wernecke2017test}:
\begin{equation}
 \lim\limits_{t\to\infty}C(t)\ =\ \left\{ 
\begin{array}{cl} 
1 & \text{regular motion} \\  
0 & \text{chaotic attractor} \end{array}
\right.\,.
\label{eq:corrcond} 
\end{equation}
The long-term limit is usually approximated by the Lyapunov 
prediction time $T_\lambda$ (cf.~Sect.~\ref{sec:lyapunovpredictiontime}),
which is inversely proportional
to the maximal Lyapunov exponent (cf.~Sect.~\ref{sec:lyapunov}),
since $T_\lambda$ provides an estimate for the time needed for 
the exponential divergence of two trajectories to become sizable.
 
\begin{figure*}[t]\centering
\includegraphics[width=1\textwidth]{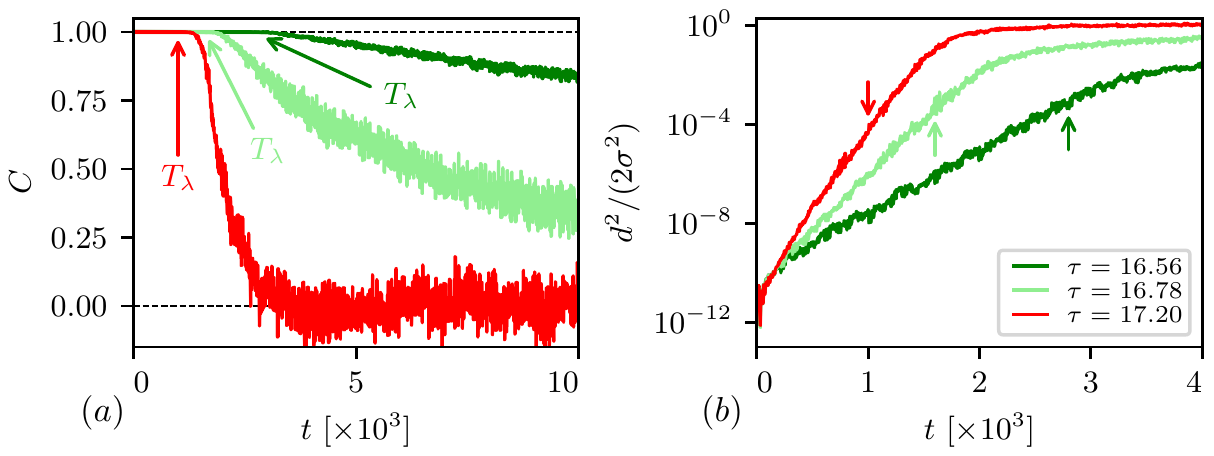}
\caption[Cross-correlation of chaotic and PPC attractors over time]
{\label{fig:correlation}
The cross-correlation $C$ of initially close ($\delta=10^{-6}$) 
pairs of trajectories for the chaotic attractors of the
Mackey-Glass system shown in Figs.~\ref{fig:mgchaos} 
and \ref{fig:mgppc}
(dark green: $\tau=16.56$, $T_\lambda=2791$;
light green: $\tau=16.78$, $T_\lambda=1588$;
red: $\tau=17.20$, $T_\lambda=1071$).
An average over $100$ initial functions has been performed,
with the arrows indicating the corresponding Lyapunov prediction 
times $T_\lambda$ (cf.~Sect.~\ref{sec:lyapunovpredictiontime}).
Shown is ($a$) the cross-correlation (\ref{eq:correlation}) over
time and ($b$) a semi-log plot of the initial 
exponential divergence of the corresponding distance 
$d(t)$. Full decorrelation, which occurs
strictly only in the long-term limit $C(t\!\to\!\infty)\to0$,
is sizable for classical chaos (red) for $t\approx T_\lambda$. 
The final decorrelation is much slower for partially predictable chaos 
(PPC, green).
}
\end{figure*} 

In Fig.~\ref{fig:correlation} the cross-correlation $C$ 
for the chaotic attractors from Figs.~\ref{fig:mgchaos} 
and \ref{fig:mgppc} is plotted over time, with the arrows 
indicating the respective Lyapunov prediction times $T_\lambda$. 
Also presented in Fig.~\ref{fig:correlation} is the distance 
$d^2/(2\sigma^2)$ in a semi-log plot that amplifies the 
initial exponential divergence of the two trajectories.

For some attractors the exponential initial decorrelation is followed 
by a second slower phase of linear decorrelation.
The latter is due to diffusive motion of trajectories on the chaotic
attractor along the braid tracing the formerly stable
limit cycle \cite{wernecke2017test}.

\begin{itemize}
\item For the chaotic attractor with $\tau=17.20$ the exponential and
      diffusive loss of correlation happen on the same time scale,
      leading to an essentially fully uncorrelated motion when the 
      Lyapunov prediction time $T_\lambda$ is reached. See 
      Fig.~\ref{fig:correlation}.
      We term this type of behavior `classical chaos'.
\item For $\tau=16.56$ and $\tau=16.78$ only the exponential initial
      decorrelation occurs within the Lyapunov prediction time, with 
      the subsequent diffusive loss of correlation taking orders of 
      magnitudes longer. This leads to a high residual correlation even 
      after comparably long times $t\gg T_\lambda$, which implies
      that long-term coarse-grained predictions remain possible. This type 
      of behavior has been denoted `partially predictable chaos' (PPC)
      \cite{wernecke2017test}. 
\end{itemize}
The distinction between classical and partially predictable chaos 
in terms of the cross-correlation function is
\begin{align}
 C(t\!\gg\!T_\lambda)\quad
\left\{ \begin{array}{rl} 
=\!0 & \text{classical chaos}\\
>\!0 & \text{partially predictable chaos (PPC)} 
\end{array} \right.\,.
\end{align}
%

\begin{figure}[t]\centering
\includegraphics[width=0.5\textwidth]{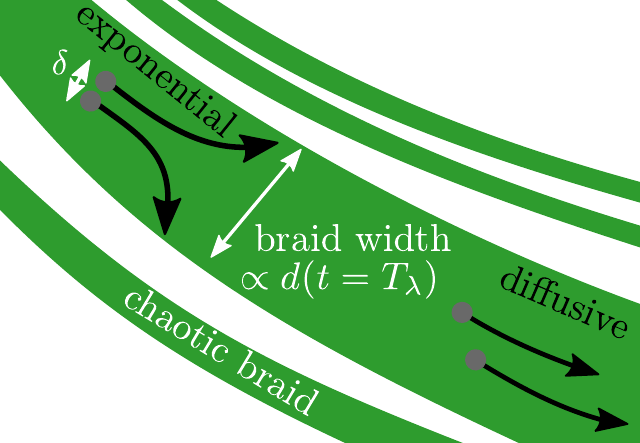}
\caption[Time-scale separation in the separation of trajectories within a chaotic braid]
{\label{fig:sketch_ppc}
Sketch of the time-scale separation characterizing the 
divergence of trajectories for partially predictable chaos (PPC). 
Two initially close states $d(t\!=\!0)=\delta$ (gray bullets) 
diverge exponentially leading to an exponential drop of correlation 
(cf.~Fig.~\ref{fig:correlation}).
This process is limited by the width of the braid (green), such 
that by the Lyapunov prediction time $d(t\!=\!T_\lambda)$ the
distance is of the order of the braid width. The residual cross-correlation 
is lost subsequently due diffusive motion of trajectories along the fractal braid.
 }
\end{figure}

The time scale separation between exponential and diffusive 
decorrelation in PPC is closely related to the topology of 
the chaotic braids, as evident from the insets of Fig.~\ref{fig:mgppc}.
The initial exponential divergence occurring mainly perpendicular 
to a braid is limited by the braid width (cf.\ Fig.~\ref{fig:sketch_ppc}),
which is therefore related to the distance $d(t\!=\!T_\lambda)$ 
of two trajectories after the Lyapunov prediction time. Distinct
fractal braids are on the other hand absent for classical chaos, 
with the consequence that the initial exponential 
decorrelation is not directly bounded by topology, see
Figs.~\ref{fig:mgchaos} and \ref{fig:correlation}. 
Classical chaos and PPC are two limiting cases, with
the distinction becoming somewhat fluid for very thick 
fractal braids.

\begin{figure*}[t]\centering
\includegraphics[width=1.\textwidth]{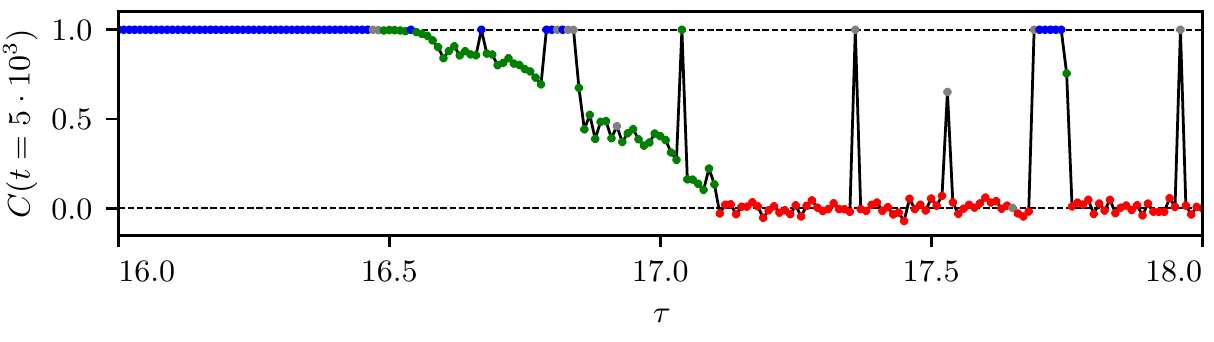}
\caption[Cross-correlation of the Mackey-Glass system as a function of the delay]
{\label{fig:correlationrange}
Long-term cross-correlation $C(t\!=\!5\!\cdot\!10^3)$ of the Mackey-Glass
system~(\ref{eq:mackeyglass}) as a function of the delay time $\tau$
and averaged over $100$ pairs of trajectories. The
initial distance is $\delta=10^{-6}$.
Regular motion (blue) is fully correlated $C=1$,
with classical chaos being characterized by a complete loss 
of correlation $C\to0$. Partially predictable
chaos (PPC, green) occurs for chaotic motion with finite 
residual long-term correlation $1>C>0$.}
\end{figure*}

PPC chaos is found for the Mackey-Glass system (\ref{eq:mackeyglass}),
e.\,g., close to the transition to chaos at a time delay $\tau\approx16.48$.
Figure~\ref{fig:correlationrange} shows the residual correlation
$C(t\!=\!5\cdot10^3)$, where $5\cdot10^3\gg T_\lambda$,
for  pairs of trajectories and as a function of the time delay $\tau$
(cf.~Sect.~\ref{sec:lyapunovpredictiontime}).


Note that the auto-correlation function \cite{thomae1981correlations,badii1988correlation},
which can be computed from a single trajectory, can be also used to describe
the decorraltion process on chaotic attractors.
However, it has been pointed out \cite{wernecke2017test} 
that it is more challenging to quantify both the initial 
decorrelation and the linear loss of correlation
in PPC through the auto-correlation function.

\begin{figure*}[t]\centering
\includegraphics[width=\textwidth]{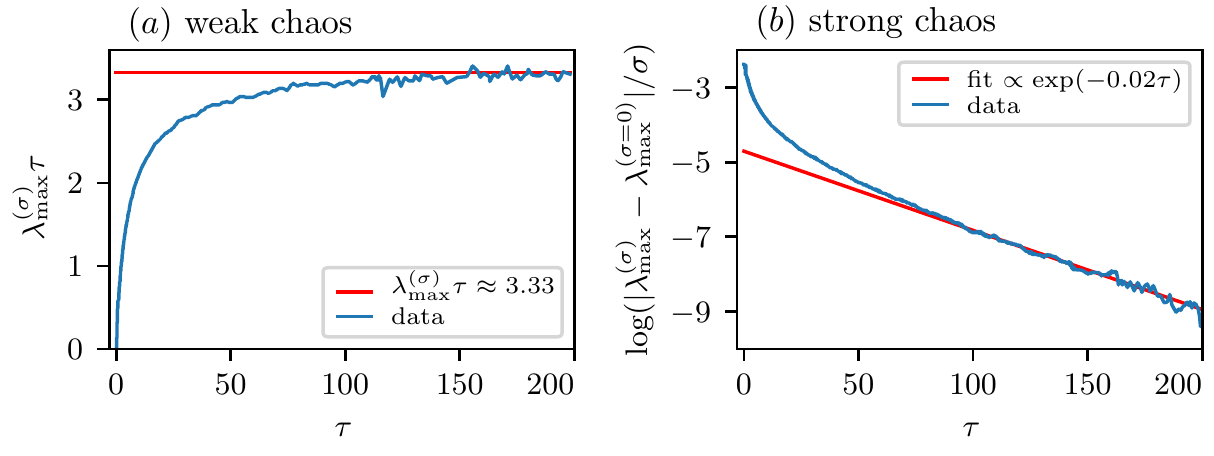}
\caption[Lyapunov exponents for weak and strong chaos ]
{\label{fig:weakchaos}
The maximal Lyapunov exponent $\lambda_\text{max}=\lambda_\text{max}^{(\sigma)}$
and the maximal instantaneous Lyapunov exponent $\lambda_\text{max}^{(\sigma=0)}$
of an attractor of the Lang-Kobayashi system~\cite{lang1980external}
(cf.~Eq.~(\ref{eq:weakchaosdde})).
($a$)~The rescaled maximal Lyapunov exponent $\lambda_\text{max}^{(\sigma)}$ 
as a function of the time delay $\tau$ for weak chaos with
coupling $\sigma=21\,\mathrm{ns}^{-1}$.
In the limit of large delays $\tau\to\infty$ the maximal Lyapunov scales
$\lambda_\text{max}^{(\sigma)}\sim1/\tau$ inversely with the delay.  
($b$)~The logarithmic difference of the maximal $\lambda_\text{max}^{(\sigma)}$ and
the instantaneous Lyapunov exponent $\lambda_\text{max}^{(\sigma=0)}$
over time delay $\tau$ for coupling $\sigma=12\,\mathrm{ns}^{-1}$.
In the regime of strong chaos the difference vanishes exponentially
in the limit $\tau\to\infty$ of large
delays. Figure replicated from \cite{heiligenthal2011strong}.
}
\end{figure*}

\subsubsection{Weak and strong chaos}

Several proposals for the distinction of weak and strong
chaos, and thus for a differentiation between different types 
of chaotic motion, have been put forward
\cite{suzuki2016periodic,rabaud1990dynamical,klages2013weak}.
For concreteness consider with
\begin{align}
\dot{\mathbf{x}}(t)&=\mathbf{F}_1\big( \mathbf{x}(t) \big)+
\sigma \mathbf{F}_2\big( \mathbf{x}(t-\tau) \big)
\label{eq:weakchaosdde}
\end{align}
a network of dynamical units
$\mathbf{x}=(x_1,\, x_2,\, \dots)$
that are coupled instantaneously through $\mathbf{F}_1(\mathbf{x}(t))$,
and delayed via $\mathbf{F}_2(\mathbf{x}(t-\tau))$
\cite{heiligenthal2011strong} (see also 
\cite{yanchuk2017spatio,d2013synchronisation}). The respective 
coupling strength is $\sigma$. Networks of this type 
are suitable for the description of chaos in coupled lasers
\cite{heiligenthal2013strong, soriano2013complex} and
for the study of delay induced chaos 
(cf.\ Sect.~\ref{sec:delayinduced}),

A distinction between weak and strong chaos can now be 
made \cite{heiligenthal2011strong} for the special case
that a fully synchronized state $\mathbf{s}(t)$, as defined 
by $x_i(t)\equiv s(t)$, is a solution of (\ref{eq:weakchaosdde}). 
The synchronized state may be stable or unstable. Stable 
synchronized states correspond to weak chaos, unstable 
synchronized states on the other side to strong chaos. 
One starts by defining two types of maximal Lyapunov 
exponents \cite{heiligenthal2011strong}:

\begin{itemize}
\item $\lambda_\text{max}^{(\sigma)}=\lambda_\text{max}$,
      which describes the divergence
      of trajectories from the synchronized state
      for the original system (\ref{eq:weakchaosdde}).
\item $\lambda_\text{max}^{(\sigma\!=\!0)}$, which describes the
      divergence of trajectories from the synchronized state 
      $\mathbf{s}(t)$ under the influence of only the instantaneous
      dynamics $\mathbf{F}_1(\mathbf{s}(t))$. Note, that $\mathbf{s}(t)$ 
      is still a solution of the full system.
\end{itemize}

The distinction of weak and strong chaos follows then from 
the comparison of the full exponent $\lambda_\text{max}^{(\sigma)}$ 
and the instantaneous maximal Lyapunov exponent 
$\lambda_\text{max}^{(\sigma\!=\!0)}$: 
\begin{itemize}
\item \textbf{Weak chaos}: For weak chaos the instantaneous 
      Lyapunov exponent is negative, $\lambda_\text{max}^{(\sigma\!=\!0)}<0$,
      indicating that the evolution of perturbations at $\sigma=0$
      is stable. The overall dynamics is at the same time unstable
      due to a positive full exponent, $\lambda_\text{max}^{(\sigma)}>0$.  
      The synchronized state is then a stable but chaotic solution
      of (\ref{eq:weakchaosdde}).
 \item \textbf{Strong chaos}: For strong chaos both the instantaneous 
       and the full maximal Lyapunov exponents are positive,
       $\lambda_\text{max}^{(\sigma\!=\!0)}>0$ and $\lambda_\text{max}^{(\sigma)}>0$.
       The system then settle into a global chaotic state, which is
       however not given by $\mathbf{s}(t)$.
\end{itemize}
Strong and weak chaos differ furthermore by their
Lyapunov divergence times $T_\lambda\sim 1/\lambda_\text{max}$
(cf.~Sect.~\ref{sec:lyapunovpredictiontime}),
with the scaling $T_\lambda\sim\tau^\eta$ for large time delays $\tau\to\infty$,
where $\eta=1$ for strong chaos
and $\eta=0$ for weak chaos 
\cite{heiligenthal2011strong} (cf.~Fig.~\ref{fig:weakchaos}).

According to this classification scheme, the Mackey-Glass
system (\ref{eq:mackeyglass}), which has a negative 
instantaneous Lyapunov exponent 
$\lambda_\text{max}^{(\sigma\!=\!0)}=b<0$, exhibits only weak 
chaos. Note that the coupling constant $a$ corresponds 
here to $\sigma$ and that bounded solutions need negative $b$.
Vice versa, the difference between the attractors shown in
Fig.~\ref{fig:mgchaos} and Fig.~\ref{fig:mgppc}
cannot be explained in terms of weak and strong chaos.

\subsubsection{Intermittent and laminar chaos}
\label{sec:laminarchaos}
\begin{figure*}[t]\centering
 \includegraphics[width=1.\textwidth]{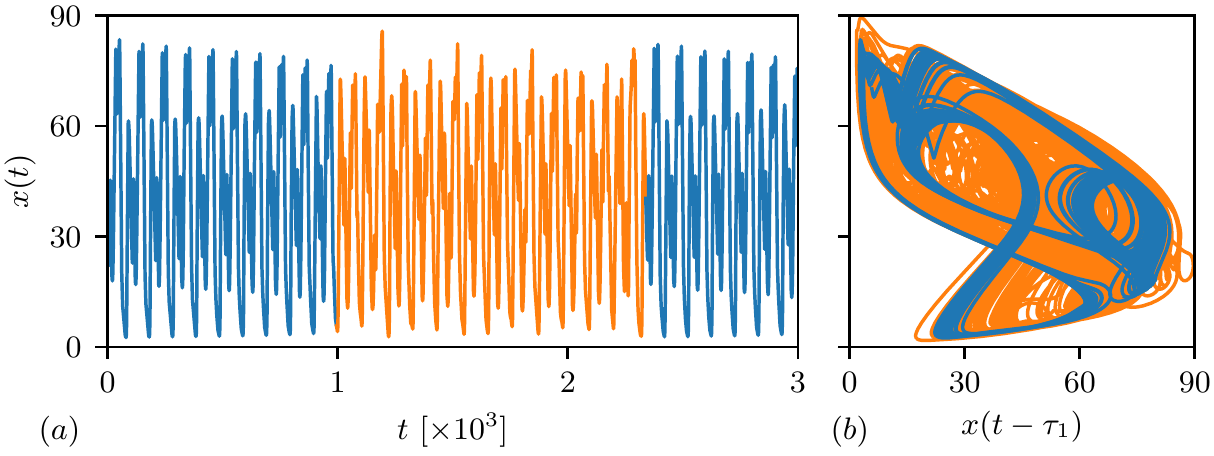}
\caption[Intermittent chaos]
{\label{fig:intermittentchaos}
The solution~$x(t)$ of a delayed feedback system with multiple time delays
as defined by Eq.~(\ref{eq:intermittentdde})
for parameters $n=6$, $m=1$ and $\xi=38$, coupling strengths 
$k=0.2$ and $g=50$ and time delays $\tau_1=26$ and $\tau_2=26.25$.
($a$)~A trajectory $x(t)$ showing intermittent chaos. The
almost period dynamics (blue) is interseeded by chaotic bursts (orange).
($b$)~Projection of the trajectory $x(t)$ shown in panel~($a$) 
with respect to the delayed state $x(t-\tau_1)$.
Highlighted (blue) is the braid of the quasi-periodic motion, which the
solution follows most of the time. The chaotic bursts lead the system 
intermittently away from the braid.
Figure replicated from \cite{suzuki2016periodic}.
}
\end{figure*}

Intermittent chaos is a type of chaos known from 
non-delayed systems \cite{schuster2006deterministic, sandor2015versatile}.  
It is also observed in time delay systems 
\cite{parthimos2001universal,hamilton1992intermittently}, e.\,g.\ 
in models describing gene regulation networks \cite{suzuki2016periodic}.
Consider the case that the delay term is with
\begin{align}
 \dot{x}(t)&=-k\,x(t)+g\,f_1\big(x(t-\tau_1)\big)\,f_2\big(x(t-\tau_2)\big)
\label{eq:intermittentdde}
\end{align}
a product of a self-inhibitory and a self-activation term, 
$f_1$ and $f_2$, acting respectively with fixed but 
distinct delays $\tau_1$ and $\tau_2$.
The coupling constants $k$ and $g$ determine the 
respective influence of the instantaneous
and the delayed feedback on the dynamics.

Choosing  Hill functions \cite{gesztelyi2012hill}
\begin{equation}
 f_1(x)=1/\big(1+(x/\xi)^{n}\big)\,,\qquad f_2(x)=1-1/\big(1+(x/\xi)^{m}\big)
\end{equation}
with parameters $n$, $m$ and $\xi$ for the activation and 
inhibition function \cite{suzuki2016periodic}, the 
solutions of (\ref{eq:intermittentdde}) show intermittent chaos,
which is in this case characterized by quasi-periodic dynamics interseeded 
by chaotic bursts. A typical trajectory is presented in 
Fig.~\ref{fig:intermittentchaos}.

\begin{figure*}[t]\centering
\includegraphics[width=\textwidth]{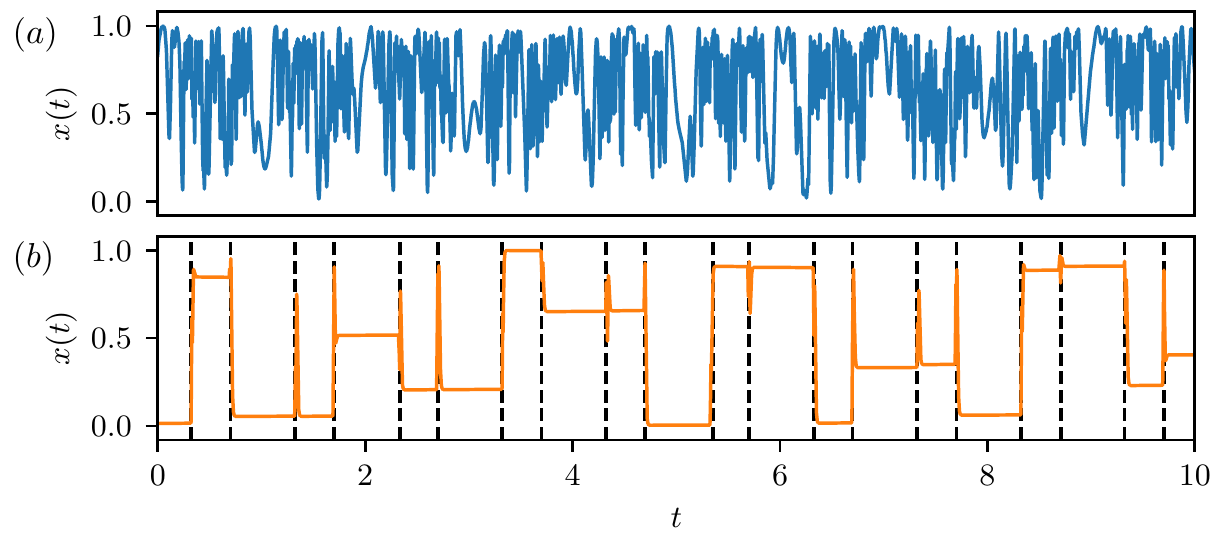}
\caption{\label{fig:laminarchaos}
Chaotic solutions $x(t)$ of (\ref{eq:laminarchaosdde}), as
a function of time $t$, for $A=0.9/(2\pi)$ and $T=200$. 
The corresponding Lyapunov exponents are shown in Fig.~\ref{fig:dissipativedelay}.
($a$)~Turbulent chaos for a conservative delay with $\tau_\text{o}=1.54$.
($b$)~Laminar chaos for dissipative delay with $\tau_\text{o}=1.51$.
The solution jumps between different laminar levels, where
the height $h_n$ of the levels and the switching times $T_n$ 
(indicated by vertical dashed lines) can be predicted by
an appropriate mapping. Figure replicated from \cite{muller2018laminar}.  
}
\end{figure*}


Laminar chaos is on the other side closely related to 
the concept of dissipative time-varying delay
\cite{otto2017universal} (cf.~Sect.~\ref{sec:dissipativedelay}). 
An example of a system with time-varying feedback for
which laminar chaos is observed is \cite{muller2018laminar}
\begin{align}
\frac{1}{T}\dot{x}(t)&=-x(t)+ 4x\big(R(t)\big) \Big(1-x\big(R(t)\big)\Big),
\quad\quad
R(t)= t-\tau_\text{o}-A\sin(2\pi t)\,,
\label{eq:laminarchaosdde} 
\end{align}
where $T$ is the overall time-scale. The access function $R(t)$,
which enters (\ref{eq:laminarchaosdde}) via a logistic
feedback coupling, incorporates here a superposition of 
a constant delay $\tau_\text{o}$ and a sinusoidal contribution of amplitude $A$.
Depending on the parameters, the dynamics may jump
between constant plateaus of laminar motion, as illustrated
in Fig.~\ref{fig:laminarchaos}. The system is chaotic because
both the sequence of plateau heights and the sequence
of plateau durations exhibit non-regular dynamics \cite{muller2018laminar}.
For comparison a case of classically turbulent chaotic dynamics
is shown as well.

Laminar chaos is not to be confused with intermittent chaos: 
for the first the laminar plateaus have a chaotically distributed 
height, while for the latter the chaotic bursts are framed by 
laminar or quasi-periodic oscillations of similar amplitudes.

\subsubsection{Transient chaos}
\label{sec:transientchaos}

\begin{figure}[t]\centering
\includegraphics[width=\textwidth]{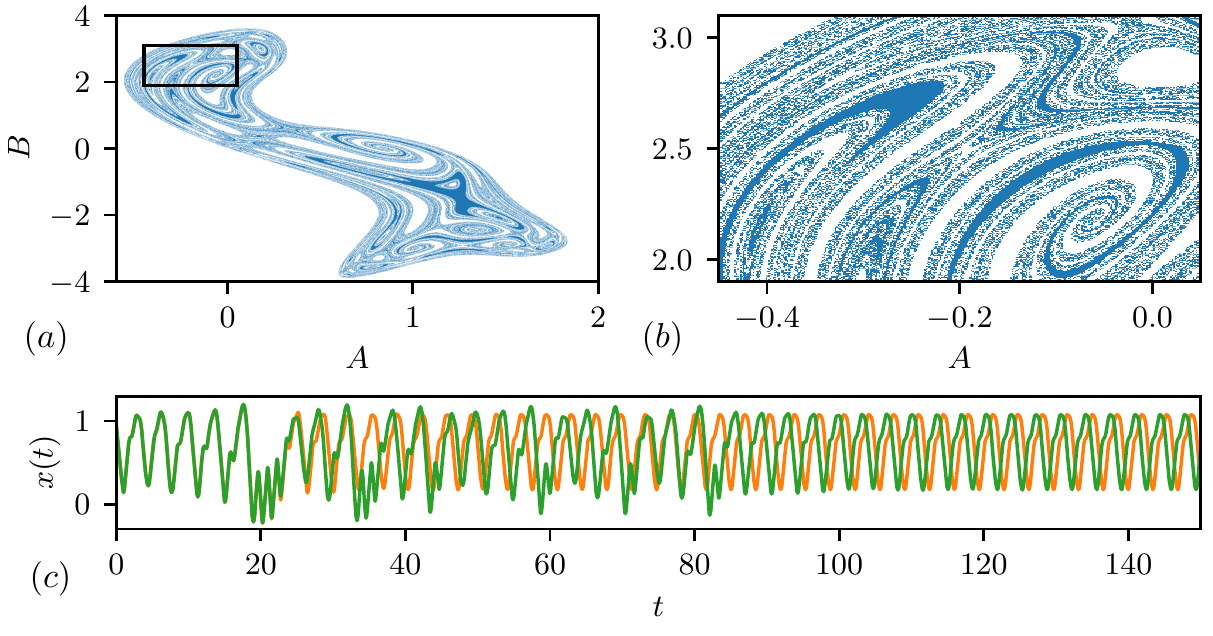}
\caption{
\label{fig:transientchaos}
($a$)~Basin of attraction of the delayed logistic Eq.~(\ref{eq:delayedlogistic})
for $\tau=1$ and $\beta=6.16$, with $A$ being the offset and $B$ the frequency
of the initial function~(\ref{eq:initialsinus}) on a $2^{11}\times2^{11}$ grid.
Parameter pairs $(A,B)$ marked blue correspond to an initial function
for which the system (\ref{eq:delayedlogistic}) converges to
a periodic attractor. Else the motion diverges as $\lim_{t\to\infty}x(t)=-\infty$.
($b$)~Magnifying the parameter region marked by the black square in panel~($a$)
reveals the fractal character of the basin of attraction.
Its fractal dimension is found to be $D_\text{f}=1.85 \pm 0.01$
(cf.~Sect.~\ref{sec:fractaldimension}).
($c$)~Two solutions of DDE~(\ref{eq:delayedlogistic}) for $\tau=1$ and 
$\beta=6.16$, starting at initial functions~(\ref{eq:initialsinus}),
with parameters $B_1=-0.95$ and $A_1=0.947$ and respectively with
$B_2=B_1$ and $A_2=A_1+10^{-4}$.
The trajectories experience transient chaos before joining the 
same periodic attractor, albeit with a phase shift.
Figure replicated from \cite{taylor2007approximating}.
}
\end{figure}

The chaotic attractors discussed in Sects.~\ref{sec:delayinduced} 
to \ref{sec:laminarchaos} are asymptotically stable, i.\,e.\ 
the dynamics settles onto the attracting set in the long-term 
limit $t\to\infty$. However, it is known that (asymptotically) 
unstable, fractal sets in the phase space of a system, 
so-called chaotic saddles \cite{nusse1989procedure,sweet2001stagger},
can cause initially close-by trajectories to decorrelate. The motion 
of a trajectory in the vicinity of a chaotic saddle is termed
transient chaos \cite{kantz1985repellers,battelino1988multiple,lai2011transient}. 
Transiently chaotic motion is furthermore accompanied by fractal 
basin boundaries in the phase space of the system.
In the case of a time delay system this is reflected as a 
fine-grained subdivision of the space of initial functions.
Small changes of the initial condition may then lead to different 
asymptotic attractors.

For DDE a trajectory is uniquely determined by an initial 
function (\ref{eq:initialfunction}), as defined on an initial 
time interval. A practical way to scan the space of possible 
initial functions is to sub-sample using functions parameterized 
by a finite number of parameters
\cite{taylor2007approximating,foss1996multistability}.
An arbitrary but suitable choice is 
\begin{equation}
\varphi(t)=A+\sin(Bt) \qquad \text{ for } t\in[-\tau,0]\,,
\label{eq:initialsinus}
\end{equation}
where the initial function $\varphi(t)$ is parameterized by 
an offset $A$ and a sinusoidal with frequency $B$. Using
(\ref{eq:initialsinus}) for the delayed logistic equation
\begin{equation}
\dot{x}(t)=-x(t)+\beta x(t-\tau)\big( 1-x(t-\tau) \big)
\label{eq:delayedlogistic}
\end{equation}
with a fixed time delay $\tau=1$ and a coupling strength 
$\beta=6.16$ one finds that for almost any pairs 
$(A,B)$ of parameters entering $\varphi(t)$ via
(\ref{eq:initialsinus}) the motion is not bound,
that is $\lim_{t\to\infty}x(t)=-\infty$. Only for 
certain combinations of parameters from a fractal set 
in the parameter set shown in Fig.~\ref{fig:transientchaos}
the motion settles to a periodic attractor in the long term.
As argued in \cite{taylor2007approximating}, the presence 
of a chaotic saddle induces transient chaos
(cf.~Fig.~\ref{fig:transientchaos}).


\begin{figure*}[t]\centering
\includegraphics[width=1.\textwidth]{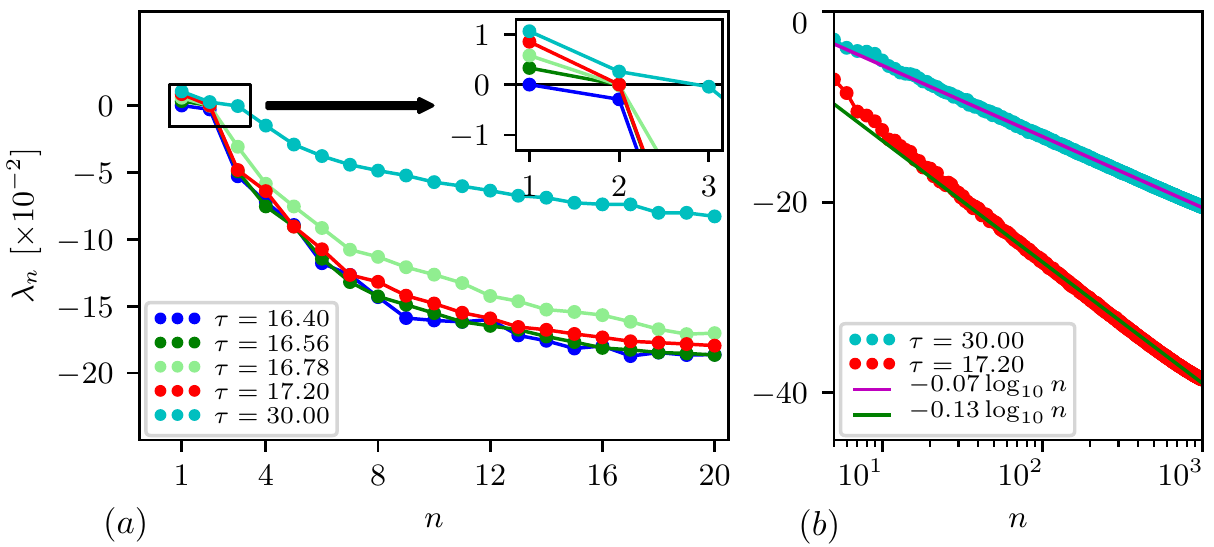}
\caption{\label{fig:lyapspecsmall}
The largest global Lyapunov exponents $\lambda_n$
of the Mackey-Glass system (\ref{eq:mackeyglass}), as
computed with Benettin's method and for different delays 
$\tau$ (cf.~Sect.~\ref{sec:benettin}).
($a$) As in Figs.~\ref{fig:mgchaos} and \ref{fig:mgppc},
for regular motion ($\tau=16.40$, blue), for  PPC 
($\tau=16,56$, green; $\tau=16.78$, light green) and 
classical chaos ($\tau=17.20$, red).
The hyper-chaotic attractor ($\tau=30.00$, cyan) has two positive
exponents.
($b$) A semi-log plot of the extended spectrum for the chaotic 
attractors shown in Fig.~\ref{fig:mgchaos}. The fits indicate 
logarithmic scaling $\lambda_n\sim-\log n$ in the limit $n\to\infty$,
where $n$ is the index.}
\end{figure*}

\begin{table*}[b]\centering
\caption{\label{tab:lyapunov}  
Scheme for identifying different types of dynamics using 
the number of positive and zero global Lyapunov exponents 
$\lambda_\text{max}=\lambda_1\geq\lambda_2\geq\ldots\,$
\cite{wolf1985determining}.
 }
 \begin{tabular}{lllc}
  \toprule
  &\multicolumn{3}{c}{Lyapunov exponents}\\
  dynamics&positive&zero&largest neg.\\
  \midrule
  stable fixed point&&&$\lambda_\text{max}<0$\\
  limit cycle&&$\lambda_\text{max}=0$&$\lambda_{2\phantom{+1}}<0$\\
  hypertorus ($d$ dim.)      &&$\lambda_\text{max},\ldots,\lambda_d=0$&$\lambda_{d+1}<0$\\
  chaos      &$\lambda_\text{max}>0$&$\lambda_{2\phantom{+1}}=0$&$\lambda_{3\phantom{+1}}<0$\\
  hyperchaos &$\lambda_\text{max},\ldots,\lambda_k>0$&$\lambda_{k+1}=0$&$\lambda_{k+2}<0$\\
  \bottomrule
 \end{tabular}
\end{table*}

\subsection{Lyapunov spectrum}
\label{sec:lyapunov_spectrum}

Chaotic attractors are considered to be strange in the sense
that they are overall contracting
\cite{eckmann1985ergodic,pikovsky2016lyapunov}, being characterized 
on the other hand by at least one positive global Lyapunov exponent,
as defined in Sect.~\ref{sec:lyapunov}.

As an illustration we present in Fig.~\ref{fig:lyapspecsmall}
the spectrum of global Lyapunov exponents of the Mackey-Glass
system (\ref{eq:mackeyglass}) for different values of the time 
delay $\tau$. The largest Lyapunov exponent vanishes
for regular motion (limit cycles), for which an initial 
deviation $\delta$ does not grow nor vanish on the average.
This is consistent with the definition of the maximal Lyapunov
as an average over the attractor. The spectrum of Lyapunov 
exponents is in contrast very broad on short time scales 
\cite{wernecke2017test}. 

For chaotic motion at least one exponent is positive,
$\lambda_\text{max}>0$, with chaotic motions with two
or more positive exponents being termed 
hyperchaotic \cite{rossler1979equation,wolf1985determining}. 
This is the case in Fig.~\ref{fig:lyapspecsmall} for
$\tau=30.00$. Note that all strange attractors have
in addition one vanishing exponent describing the neutral 
flow along the trajectory. A categorization of the chaotic 
dynamics derived from the spectrum of Lyapunov exponents 
is given in Table~\ref{tab:lyapunov}.

It can be shown analytically \cite{farmer1982chaotic}, that 
the Lyapunov spectrum of a DDE with linear dependence on the 
instantaneous state $x(t)$ and a single constant time delay 
scales logarithmically, $\lambda_n\sim-\log n$, as a function
of the index $n$, when  $n\to\infty$. This results holds
for the two Lyapunov spectra of the Mackey-Glass system
shown in Fig.~\ref{fig:lyapspecsmall}. The scaling 
is in contrast linear for DDE with time-varying $\tau=\tau(t)$ 
that are dissipative \cite{otto2017universal}, as discussed 
in Sect.~\ref{sec:timevardelay}.

\begin{figure*}[t]\centering
\includegraphics[width=\textwidth]{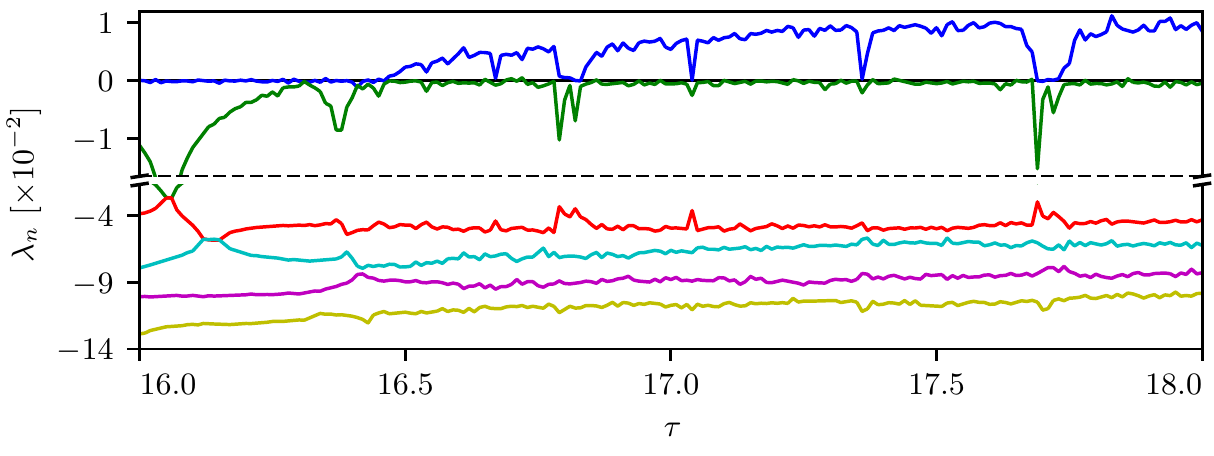}
\caption{\label{fig:lyapspec}
As a function of the delay $\tau$, the six largest Lyapunov 
exponents $\lambda_n$ of the Mackey-Glass 
system (\ref{eq:mackeyglass}). Note the change in the 
scaling of the vertical axis indicated by the dashed horizontal 
line.
}
\end{figure*}

\subsection{Lyapunov prediction time}
\label{sec:lyapunovpredictiontime}

For the Mackey-Glass system (\ref{eq:mackeyglass}), the 
evolution of the six largest global Lyapunov exponents 
as a function of the time delay is presented in Fig.~\ref{fig:lyapspec}.
The maximal exponent $\lambda_\text{max}$ changes from 
zero to a positive value at $\tau\approx16.45$, the classical
indicator of the transition from regular motion to chaos.

The maximal Lyapunov exponent~$\lambda_\text{max}$ describes 
by definition the maximum rate of divergence, or the minimum
rate of convergence (for positive and respectively for negative 
exponents). With the distance of two trajectories
scaling as $\sim\delta\exp(\lambda_\text{max} t)$,
one defines the Lyapunov prediction time $T_\lambda$
as
\begin{align}
T_\lambda(\delta,d_\text{p}) &=
\frac{1}{\lambda_\text{max}}\log\left(\frac{d_\text{p}}{\delta}\right)\,.
\label{eq:lyappredic}
\end{align}
It quantifies the time it takes the exponential divergence 
of two trajectories with initial distance $\delta$ 
to reach a final distance $d_\text{p}$ 
(cf.~Fig.~\ref{fig:lyapfromdist}). The exact values
for $d_\text{p}$ and $\delta$ are not critical, due to the
logarithmic discounting in (\ref{eq:lyappredic}).

In Table~\ref{tab:mglyapunov} the four largest Lyapunov exponents 
for the attractors of the Mackey-Glass system shown in 
Figs.~\ref{fig:mgchaos} and \ref{fig:mgppc} are listed
together with the corresponding Lyapunov prediction times 
$T_\lambda$. One finds that the thin chaotic braids of PPC 
also lead to longer predictability in the regime of 
exponential divergence (cf.~Sect.~\ref{sec:ppc}).

\begin{table*}[b!]\centering
\caption{\label{tab:mglyapunov}
The largest four Lyapunov exponents and the Lyapunov prediction 
time $T_\lambda$, as defined by (\ref{eq:lyappredic}), for the 
attractors of the Mackey-Glass system~(\ref{eq:mackeyglass}) 
shown in Figs.~\ref{fig:mgchaos} and \ref{fig:mgppc}. The
parameters entering (\ref{eq:lyappredic}) are $d_\text{p}=10^{-2}$ 
and $\delta=10^{-6}$. For orientation the zero Lyapunov exponent 
is printed in bold.}
 \begin{tabular}{lcccccc}
  \toprule
  \multicolumn{2}{l}{$\tau$}&$16.40$&$16.56$&$16.78$&$17.20$&$30.00$\\
  \multicolumn{2}{l}{dynamics}&regular&PPC&PPC&chaos&hyperchaos\\
  \multicolumn{2}{l}{$T_\lambda$}&-&$2791$&$1588$&$1071$&$\phantom{0}867$\\
  \midrule
  $\lambda_\text{max}$& [$\times10^{-2}$] &$\phantom{-}\mathbf{0.0}$&$
\phantom{-}0.3$&$\phantom{-}0.6$&$\phantom{-}0.9$&$\phantom{-}1.1$\\
  $\lambda_2$& [$\times10^{-2}$]&$-0.3$&$\phantom{-}\mathbf{0.0}$&$\phantom{-}\mathbf{0.0}$&$\phantom{-}\mathbf{0.0}$
&$\phantom{-}0.3$\\
  $\lambda_3$& [$\times10^{-2}$]&$-5.3$&$-5.0$&$-5.3$&$-4.8$&$\phantom{-}\mathbf{0.0}$\\
  $\lambda_4$& [$\times10^{-2}$]&$-7.2$&$-7.4$&$-6.7$&$-6.4$&$-1.5$\\
  \bottomrule
 \end{tabular}
\end{table*}

\subsection{Phase space contraction rate}
\label{sec:contractionrate}

The phase space contraction rate~$\kappa$ is an effective tool to 
quantify the behavior of the flow in finite dimensional continuous-time 
systems \cite{gros2015complex}. It describes the evolution of a volume 
element $V$ in the phase space over time
\begin{align}
V(t)&=V_\text{o}\;\e^{\kappa t}\,,
\end{align}
where the initial volume is denoted $V_\text{o}$ \cite{mori1980fractal}.
The sign of the phase space contraction rate indicates whether a system is
dissipative, $\kappa<0$, conservative, $\kappa=0$, or whether energy
is taken up when $\kappa>0$. The contraction rate is a local quantity 
that may vary strongly within phase space. For stable limit cycles and 
chaotic attractors the contraction rate needs to be negative when averaged 
over the attracting set, but not locally \cite{daems1999entropy}.

\begin{figure*}[t]\centering
 \includegraphics[width=1\textwidth]{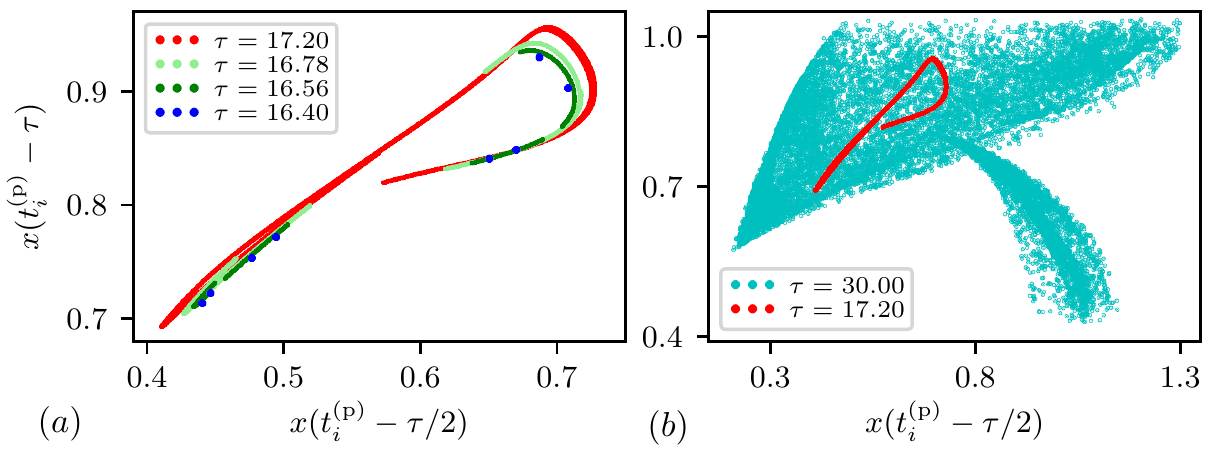}
\caption{\label{fig:poincaresection}
Poincar\'{e} sections (defined by $x_\text{p}=0.85$ in
Eq.~(\ref{poincareSection})) for the Mackey-Glass system 
(\ref{eq:mackeyglass}), sampled over time $t\in[0,10^4]$ 
and projected to states $x(t-\tau/2)$ and $x(t-\tau)$, 
at the intersection times $t\to t_i^{(p)}$.
($a$) The periodic trajectory ($\tau=16.40$, blue) crosses 
the Poincar\'{e} plane eight times, whereas one observes
extended fractal sets for PPC ($\tau=16.56$, green; 
$\tau=16.78$, light-green). For classical chaos ($\tau=17.20$, red) 
the fractal set has a larger extent than for PPC.
($b$) The Poincar\'{e} sections of classical chaos ($\tau=17.20$, red),
as in ($a$), and for hyperchaos ($\tau=30.00$, cyan).
Compare Figs.~\ref{fig:mgchaos} and \ref{fig:mgppc} and
Table~\ref{tab:mglyapunov}.
}
\end{figure*}

For a finite dimensional system the contraction rate 
is given by the sum of local Lyapunov exponents, viz as
$\kappa=\sum_n\Lambda_n$. The situation is less clear 
for infinite dimensional time delay systems, for which 
the number of negative Lyapunov exponents diverges
\cite{farmer1982chaotic}, as discussed in Sect.~\ref{sec:lyapunov},
as $\Lambda_n\to-\log n$ for $n\to\infty$. In the 
phase space of state histories the contraction rate is 
therefore formally diverging,
\begin{align}
\kappa&=\lim\limits_{N\to\infty}\sum\limits_{n=1}^{N}\Lambda_n
\ =\ -\infty\,,
\end{align}
and hence not well defined for time delay systems.

\subsection{Poincar\'{e} section}
\label{sec:poincare}

A widely used tool for the analysis of the flow in
reduced dimensions is the Poincar\'{e} section
\cite{poincare1890probleme,gros2015complex}. For a
dynamical system with dimension $N$ the Poincar\'{e} 
hyperplane $P$ has dimension $N-1$, which is still
infinite for a DDE, for which the phase space is
given by the formally infinite-dimensional
space of state histories $\mathbf{X}(t)=\{x(t^\prime)\}$,
where $t^\prime \in[t-\tau, t]$
(cf.\ Sect.~\ref{sect_state_histories}).
The intersections of a trajectory
$\mathbf{X}(t)$ in the space of state histories with the
selected hyperplane $P$ defines via
\begin{equation}
\mathbf{X}^{(\text{p})}_i\equiv \mathbf{X}(t^{(\text{p})}_i),
\qquad\quad
 \mathbf{X}(t^{(\text{p})}_i) \in P,
\qquad\quad
\mathbf{X}^{(\text{p})}_i\to \mathbf{X}^{(\text{p})}_{i+1}\,,
\end{equation}
a map $\mathbf{X}^{(\text{p})}_i\to \mathbf{X}^{(\text{p})}_{i+1}$
between consecutive crossings. A convenient way to define the 
intersections, and the respective crossing times $t^{(\text{p})}_i$, is
to choose a value $x_\text{p}$, such that
\begin{equation}
x^{(\text{p})}_i\equiv x(t^{(\text{p})}_i)=x_\text{p}
\label{poincareSection}
\end{equation}
holds for the trajectory $x(t)$ in configuration space.

The map between consecutive intersections, also called 
first recurrence map, defined by the Poincar\'{e} section 
can be studied also in configuration space, 
$x^{(\text{p})}_i\to x^{(\text{p})}_{i+1}$. Apart from the
location, one may also consider the direction
of the intersection and restrict, as it is usually
done, the Poincar\'{e} map to consecutive intersections
characterized by the same direction.

For graphical illustrations in two dimensions it is custom
to select two states from the state histories that are 
separated in time, such as $x(t-\tau_1)$ and $x(t-\tau_2)$,
as representatives of the state histories defined by
the intersection of the trajectory with the Poincar\'{e} hyperplane.
For a system with a fixed time delay $\tau$, a convenient
choice for the Poincar\'{e} section is 
$\tau_1<\tau_2=\tau$.

\begin{figure*}[t]\centering
\includegraphics[width=1\textwidth]{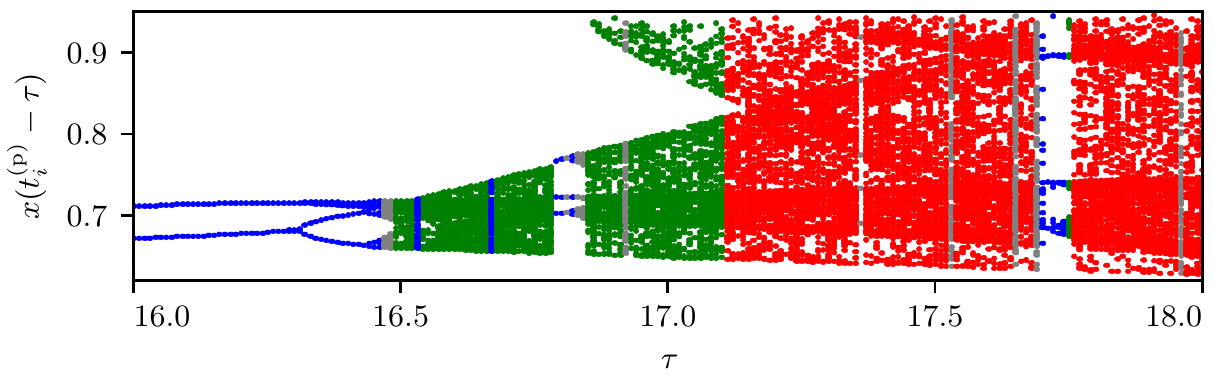}
\caption{\label{fig:poincarerange}
Poincar\'{e} sections of the Mackey-Glass system~(\ref{eq:mackeyglass}) for a
range of time delays $\tau$, as projected to $x(t^{(\text{p})}_i-\tau)$
(for details see Fig.~\ref{fig:poincaresection}). The transition from periodic
motion (blue) to PPC (green) occurs due to a period-doubling cascade.
PPC is distinguished from classical chaos (red) by its finite residual
correlation (cf.~Fig.~\ref{fig:correlationrange}).
}
\end{figure*}

In Fig.~\ref{fig:poincaresection} the Poincar\'{e} section for 
the attractors of the Mackey-Glass system~(\ref{eq:mackeyglass}) 
shown in Figs.~\ref{fig:mgchaos} and \ref{fig:mgppc} are compared. 
Note that periodic motion, which corresponds to fixed points of the 
Poincar\'{e} map, may also be of higher period, like
$x^{(\text{p})}_1\to\ldots\to x^{(\text{p})}_8\to x^{(\text{p})}_1$.
Partially predictable and classical chaotic attractors form on
the other hand extended sets resembling thin filaments in the 
projection of the Poincar\'{e} hyperplane, which can be shown 
to be self-similar \cite{eckmann1985ergodic,mandelbrot1982fractal}. 
Moreover, one observes that hyperchaotic attractors tend to be more 
space filling in terms of the Poincar\'{e} section
(cf.~Sect.~\ref{sec:dimensions}).

In Fig.~\ref{fig:poincarerange} a color-coded bifurcation diagram 
of the Mackey-Glass system generated using a one dimensional 
projection of the Poincar\'{e} section is presented. The cascade 
of period-doubling bifurcations
\cite{yorke1985period,sander2011period} (also called Brunovsky bifurcation
\cite{brunovsky1971symposium}) leading to partially predictable
chaos (PPC) upon increasing the time delay $\tau$ is evident, 
with the phase of PPC being interseeded 
by periodic windows. The transition from PPC to
classical chaos then induces a fast drop in correlations,
as detailed out in Sect.~\ref{sec:ppc}.

\begin{figure}[t]\centering
\includegraphics[width=0.60\textwidth]{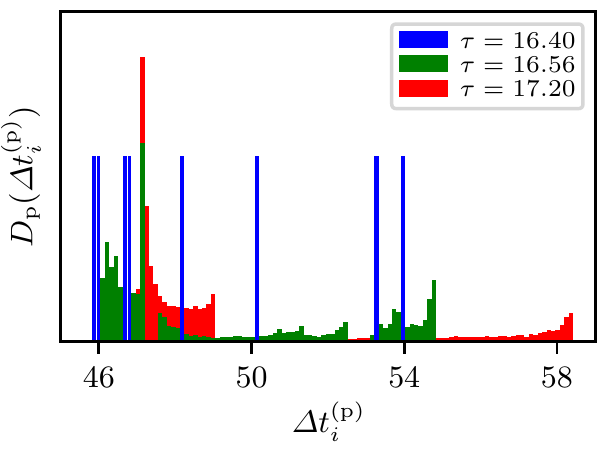}
\caption{\label{fig:crossingtimes}
For different attractors of the Mackey-Glass system~(\ref{eq:mackeyglass}),
the distribution $D_\text{p}$ (in arbitrary units) of the time intervals
$\varDelta t^{(\text{p})}_i$ between consecutive Poincar\'{e} sections
(cf.~Fig.~\ref{fig:poincaresection}). $120$ bins have been used. 
The periodic attractor ($\tau=16.40$, blue) cuts the Poincar\'{e} 
plane eight times (cf.~Fig.~\ref{fig:poincaresection})
leading to a distribution that consists of peaks of equal height.
For partially predictable (PPC, $\tau=16.56$, green) and 
classical chaos ($\tau=17.20$, red) the distribution
widens around the peaks.}
\end{figure}

Another aspect of the Poincar\'{e} map involves the time intervals 
between consecutive sections, the recurrence time
\cite{frisch1956poincare,gao1999recurrence}:
\begin{align}
 \varDelta t^{(\text{p})}_i &= t^{(\text{p})}_{i+1}-t^{(\text{p})}_{i}\,.
\end{align}
The distribution of the recurrence times of three attractors 
is presented in Fig.~\ref{fig:crossingtimes}. As the regular 
motion crosses the Poincar\'{e} plane periodically, the inter-section 
times are discrete peaks of equal probability.

For partially predictable chaos the distribution is blurred, 
with some residual resemblance to the original periodic peaks.
As the topology of the classical chaotic state deviates from periodic 
and PPC attractors, the distribution becomes more wide-spread.

\subsection{The power spectrum of attractors}

\begin{figure*}[t]\centering
\includegraphics[width=\textwidth]{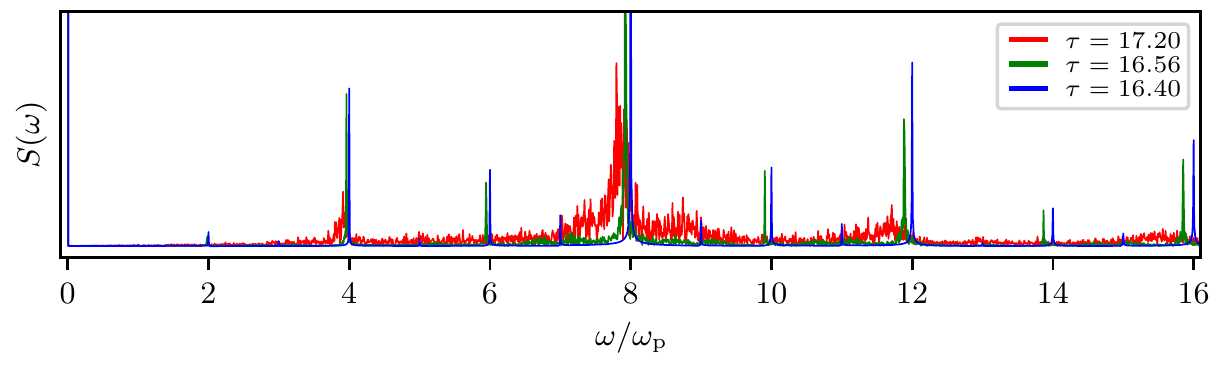}
\caption{\label{fig:fourierspectrum}
The spectral density $S(\omega)$ (in arbitrary units) as a function 
of the frequency $\omega$, of different attractors of the 
Mackey-Glass system (\ref{eq:mackeyglass}). The data was obtained 
by sampling trajectories over $t\in[0,5\cdot10^4]$, with a resolution of 
$\varDelta t=10^{-1}$. The frequency is normalized by the frequency
$\omega_\text{p}\approx0.016$ of the periodic attractor found for 
$\tau=16.40$.
For periodic motion ($\tau=16.40$, blue) the distribution consists of 
isolated peaks, which widen when going from PPC ($\tau=16.56$, green) to 
classical chaos ($\tau=17.20$, red).
}
\end{figure*}

In addition to the distribution of return times in the 
Poincar\'{e} plane, the power spectrum $S(\omega)$ (the 
spectral density) of an attractor can be used to 
characterize classes of distinct time delay dynamics
\cite{conte2017elementary,gwinn1986frequency}.
It is evaluated from the Fourier transformation 
$\hat{x}(\omega)$ of a trajectory $x(t)$ as
\begin{equation}
\hat{x}(\omega)=\int\limits_{-\infty}^\infty\!\!\!\mathrm{d}t\; x(t)\;
\e^{-\imath\omega t}\,,\quad\qquad
S(\omega)=\lVert \hat{x}(\omega) \rVert^2\,, 
\label{eq:fourierspec}
\end{equation}
which is in practice evaluated using numerical tools, such as 
the Fast Fourier Transformation \cite{press1989numerical,boyd2001chebyshev}.  
For comparison, Fig.~\ref{fig:fourierspectrum} shows the power
spectra of a periodic, a partially predictable and a classical 
chaotic attractor of the Mackey-Glass system (\ref{eq:mackeyglass}).

The frequency has been rescaled in Fig.~\ref{fig:fourierspectrum}
by the frequency $\omega_\text{p}=0.016$ of the periodic trajectory, 
which corresponds to the period $T_\text{p}=2\pi/\omega_\text{p}\approx390$. 
As a consequence of the eightfold winding of the limit cycle the 
main peak in the corresponding power spectrum occurs at 
$\omega_\text{qp}=8\,\omega_\text{p}$, with a
winding time of $T_\text{qp}=T_\text{p}/8\approx48.8$
(cf.~Fig.~\ref{fig:mgppc} and Sect.~\ref{sec:poincare}).
The remainder of the spectrum of the limit cycle consists 
of sharp peaks at integer multiples of the frequency $\omega_\text{p}$.

The peaks in the spectral density of the PPC attractor shown in
Fig.~\ref{fig:fourierspectrum} overlap with the spectrum of 
the periodic attractor, having in addition smaller contributions 
close to the main peak. This behavior results from the fact that 
the topology of the partially predictable chaotic attractor resembles the 
topology of the former limit cycle (cf.~Fig.~\ref{fig:mgppc}). For 
the frequency spectrum of the classical chaotic attractor one can 
also observe major contributions close to the frequencies of the 
periodic orbit, this time however with a substantial spread 
\cite{glazier1988quasi,gwinn1986frequency}.

\begin{table*}[b]\centering
\caption{\label{tab:dimensions}
Comparing the Mori dimension $D_\text{M}$ and the Kaplan-Yorke dimension
$D_\text{KY}$ to the fractal dimension $D_\text{f}$ and the correlation
dimension $D_\text{c}$ of the attractors shown in Figs.~(\ref{fig:mgchaos}) and
(\ref{fig:mgppc}). PPC stands for partially predictable chaos.
}
 \begin{tabular}{cccccl}
  \toprule
  $\tau$&$D_\text{M}$&$D_\text{KY}$&$D_\text{f}$&$D_\text{c}$&dynamics\\
  \midrule
  $16.40$&$1$&$1.094$&$1.008\pm0.014$&$1.000\pm0.001$&regular\\
  $16.56$&$2$&$2.061$&$1.968\pm0.005$&$1.964\pm0.003$&PPC\\
  $16.78$&$2$&$2.110$&$2.040\pm0.007$&$2.070\pm0.005$&PPC\\
  $17.20$&$2$&$2.180$&$2.133\pm0.009$&$2.146\pm0.003$&classical chaos\\
  $30.00$&$3$&$3.707$&$3.258\pm0.026$&$3.197\pm0.022$&hyperchaos\\
  \bottomrule
 \end{tabular}
\end{table*}

\subsection{The dimension of attractors}
\label{sec:dimensions}

An interesting point when investigating chaotic dynamics
is the dimension of the attracting set of points in
phase space. Different measures describing the number of 
independent dimensions needed for embedding the attractor,
based either on the geometric properties \cite{mandelbrot1982fractal}, 
on the change of the entropy \cite{benettin1976kolmogorov}, and 
on the correlation of trajectories on the attractor
\cite{grassberger1983characterization}, have been proposed
in this context. The embedding dimension is of particular relevance
for infinite dimensional systems, such as a DDE, as it determines
the number of time delays $\tau_i\in[0,\tau]$ needed to span 
a minimal Poincar\'{e} hypercube $\{x(t-\tau_i)\}$.
An attractor can then be studied without information loss
via its projection onto the minimal Poincar\'{e} hypercube.

In this section several different definitions for the dimension 
of an attractor are reviewed, of which two are computed from 
the Lyapunov spectrum (cf.~Sect.~\ref{sec:lyapunov}), with the 
remaining two definitions retrieving geometric information from 
Poincar\'{e} sections. An overview of the respective estimates
for the attractors shown in Figs.~(\ref{fig:mgchaos}) and
(\ref{fig:mgppc}) is given in Table~\ref{tab:dimensions},
as discussed below.

\begin{figure*}[t]\centering
\includegraphics[width=\textwidth]{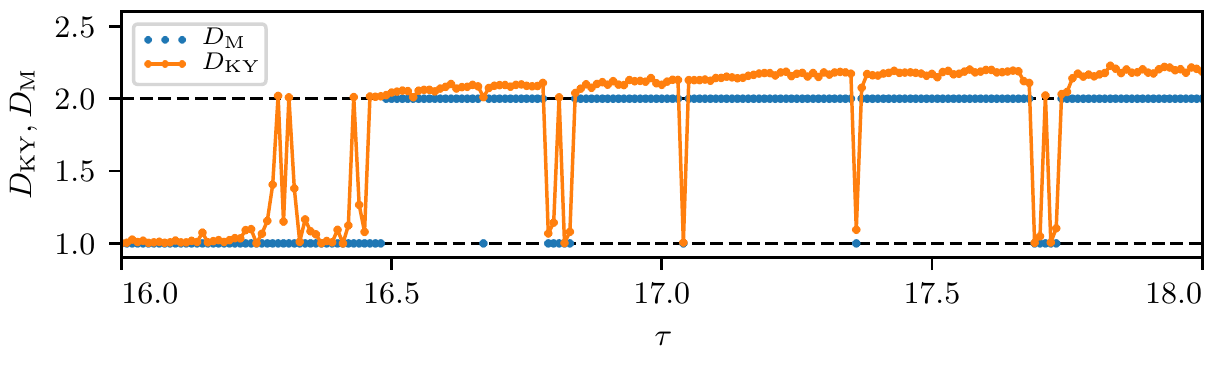}
\caption{\label{fig:dimensions}
The Mori dimension $D_\text{M}$ and the Kaplan-Yorke dimension 
$D_\text{KY}$, defined respectively by (\ref{eq:mori}) and
(\ref{eq:kaplanyorke}), for the Mackey-Glass system 
(\ref{eq:mackeyglass}). The estimates have been computed from 
the Lyapunov spectrum as a function of the time delay $\tau$
(cf.~Fig.~\ref{fig:lyapspec}). For time delay systems the 
Mori dimension $D_\text{M}$ attains only integer values, 
which is the number of dimensions for embedding the Poincar\'{e} 
section of the corresponding attractor. The Kaplan-Yorke 
dimension $D_\text{KY}\geq D_\text{M}$ is in contrast able to
probe the fractal character of the attractors.
}
\end{figure*}

\subsubsection{Mori dimension}

The Mori dimension $D_\text{M}$ is calculated from the ordered 
spectrum of global Lyapunov exponents $\lambda_n$ via 
\cite{farmer1982chaotic,mori1980fractal} 
\begin{align}
 D_\text{M}&=p+\frac{\sum_{\lambda_n>0}\,\lambda_n}
{\sum_{\lambda_n<0}\,\lvert\lambda_n\rvert}\,,
\qquad\quad
\lambda_\text{max}=\lambda_1\geq\lambda_2\geq\ldots
\label{eq:mori}
\end{align}
where $p$ denotes the number of non-negative Lyapunov exponents
$\lambda_n\geq0$. It is constructed to weigh the contribution 
of expanding dimensions $\lambda_n>0$ with respect to
the contribution of contracting dimensions $\lambda_n<0$.

For time delay systems the Mori dimension reduces to $D_\text{M}=p$,
due to the fact that the spectrum of negative exponents is not
integrable, viz that $\lambda_n\sim-\log n$ for $n\to\infty$ 
(cf.\ Sect.~\ref{sec:lyapunov}). The results for the Mori dimension
of the Mackey-Glass system (\ref{eq:mackeyglass}) are given 
in Fig.~\ref{fig:dimensions} as function of the time delay $\tau$,
see also Fig.~\ref{fig:lyapspec}. The Mori dimension is 
$D_\text{M}=1$ for limit cycles and
$D_\text{M}=2$ for both classical and partially predictable chaos,
increasing further for hyperchaos. A comparison is presented
in Table~\ref{tab:dimensions}.

\subsubsection{Kaplan-Yorke dimension}

The Kaplan-Yorke dimension
\cite{farmer1982chaotic,grassberger1984dimensions}, originally 
also called Lyapunov dimension \cite{frederickson1983liapunov}, 
is defined by
\begin{align}
 D_\text{KY}&=j+\frac{\sum_{n=1}^{j}\lambda_n}{\lvert\lambda_{j+1}\rvert}\,,
\label{eq:kaplanyorke}
\end{align}
which resembles the definition of the Mori dimension 
(\ref{eq:mori}). Here $j$ is the largest index for which the 
sum of Lyapunov exponents is not negative:
\begin{align}
 &\sum\limits_{n=1}^{j}\lambda_n\geq0 \qquad \text{and} \qquad
 \sum\limits_{n=1}^{j}\lambda_n+\lambda_{j+1}<0 \label{eq:kydimcond}\,.
\end{align} 
The first sum in (\ref{eq:kydimcond}) takes into account 
the $j$ largest dimensions describing the overall expansion 
of the system, that is the maximal number of exponents for which
the phase volume expansion, as defined in Sect.~\ref{sec:contractionrate},
is still positive. 

With the second term in (\ref{eq:kaplanyorke}) the non-integer
part of the fractal dimension of a chaotic attractor is estimated
as the ratio of the phase volume expansion generated by the $j$
largest exponents, $\sum_{i=1}^{j}\lambda_n$, and the magnitude
of the contraction rate due to the next largest exponent, $\lambda_{j+1}$.
From the second condition in (\ref{eq:kydimcond}) one infers that 
$j\leq D_\text{KY}<j+1$.  

The Kaplan-Yorke dimension is used to characterize attractors in
instantaneous and delayed systems \cite{sano1985measurement},
e.\,g., when modelling turning processes~\cite{palmai2013effects}.

With $p$ being the number of non-negative Lyapunov exponents, 
it follows that $p\leq j$ and consequently that the Mori 
dimension is a lower bound for the Kaplan-Yorke dimension,
$D_\text{M} \leq D_\text{KY}$. This relation shows up in 
Fig.~\ref{fig:dimensions}, where both estimates are presented 
in comparison. The Mori and the Kaplan-Yorke dimension take 
the same value $D_\text{M}=D_\text{KY}=1$ when the underlying 
motion is periodic (limit cycle). For chaos the Kaplan-Yorke 
dimension is fractal and hence larger, $D_\text{KY}>D_\text{M}$. 
See also Fig.~\ref{fig:lyapspec}. The Kaplan-Yorke dimension 
does however not distinguish qualitatively between classical 
and partially predictable chaos
(cf.\ Table~\ref{tab:dimensions} and Fig.~\ref{fig:correlationrange}).

\begin{figure*}[t]\centering
\includegraphics[width=0.55\textwidth]{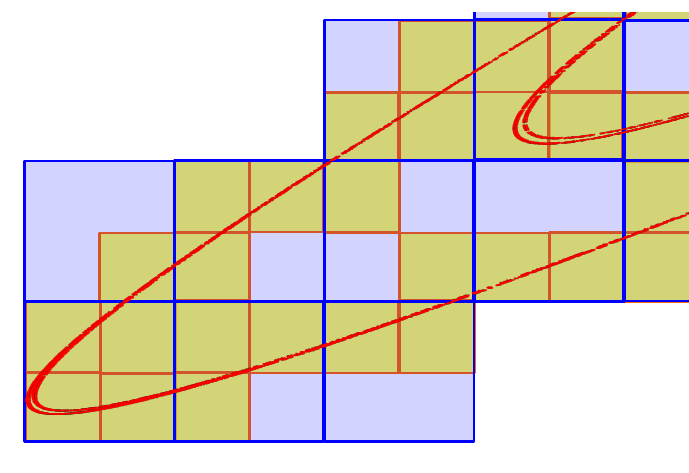}
\caption{\label{fig:sketchboxcounting}
Sketch of the box-counting method for computing the fractal dimension $D_\text{f}$
of a trajectory (red lines), as projected to a two-dimensional plane
(cf.~Fig.~\ref{fig:poincaresection}), by counting the number of boxes 
$N(\epsilon)$ (orange squares) it takes to cover the attracting set 
in relation to the box size $\epsilon$.
}
\end{figure*}

\subsubsection{Fractal dimension}
\label{sec:fractaldimension}

The fractal dimension $D_\text{f}$ measures the space-filling 
capacity of a geometric set \cite{farmer1983dimension},
or of a set of points embedded in a $D\geq D_\text{f}$ dimensional 
space \cite{mandelbrot1982fractal}, e.\,g.\ such as a time series 
$\left\{\mathbf{x}_i=x(t_i)\right\}$ sampled equidistant 
in time from a trajectory $x(t)$. It is effectively
defined by the scaling exponent of the number $N(\epsilon)$ 
of $D$ dimensional boxes with box size $\epsilon$ needed to cover 
the set in the limit of small boxes:
\begin{align}
 N(\epsilon) &\sim \epsilon^{-D_\text{f}} \qquad \text{for} \qquad \epsilon\to0\,.
\end{align}
Equivalently one has
\begin{align}
 D_\text{f}&=-\lim\limits_{\epsilon\to0}\frac{\log N(\epsilon)}{\log \epsilon}\,.
\end{align}
The method, which is also called \text{box-counting}, is illustrated in
Fig.~\ref{fig:sketchboxcounting} for the two-dimensional projection
of an attracting set. For a simple geometric object the fractal dimension 
is integer, as it corresponds to the number of linearly independent 
vectors needed to span the object. However, for objects with a more 
complicated, e.\,g.\ fractal structure, such as the Poincar\'{e} 
section of chaotic attractors, the fractal dimension attains non-integer
values. It has been conjectured that the fractal and the Kaplan-Yorke 
dimension may coincide \cite{frederickson1983liapunov}.

In order to determine the fractal dimension of an attractor one usually
considers two options: either retrieving the fractal dimension from a
trajectory on the attractor; alternatively one performs the box-counting on the
Poincar\'{e} section of the trajectory and determines the fractal dimension of
the sections, which neglects by definition of the section one dimension.
With the Poincar{\'e} hyperplane of a DDE being infinite-dimensional,
one then works with a projection, with the dimension $D$ of the
projection being large enough to embed the attractor in question, 
that is at least the overall embedding dimension minus one.

There are different approaches for embedding an infinite-dimensional
attractor in a time delay system to a
space spanned by $D<\infty$ dimensions.
The so-called time delay embedding or Takens' embedding is one of the most
widely used techniques \cite{takens1981detecting,sauer1991embedology}.
In practice one selects $D$ time delays $\tau_1,\tau_2,\ldots,\tau_D$
for the embedding, such that
$\mathbf{x}_i=(x(t_i-\tau_1),x(t_i-\tau_2),\ldots,x(t_i-\tau_D))$ corresponds to
the projection of the time series sampling the attractor.
A convenient choice for the embedding delays is
$\tau=\tau_1>\tau_2>\ldots>\tau_D>0$, where $\tau>0$ denotes the
delay of the system.
Note that Takens embedding does not require the underlying dynamics to be
delayed.
It rather samples past states in order to describe a system's state.
How to find the minimal embedding dimension, i.\,e.\ how to
determine the smallest possible $D$ for which the embedded 
attractor has the same features as the original dynamics, 
is a problem that has been studied extensively 
\cite{kennel1992determining,cao1997practical}.

\begin{figure*}[t]\centering
\includegraphics[width=0.98\textwidth]{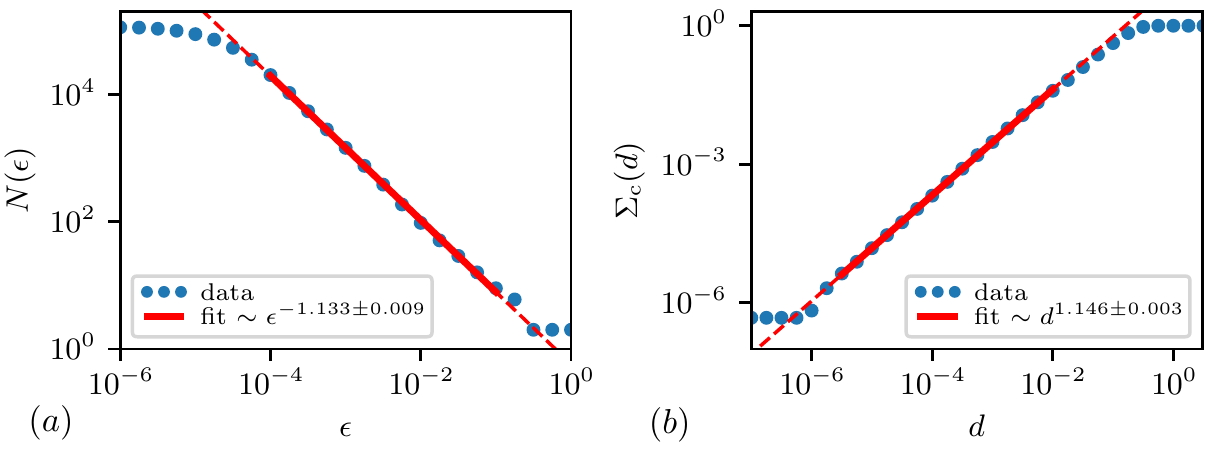}
\caption{\label{fig:boxcorrdim}
($a$)~Number of boxes
$N(\epsilon)$ over the box size $\epsilon$ (blue bullets) for the Poincar\'{e} 
section of the chaotic attractor $\tau=17.20$ embeded into two dimensions as
presented in Fig.~\ref{fig:poincaresection} on log-log axes. The slope 
of the fit (red line) yields the estimate
$D_\text{f}=1+1.133\pm0.009$ for the fractal 
dimension $D_\text{f}$ of the attractor.
($b$)~The correlation integral $\Sigma_\text{c}(d)$, as defined 
by Eq.~(\ref{eq:corrint}), as a function of the distance $d$ (blue bullets) 
for the chaotic attractor $\tau=17.20$ shown in 
Fig.~\ref{fig:poincaresection}, on log-log axes.
The slope of the fit (red line) yields an estimate
$D_\text{c}=1+1.146\pm0.003$ for the correlation
dimension of the attractor.
}
\end{figure*}

For the box counting of the chaotic attractor in the Mackey-Glass system
we use the time series of the states obtained from the
Poincar\'{e} section in Fig.~\ref{fig:poincaresection},
i.\,e.\ $\mathbf{x}_i=(x(t^{(\text{p})}_i-\tau_1),\ldots,x(t^{(\text{p})}_i-\tau_D))$.
The result is plotted in Fig.~\ref{fig:boxcorrdim}, 
with both the number of boxes $N(\epsilon)$ and the box size $\epsilon$ 
being logarithmic. From a linear fit one retrieves the 
exponent of the fractal dimension of the Poincar\'{e} 
section, which is here $D_\text{f}^{(2)}\approx1.13$.
The fractal dimension of the attractor, which is shown in 
Fig.~\ref{fig:mgchaos}, is in consequence $D_\text{f}\approx2.13$. 
The range of box sizes for which the linear fit holds is 
$\epsilon>10^{-4}$, due to the circumstance that the number of 
points in the Poincar\'{e} section is limited 
to $\sim\!10^5$, for computational reasons.

A comparison of the estimates for the fractal and the Kaplan-Yorke
dimension is given in Table~\ref{tab:dimensions}. For regular 
motion and classical chaos the results are in good agreement, 
though for PPC a substantial quantitative discrepancy 
is observed. We note that the dimension used for embedding the 
Poincar\'{e} section has been selected to be the Mori dimension. 
The fractal dimension would however 
change only for an embedding with of insufficient dimension.

\begin{table*}[b!]\centering
\caption{\label{tab:binarytest}
Comparing the exponent $K$ from the Gottwald-Melbourne $0-1$ test and the
distance scaling exponent $\nu$ of the Mackey-Glass
system~(\ref{eq:mackeyglass}) for different delays $\tau$.  $K$ is computed
from $5\cdot10^3$ points sampled with step size $\varDelta t=10$ and $\zeta=2$
in (\ref{eq:gottwaldangular}) (cf.~Fig.~\ref{fig:gottwald}). 
The distance scaling exponent $\nu$ is averaged over $100$ pairs 
trajectories starting from initial distance
$\delta=10^{-6}$ (cf.\ Fig.~\ref{fig:distscaling}).
 }
 \begin{tabular}{cccl}
  \toprule
  $\tau$&$\nu$&$K$&dynamics\\
  \midrule
  $16.40$&$0.997\pm0.003$&$0.01 \pm 0.04$&regular\\
  $16.56$&$0.021\pm0.005$&$0.70 \pm 0.02$&PPC\\
  $16.78$&$0.015\pm0.006$&$0.61 \pm 0.02$&PPC\\
  $17.20$&$0.001\pm0.003$&$0.97 \pm 0.01$&classical chaos\\
  $30.00$&$0.001\pm0.004$&$1.04 \pm 0.01$&hyperchaos\\
  \bottomrule
 \end{tabular}
\end{table*}

\subsubsection{Correlation dimension}

An alternative to box counting is the correlation integral $\Sigma_\text{c}(d)$
\begin{align}
\Sigma_\text{c}(d)&=\lim\limits_{N\to\infty}\frac{1}{N^2}\sum\limits_{i,j=0}^N\theta
\big( d-\lVert \mathbf{x}_i-\mathbf{x}_j\rVert \big)\,,
\label{eq:corrint}
\end{align}
which depends on the distance $d$ and where $\theta$ denotes
the Heaviside function. The correlation integral measures the 
spatial correlation of a set of $N$ points $\left\{\mathbf{x}_i\right\}$ 
sampled equidistant in time from the trajectory of an attractor or a set of
points in a Poincar\'{e} section.

The correlation dimension $D_\text{c}$ is defined from the scaling 
of the correlation integral $\Sigma_\text{c}(d)$ with the distance $d$ in the 
limit of small distances
\cite{grassberger1983measuring,grassberger1983characterization},
\begin{align}
D_\text{c}&=\lim\limits_{d\to0}\frac{\log \Sigma_\text{c}(d)}{\log d}\,.
\label{eq:corrdim}
\end{align}
An example is shown in Fig.~\ref{fig:boxcorrdim}, where the
correlation integral has been computed for $\sim10^5$ points from 
the projected Poincar\'{e} section of the chaotic attractor $\tau=17.20$ 
shown in Fig.~\ref{fig:poincaresection}. From the linear fit to 
the log-log representation one obtains $D_\text{c}^{(2)}\approx1.15$ 
for the Poincar\'{e} section and thus $D_\text{c}\approx2.15$ for 
the trajectory of the chaotic attractor. The fractal dimension 
$D_\text{f}$ has been shown to be an upper bound for the correlation 
dimension \cite{grassberger1983measuring}
(cf.\ Table~\ref{tab:dimensions}).


\subsection{Binary tests for identifying chaos}

The measures described hitherto are capable of characterizing 
different types of dynamics in a quantitative manner. However, 
quantities such as the maximal 
Lyapunov exponent and the fractal dimension change
continuously between regular motion and chaotic sates. 
It is numerically therefore challenging to detect a qualitative
difference in the vicinity of the transition.

In this section we present two alternative methods, which
are based respectively on the computation of distinct scaling 
exponents and which hence are capable of identifying chaos in 
a binary manner. A comparison of the corresponding results for 
the Mackey-Glass system for selected time delays is
presented in Table~\ref{tab:binarytest}.

\begin{figure*}[t]\centering
\includegraphics[width=\textwidth]{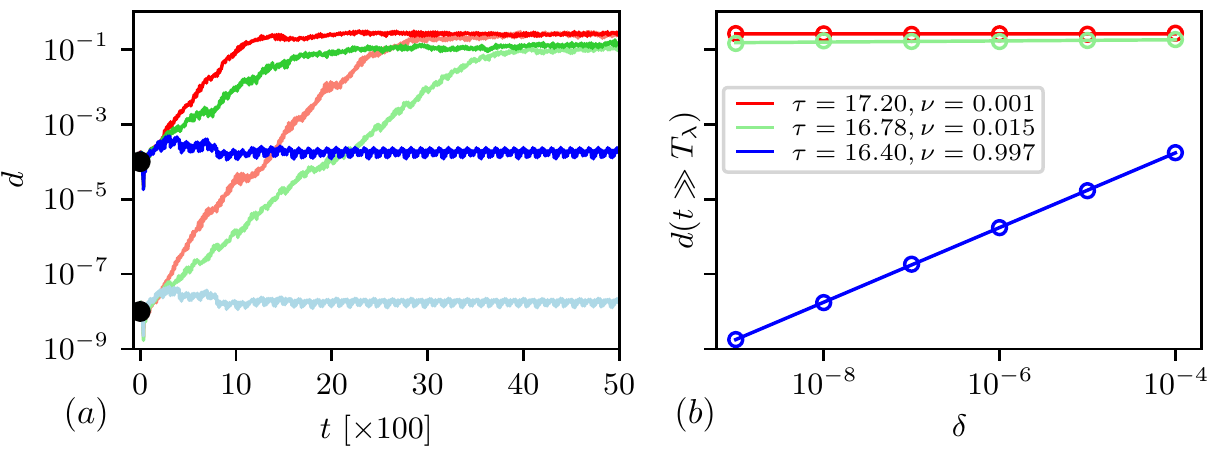}
\caption{\label{fig:distscaling}
The evolution of the distance $d=d(t)$ between pairs of trajectories
in the Mackey-Glass system~(\ref{eq:mackeyglass}). For each
time delay $\tau$ and initial distance $\delta$ an average over 
$100$ pairs has been performed.
($a$)~For initial distances $\delta=10^{-8}$ (light colors)
and  $\delta=10^{-4}$ (dark colors), as marked by black bullets.
For PPC ($\tau=16.78$, green) and classical chaos ($\tau=17.20$, red), 
the distance grows exponentially as a function of time $t$, with the 
saturation plateau $d(t\to\infty)\sim\sigma$ being of order of the 
attractor size $\sigma$, independently of the initial distance $\delta$.
For regular motion ($\tau=16.40$, blue), the long-term distance depends 
on the initial separation $\delta$.
($b$)~Log-log plot of the long-term distance plateau $d(t\!\gg\!T_\lambda)$ 
as a function of the initial separation $\delta$, with the lines corresponding
to linear fits. The dependence is linear for regular motion (blue), being
near to constant on the other side for both partially predictable
(green) and classical chaos (red).
}
\end{figure*}

\subsubsection{Cross-distance scaling exponent}

In the vicinity of a chaotic attractor the divergence of two 
trajectories $x_\text{o}(t)$ and $x_1(t)$ with a small initial 
distance $d(t\!=\!0)=\delta\ll\sigma$ is exponential for 
$t<T_\lambda$. Here we have denoted with $\sigma^2$ the variance 
of the attractor. For systems characterized by attractors 
confined in a finite volume element of the phase space, viz 
when $\sigma<\infty$, the cross-distance $d(t)$ of a pair 
of trajectories reaches a saturation level, 
$d(t\!\gg\!T_\lambda)\approx\text{const.}$, after the initial
divergence. 

As an example we present in Fig.~\ref{fig:distscaling} 
the evolution of the inter-trajectory distance $d(t)$ 
of the Mackey-Glass system (\ref{eq:mackeyglass}) for 
different parameters and initial distances. Due to 
the finite size of the attractor the long-term distance 
is independent of the initial conditions $\delta$ for both
classical and partially predictable chaos, one hence 
finds that $d(t\gg T_\lambda)\sim\sigma$. This saturation 
is a consequence of the decorrelation of pairs of 
trajectories occurring in the vicinity of chaotic attractors,
as discussed in Sect.~\ref{sec:ppc}. See also
the decorrelation condition (\ref{eq:corrcond}).

On the other hand, in the case of periodic motion, the
long-term distance varies linearly with the initial 
distance \cite{wernecke2017test}, as shown in 
Fig.~\ref{fig:distscaling}.
Introducing the cross-distance scaling exponent $\nu$,
one can summarize the scaling relation as 
\begin{align}
d(t\!\gg\! T_\lambda)&\propto\delta^\nu\,,
\qquad\text{where}\qquad \left\{ \begin{array}{cl}
\nu=1 & \text{ for regular motion}\\ \nu=0
& \text{ for chaos}  \end{array} \right.
\label{eq:distexp}
\end{align}
The scaling exponent attains in general only two values, 
$\nu\in\{0,1\}$, quali\-fying hence as a binary indicator 
for chaos and, respectively, for regular motion, as 
evident from Fig.~\ref{fig:distscaling}. Binary classification 
using (\ref{eq:distexp}) works also for hyperchaos 
(cf.\ Table~\ref{tab:binarytest}).

\begin{figure*}[t]\centering
\includegraphics[width=\textwidth]{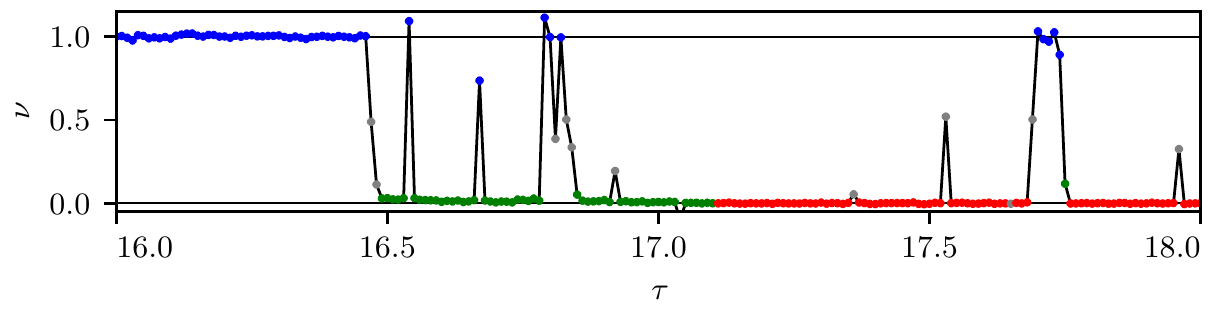}
\caption{\label{fig:distscalingrange}
The cross-distance scaling exponent $\nu$~(\ref{eq:distexp}) 
for the Mackey-Glass system~(\ref{eq:mackeyglass}). The 
exponent distinguishes in a binary manner regular motion 
($\nu\approx1$, blue), when changing the time delay $\tau$, 
from chaos ($\nu\approx0$, green, red). A sharp drop
marks the transition from regular motion (blue) to 
partially predictable chaos (green). The exponents
are obtained by fitting the long-term inter-pair distance 
plateaus as a function of the initial distances $\delta$,
with $\delta\in[10^{-9}, 10^{-4}]$, when 
averaged over $100$ pairs of trajectories 
(cf.\ Fig.~\ref{fig:distscaling}).
}
\end{figure*}

The binary character of the cross-distance scaling exponent 
can be seen also in the parameter scan presented in 
Fig.~\ref{fig:distscalingrange}. The scaling exponent 
$\nu$ allows therefore to determine the transition between 
regular motion and chaos, as well as the presence
of periodic windows.

\begin{figure*}[t]\centering
\includegraphics[width=1.\textwidth]{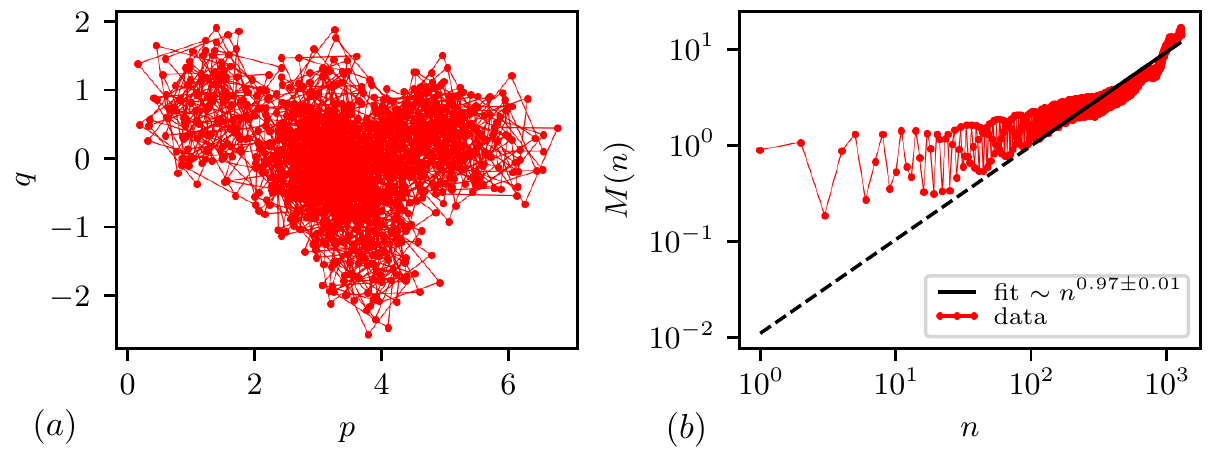}
\caption{\label{fig:gottwald}
The Gottwald-Melbourn $0-1$ test for a chaotic attractor of the Mackey-Glass
system~(\ref{eq:mackeyglass}) with $\tau=17.20$, using $\phi_n=x(t_n)$ 
(extracted from $10^3$ points sampled with step size $t_{n+1}-t_n=10$)
as the driving time series for the measuring device (\ref{eq:gottwaldangular}).
Here $\zeta=2$.
($a$)~The motion in phase space $p,\,q$ resembles a diffusive process.
($b$)~Log-log plot of the mean-square deviation $M$, as defined by 
(\ref{eq:gottwaldmsd}), as function of $n$. The close to linear growth
exponent~$K\approx0.97$ correctly indicates chaotic motion.
}
\end{figure*}

\subsubsection{Gottwald-Melbourn test}

For the Gottwald-Melbourne test one uses a time series 
characterizing the attractor under consideration to
drive a dynamical system, which serves hence as a
`measuring device' \cite{gottwald2004new,gottwald2005testing}. 
We discuss here the case of a scalar time series
$\{\phi_j\}_{1\leq j\leq N}$, which may be extracted, e.\,g.,
form a scalar projection of a given trajectory. This time 
series is used to drive the evolution of a two-dimensional mapping:
\begin{align}
p(n+1)&=p(n)+\phi_n\cos(n\zeta)\,, \qquad
q(n+1)=q(n)+\phi_n\sin(n\zeta)\,,
\label{eq:gottwaldangular}
\end{align}
where $\zeta>0$ corresponds to a constant angular velocity
and $p$ and $q$ to the map coordinates.
Of interest is the mean-square displacement (MSD)
\begin{align}
M(n)&=\lim\limits_{N\to\infty}\frac{1}{N}
\sum\limits_{j=1}^{N}\Big(\big( p(j+n)-p(j) \big)^2 
+ \big( q(j+n)-q(j) \big)^2\Big)\,,
\label{eq:gottwaldmsd}
\end{align}
which reflects the properties of the driving time series
through the mapping~(\ref{eq:gottwaldangular}). It has
been proposed \cite{gottwald2004new,gottwald2005testing},
that the MSD is constant when the driving time series 
describes regular motion, growing on the other hand 
linearly with $n$ for irregular behavior. This would
imply the binary growth rate
\begin{align}
 K&=\lim\limits_{n\to\infty}\frac{\log M(n)}{\log
n}=\left\{\begin{array}{cl}0&\text{ for regular motion}\\1&\text{ for chaotic
motion}\end{array}\right.	\label{eq:gottwaldexp}\,.  \end{align}
As an example we present in Fig.~\ref{fig:gottwald} the $p,\,q$ 
phase plane plot together with the MSD, the latter as function 
of iteration number $n$, for a chaotic attractor of the Mackey-Glass 
system (\ref{eq:mackeyglass}). 
The representation in the phase space of the `measuring device'
(\ref{eq:gottwaldangular}) resembles a diffusion process. 
From a linear fit to the MSD in Fig.~\ref{fig:gottwald} one obtains 
a close to linear growth rate $K\approx1$, which correctly indicates chaotic motion.

The Gottwald-Melbourne test is an interesting approach, which can 
be used at times to effectively identify chaos in time delay systems 
\cite{litak2012nonlinear}. It is however also known to yield ambiguous results 
in some particular cases \cite{litak2009identification, wernecke2017test}.
The results presented for different attractors of the Mackey-Glass system in
Table~\ref{tab:binarytest} yield correct results for periodic motion,
classical chaos and hyperchaos.
Though for PPC the results are ambiguous, which might hint at an insufficient,
i.\,e.\ too fine, sampling rate (see also \cite{gottwald2009implementation}).

\begin{figure*}[tbh!]\centering
\includegraphics[width=1.0\textwidth]{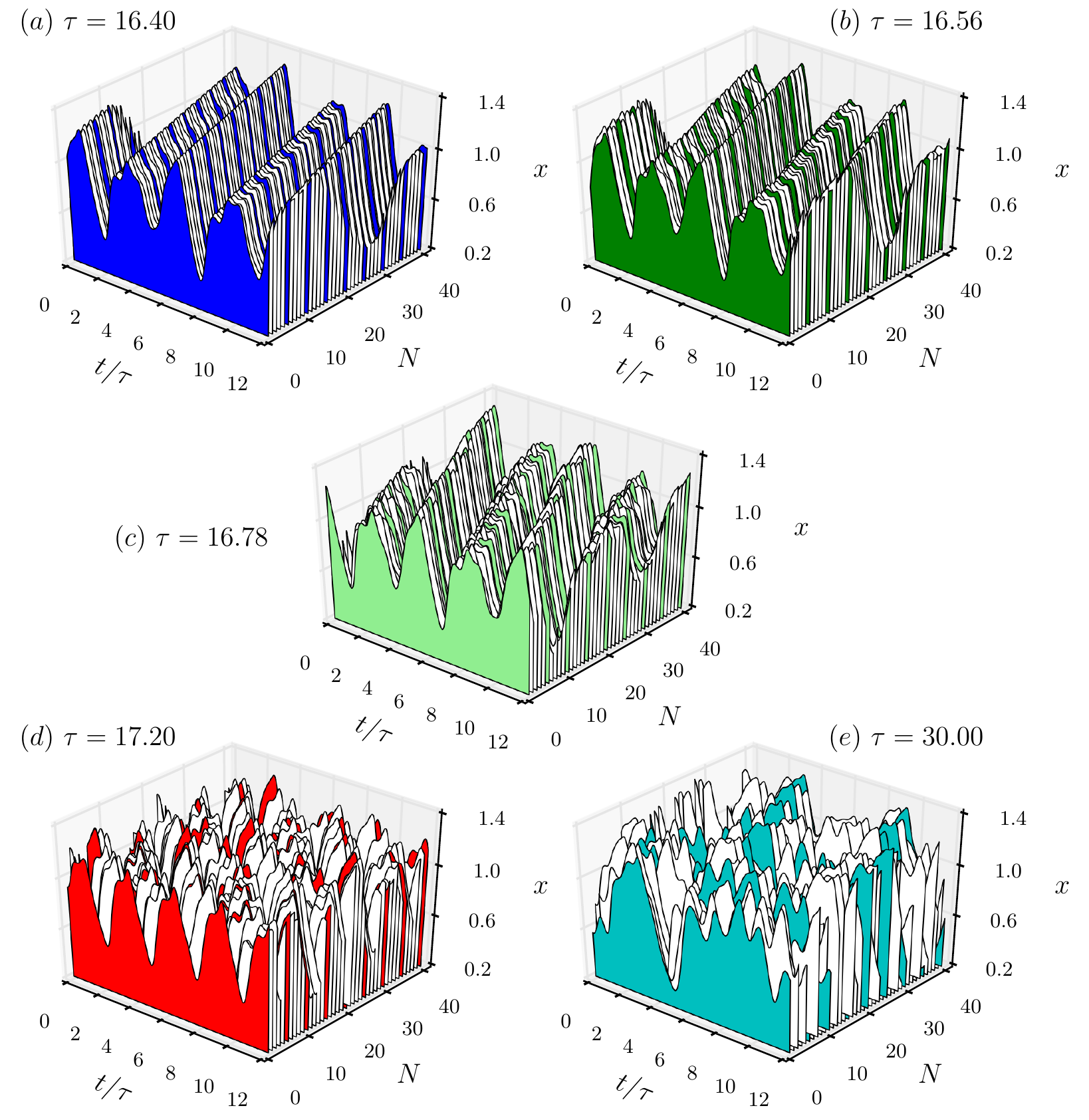}
\caption{\label{fig:spacetime}
Space-time representation of the Mackey-Glass system~(\ref{eq:mackeyglass})
for different values of the delay time $\tau$ following \cite{arecchi1992two}.
In each panel a single trajectory $x(t)$ corresponding to an attractor 
is split along the time axis $t$ into $40$ slices and numbered by
the quasi special dimension $N$, see (\ref{X_N_slice}).
Time is rescaled by the delay time $\tau$ and every fifth 
slice is colored for better visibility. Shown is
($a$)~a periodic orbit for $\tau=16.40$,
($b$)~and ($c$)~partially predictable chaos for $\tau=16.56$ and $\tau=16.78$, 
($d$)~classical chaos for $\tau=17.20$ and ($e$)~hyperchaos for $\tau=30.00$.
}
\end{figure*}

\subsection{Space-time interpretation of time delay systems}

Time delay systems of scalar variables can be interpreted in 
terms of two-dimensional space-time coordinates \cite{arecchi1992two}, 
a visualization technique that helps at times when investigating
complex dynamical patterns \cite{yanchuk2017spatio}.
Within this approach, a scalar trajectory $x(t)$ is cut into slices,
\begin{equation}
X(N)=\{ x(t) \,:\, t\in[NT,(N+1)T] \}\,,
\label{X_N_slice}
\end{equation}
of length $T$, which is usually assumed to be a multiple of
the time delay $\tau$. Each point of the trajectory 
$X(N,t)=x(NT+t)$ is parametrized by the slicing
index $N$ and the time $t\in[0,T]$ within one slice.

In Fig.~\ref{fig:spacetime} the space-time representation 
of a limit cycle and of partially predicable and classical 
chaotic states are shown together with a hyperchaotic 
trajectory of the Mackey-Glass system~(\ref{eq:mackeyglass}).
The periodic motion appears as perfectly regular wave fronts, with
PPC showing slight modulations. For classical chaotic and hyperchaotic 
motion, the space-time representation is instead irregular.

The space-time representation allows for regularities or irregular
patterns to be identified by visual inspections.  Thus, it is used to analyze
pulse trains from laser cavities \cite{terrien2018pulse}, spatio-temporal
pattern formation in systems with multiple delays \cite{yanchuk2014pattern}, 
and for the identification of chimera states in time delay systems  
\cite{larger2013virtual}.

\section{Numerical treatment}
\label{sec:numerics}

\begin{figure*}[t]\centering
\includegraphics[width=0.98\textwidth]{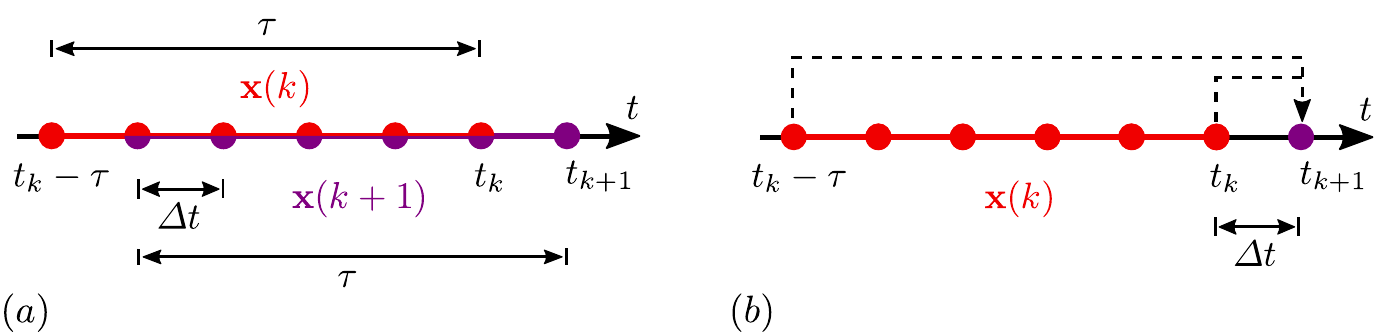}
\caption{\label{fig:sketch_numerics}
($a$) Illustration of the discretized state history $\mathbf{x}(k)$ that
approximates the history of a time delay system on the interval
$t\in[t_k-\tau,t_k]$, with the $N>0$ steps being equidistantly spaced 
in time by $\varDelta t$. 
($b$) The Euler algorithm for the numerical integration
(\ref{eq:euler}) approximates the next state $x(t_{k+1})$ combining 
the information of the first and last state in the state 
history $\mathbf{x}(k)$.
 }
\end{figure*}

In this section, which is concerned with the numerical treatment 
of delay differential equations (DDE), we restrict ourselves
for the sake of simplicity to autonomous DDE with constant 
time delay $\tau>0$ and generic flow $f$,
\begin{align}
\dot{x}(t)&=f\big(x(t), x(t-\tau)\big)\label{eq:numdde}~,
\end{align}
where $x=x(t)$ is the scalar state of the system parametrized by time 
$t$.
For an ordinary differential equation (ODE), a state in the phase space 
of the system determines the time evolution uniquely. Discretizing
time, the full information about a system with fixed time delay $\tau$ 
at time $t=t_k$ is contained in contrast in the system's state history, 
i.\ e.\ the states $x(t)$ on the whole interval $t\in[t_k-\tau,t_k]$.
A discretization into $N>0$ equally spaced time steps, i.\,e.\ $N-1$ time
intervals, therefore leads to a step-size $\varDelta t=\tau/(N-1)$,
with the discretized state history taking the form
\begin{align}
\label{4_discrete_state_histories}
 \mathbf{x}(k)&=\left\{ x(t_k-\tau), x(t_k-\tau+\varDelta
t),\ldots,x(t_k-\varDelta t),x(t_k) \right\}\\[0.5ex]
&=\{ x_\text{o}(k), x_1(k), \ldots, x_{N-2}(k), x_{N-1}(k) \}\,.
\end{align}
Note that we used capital letters in Sect.~\ref{sect_state_histories}
to denote state histories which are not discrete, like in 
(\ref{4_discrete_state_histories}), but continuous in time.
The subscript index of $x_j$ indicates $x_j$ is the
$j$th element of a vector, namely that $\mathbf{x}\in\mathbb{R}^N$

Two consecutive discretized time steps are linked by $t_{k+1}=t_k+\varDelta t$,
which implies that the state history vectors $\mathbf{x}(k)$ and $\mathbf{x}(k+1)$ 
differ only with respect to the last element $x(t_{k+1})$ 
(cf.\ Fig.~\ref{fig:sketch_numerics}).

\subsection{Numerical integration}

Numerical methods approximate the exact solution $\hat{x}(t)$ of an
DDE that is determined by an initial function $\varphi(t)$ given 
on the time interval $t\in[t_k-\tau,t_k]$ by a discrete set of points
$\{x(t_k),x(t_{k+1}),\ldots\}$ for time instances $\{t_k,t_{k+1},\ldots\}$
\cite{press1989numerical,atkinson2008introduction}.
At every integration step one can estimate the local numerical error
$\lvert x(t_k)-\hat{x}(t_k)\rvert$, which will generally depend on the
discretization step size $\varDelta t$.

For the purpose of numerical integration the state vector $\mathbf{x}(k)$ 
is updated to the next state vector $\mathbf{x}(k+1)$, as illustrated
in Fig.~\ref{fig:sketch_numerics}. The challenge lies 
in the discretized nature of the state history, which may not contain 
the states $x(t)$ at the times $t$ a given integration algorithm
may need when calculating the new element $x(t_{k+1})$.

\subsubsection{Euler algorithm}

The Euler integration algorithm uses the simplest 
numerical approximation of the time derivative 
occurring in a differential equation,
\begin{align}
\dot{x}(t)&=\lim\limits_{\varDelta t\to0}
\frac{x(t+\varDelta t)-x(t)}{\varDelta t}\approx
\frac{x(t+\varDelta t)-x(t)}{\varDelta t}\,.
\label{eq:eulerapprox}
\end{align}
This approximation implies that
\begin{align}
 x(t_{k+1})&= x(t_k)+F\big(x(t_k),x(t_k-\tau)\big)\varDelta t\label{eq:euler}~,
\end{align}
which requires the system's state at the previous time step $x(t_k)$ and the
delayed state $x(t_k-\tau)$ (cf.~Fig.~\ref{fig:sketch_numerics}). Thus,
for the Euler integration algorithm the discretization step size $\varDelta t$
and the delay $\tau$ must be commensurate, which is in accordance with the
choice $\tau=(N-1)\varDelta t$.

From the approximation (\ref{eq:eulerapprox}) of the time derivative the local
numerical error is $\mathcal{O}(\varDelta t^2)$. The cumulative error
of the Euler method when integrating up successively to a finite time difference
is however $\mathcal{O}(\varDelta t)$, which determines the overall numerical 
accuracy.

\begin{figure*}[t]\centering
 \includegraphics[width=0.98\textwidth]{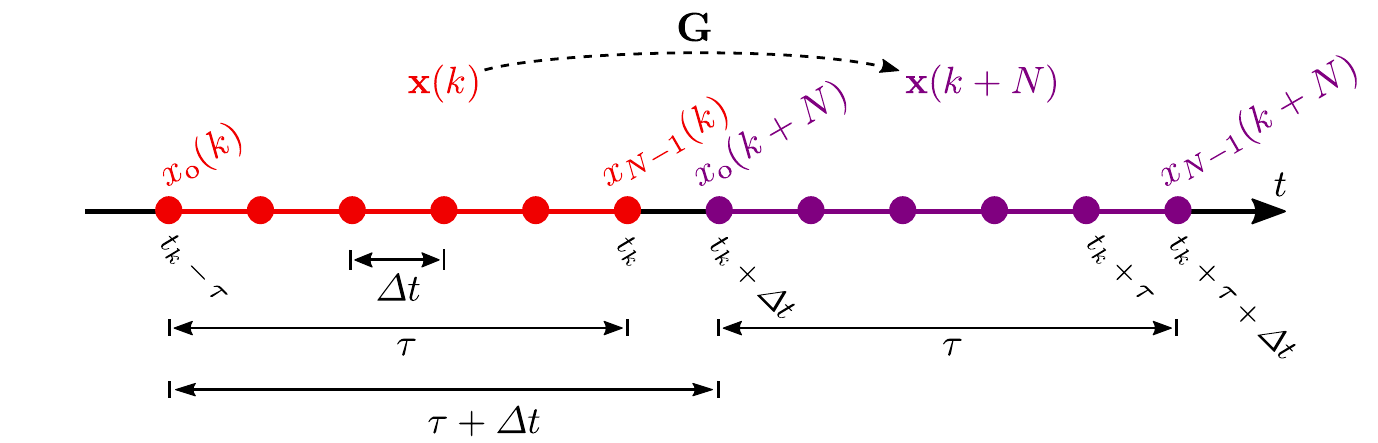}
 \caption{\label{fig:sketch_euler_map}
 The Euler map $G$, as defined by
Eqs.~(\ref{eq:eulermap}) and (\ref{eq:eulermap2}), maps the state
history~$\mathbf{x}(k)$ onto the disjoint state history $\mathbf{x}(k+N)$.
Both state histories contain the discretized states over a time span of
length $\tau$, but shifted by $\tau+\varDelta t$ with respect to each other.}
\end{figure*}

\subsubsection{Euler integration as a discrete map}
\label{sec:eulermap}

As described by \cite{farmer1982chaotic}, one can interpret the Euler algorithm
(\ref{eq:euler}) as the discrete map
\begin{align}
\mathbf{x}(k+N)&=\mathbf{G}\big(\mathbf{x}(k)\big)\,,
\label{eq:eulermap}
\end{align}
where the map $\mathbf{G}\,:\,\mathbb{R}^N\to\mathbb{R}^N$ maps the state
history $\mathbf{x}(k)$ of $N$ time steps of size $\varDelta t$, i.\,e.\ 
with $N\varDelta t=\tau+\varDelta t$, onto the disjoint state history
$\mathbf{x}(k+N)$. At first sight this approach, which is depicted in
Fig.~\ref{fig:sketch_euler_map}, seems arbitrary, but is has the advantage of
being an explicit forward recursive map once the recursive dependencies
are expanded:
\begin{equation}
\label{eq:eulermap2}
\begin{array}{rcll}
 x_\text{o}(k+N) &=& x_{N-1}(k)     
                 &+\,\varDelta t\,F\big( x_{N-1}(k), x_\text{o}(k) \big)\\[0.5ex]
 x_1(k+N)        &=& x_\text{o}(k+N)&+\,\varDelta t\,F\big( x_\text{o}(k+N), x_1(k) \big)\\[0.5ex]
 &\vdots&\\[0.5ex]
 x_{N-1}(k+N)    &=& x_{N-2}(k+N)   
                 &+\,\varDelta t\,F\big( x_{N-2}(k+N), x_{N-1}(k) \big)\,.
\end{array}
\end{equation}
Note the implicit recursion, namely that the RHS of $x_1(k+N)$ depends on
$x_\text{o}(k+N)$, and so on.
 

As an illustrative example we consider as in Sect.~\ref{sect_educational_example}
the integration of $\dot{x}(t)=-x(t-\tau)$, here with step size 
$\varDelta t=\tau/2$, which corresponds to $N=3$ steps 
per state history. The state history therefore consists of
\begin{align*}
\mathbf{x}(k)&= \big\{ x_\text{o}(k),\; x_1(k),\; x_2(k)\big\}\\
&= \big\{x(t_k-\tau),\; x(t_k-\tau/2),\; x(t_k) \big\}~,
\end{align*}
which means for the Euler map (\ref{eq:eulermap}) that one computes 
the consecutive disjoint state history
\begin{align*}
 \mathbf{x}(k+N)&=
\big\{ x_\text{o}(k+N),\; x_1(k+N),\; x_2(k+N) \big\}\\
 &= \big\{ x(t_k+\tau/2),\; x(t_k+\tau),\; x(t_k+3\tau/2) \big\}\,.
\end{align*}
The single states follow from the iterative stepwise map (\ref{eq:eulermap2}):
\begin{align}
 \begin{split}
 x_\text{o}(k+N)&= x_2(k)-\frac{\tau}{2}x_\text{o}(k)\\
 x_1(k+N)       &= x_2(k)-\frac{\tau}{2}\big( x_\text{o}(k)+x_1(k) \big)\\
 x_2(k+N)       &= x_2(k)-\frac{\tau}{2}\big( x_\text{o}(k)+x_1(k)+x_2(k) \big)  \,.
 \end{split}
\end{align}


\begin{figure*}[t]\centering
 \includegraphics[width=\textwidth]{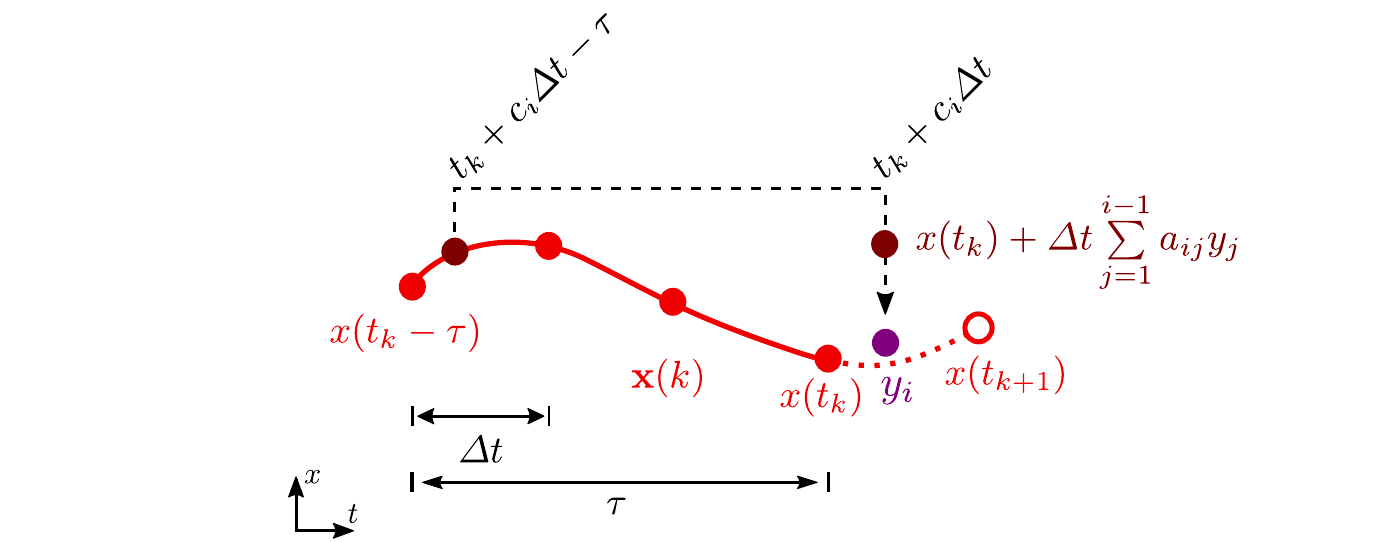}
 \caption{\label{fig:sketch_rk}
  For a Runge-Kutta integration step (\ref{eq:rk}) at $t=t_k$ one
needs to compute the intermediate states $y_i$ defined by 
Eq.~(\ref{eq:rkinter}). The flow is hence to be evaluated at 
times $t_k+c_i\varDelta t$ and at corresponding estimated points 
in phase space. The delayed contribution $x(t_k+c_i\varDelta t-\tau)$ 
to the flow are obtained correspondingly by interpolating the 
state history~$\mathbf{x}(k)$.}
\end{figure*}

\subsubsection{Explicit Runge-Kutta algorithms}

The trade-off between the integration step size and the numerical error makes
the explicit Euler integration algorithm either slow or inaccurate. Its
generalization is referred to as explicit Runge-Kutta (RK) algorithms
\cite{runge1895uber, kutta1901beitrag, nystrom1925uber}. Explicit RK algorithms
use $s>0$ intermediate sampling points $y_i$ to estimate the next state,
\begin{align}
x(t_{k+1})&=x(t_k)+\varDelta t\sum\limits_{i=1}^{s}b_iy_i~,
\label{eq:rk}
\end{align}
where the coefficients $b_i>0$ are weighting factors for the sampling
points. They are computed for $\dot x(t)=f(t,x(t),x(t-\tau))$ iteratively 
as
\begin{align} 
 y_i&=f\left(t_k+c_i\varDelta t,\; x(t_k)+
\varDelta t\sum\limits_{j=1}^{i-1} a_{ij}y_j,\; x(t_k+c_i\varDelta t-\tau) \right)\,.
\label{eq:rkinter}
\end{align}
The flow $f$ is hence evaluated at time instances $t_k+c_i\varDelta t$ in
$[t_k,t_{k+1}]$, which are determined in turn by the coefficients 
$0\leq c_i\leq1$, with the state argument of the flow being a superposition 
of previous intermediate stages, as weighted by the coefficients $a_{ij}$. 

The times at which the $y_i$ are to be evaluated, $t_k+c_i\varDelta t$,
are in general incommensurate with the underlying time discretization,
as illustrated in Fig.~\ref{fig:sketch_rk}, which means that the
state history $\mathbf{x}(k)$ needs to interpolated \cite{neves1981control}. 
However, the advantage is that an $s$ stage RK algorithm comes with 
a global numerical error of the order $\mathcal{O}(\varDelta t^p)$ with $p\leq s$,
allowing such for a faster and/or more accurate integration compared 
to the straightforward Euler method.

The coefficients $a_{ij}, b_i, c_i$ for the explicit RK algorithms are usually
written as a `Butcher tableau' \cite{butcher2016numerical}:

\begin{equation}
 \begin{array}{c|cccccc}
  c_1=0 & 0 & \multicolumn{3}{c}{\dots} & 0\\
  c_2   & a_{21} & 0\\
  c_3   & a_{31} & a_{32} & 0 & & \vdots\\
  \vdots& \vdots &        & \ddots & \ddots\\
  c_s   & a_{s1} & a_{s2}& \dots & a_{s,s-1}&0\\
  \hline
        & b_1 & b_2 & \dots & b_{s-1} & b_s
 \end{array}
\end{equation}
where the upper triangle contains only zeros for explicit RK algorithms.  The
so-called 3/8 rule \cite{kutta1901beitrag} is a fourth order ($s=4$)
Runge-Kutta method has the Butcher tableau:
\begin{equation}
 \begin{array}{c|cccc}
    0 \\
  \nicefrac{1}{3} & \phantom{-}\nicefrac{1}{3} \\
  \nicefrac{2}{3} & -\nicefrac{1}{3} & \phantom{-}1 \\
    1 & \phantom{-}1 & -1 & 1 \\
  \hline
	& \phantom{-}\nicefrac{1}{8} & \phantom{-}\nicefrac{3}{8} &
\nicefrac{3}{8} & \nicefrac{1}{8} \end{array}
\end{equation}
It is appreciated for its stability and convergence
properties \cite{kutta1901beitrag}.

\subsection{Lyapunov exponents}
\label{sect_Lyapunov_exponents}

Lyapunov exponents describe the contraction or expansion of phase space
volume associated with certain directions in phase space, or on an attractor in
particular.  While for an ordinary differential equation there is only a finite
number of Lyapunov exponents, which equals the number of dimensions of the
phase space, a time delay system has infinitely many Lyapunov exponents.
In consequence one can only approximate the $N$ largest exponents (largest by
real part) with numerical methods.  In the following three different commonly
used numerical methods for computing the largest or the $N$ largest Lyapunov
exponents are discussed.

Note that the methods for evaluating Lyapunov exponents presented in this
section are suited for smooth systems.
For non-smooth dynamical systems one typically needs 
dedicated approaches \cite{stefanski2000using,stefanski2005evaluation},
which holds also for time delay systems \cite{palmai2013effects}.

\subsubsection{Maximal Lyapunov exponent from two diverging trajectories}

\begin{figure*}[t]\centering
\includegraphics[width=0.49\textwidth]{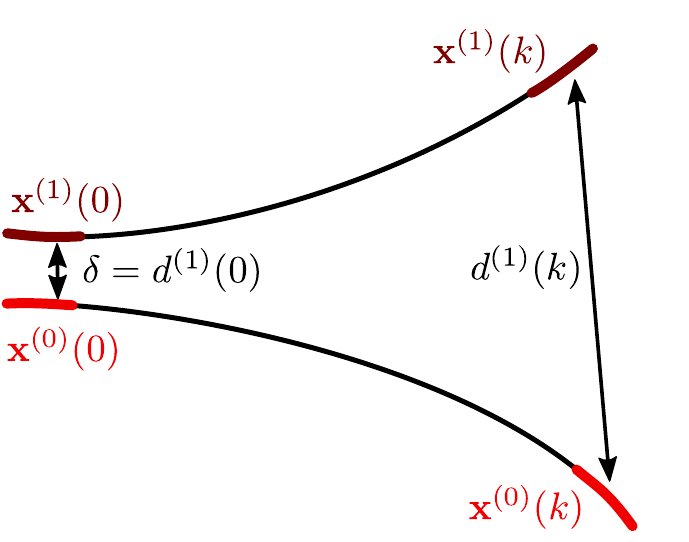}
\caption{\label{fig:lyapdiv}
Measuring the rate of divergence of two trajectories starting at initial
states $\mathbf{x}^{(0)}(0)$ and $\mathbf{x}^{(1)}(0)$ with an initial
distance $d^{(1)}(0)=\delta$ (cf.~Eq.~(\ref{eq:avdist})). The distance 
$d^{(1)}(k)$ after $k$ integration steps is used to 
compute the Lyapunov exponent via~(\ref{eq:lyapdef}). The thick segments
of length $\tau$ indicate the respective state histories.
 }
\end{figure*}

The most basic method of determining Lyapunov exponents implies measuring the
divergence rate of initially close-by trajectories \cite{wernecke2017test}. 
For the maximal Lyapunov exponent $\lambda_\text{max}=\lambda_1$ two initial
state histories $\mathbf{x}^{(0)}(0)=\left\{ x^{(0)}_\text{o}(0), \ldots,
x^{(0)}_{N-1}(0) \right\}$ and $\mathbf{x}^{(1)}(0)$ at $t=t_\text{o}$ are
chosen and evolved for $k>0$ steps, until $t_k=t_\text{o}+k\varDelta
t$, to states $\mathbf{x}^{(0)}(k)$ and $\mathbf{x}^{(1)}(k)$,
cf.~Fig.~\ref{fig:lyapdiv}. 

At every time step $t_k$ one can define the difference vector between 
the two state history vectors,
\begin{align}
 \mathbf{d}^{(1)}(k)=\mathbf{x}^{(1)}(k)-\mathbf{x}^{(0)}(k)~,
\end{align}
from which one can compute the average Euclidean distance of the state history
vectors, namely
\begin{align}
 d^{(1)}(k)&=\frac{\lVert\mathbf{d}^{(1)}(k)\rVert}{\sqrt{N}}=\left(
\frac{1}{N}\sum\limits_{i=0}^{N-1}\left( x_i^{(0)}(k)-x_i^{(1)}(k) \right)^2
\right)^{1/2}\,.
\label{eq:avdist} 
\end{align}
In contrast to the Euclidean distance between continuous state 
history vectors, as defined in Sect.~\ref{sect_state_histories},
the average distance defined by (\ref{eq:avdist}) does not diverge 
in the limit $N\to\infty$. Note also that we adapted the notation
in order to emphasis that we are working in this section with discrete
and not with continuous state histories.

The initial states $\mathbf{x}^{(0)}(0)$ and $\mathbf{x}^{(1)}(0)$ are 
normally chosen randomly in the vicinity of the attractor under 
investigation, with the initial distance $d^{(1)}(0)=\delta$ being 
small, $\delta\ll\sigma$, with respect to the variance $\sigma^2$ 
of the attractor. If $\sigma=0$, as for a fixed point, the initial 
distance should be small with respect to the microscopic length 
scales of the system. The maximal Lyapunov exponent $\lambda_\text{max}$ 
is given, as pointed out in Sect.~\ref{sec:global_maximal_Lyapunov_exponents}, 
by the divergence rate of the two trajectories,
\begin{align}
\lambda_\text{max}&=
\lim\limits_{k\to\infty}\lim\limits_{\delta\to0}\frac{1}{k\varDelta t}
\log\frac{d^{(1)}(k)}{\delta}\,,
\label{eq:lyapdef}
\end{align}
where the limit of infinitely small initial distances $\delta\to0$ 
and the long-term limit for the measurement time $k\varDelta t$
needs to be taken. For 
chaotic attractors, for which pairs of trajectories eventually 
decorrelate, one has that $d^{(1)}(k)\to\sigma$ in the limit 
$k\to\infty$ \cite{wernecke2017test}. The maximal Lyapunov exponent has 
to be evaluated accordingly for intermediate distances $d^{(1)}$, 
as defined by $\delta\ll d^{(1)}\ll \sigma$.

This expression, Eq.~(\ref{eq:lyapdef}), is an intuitive and robust 
method, it allows however to determine only a single Lyapunov
exponent, namely the maximal Lyapunov exponent. An extension 
to compute the $N$ largest Lyapunov exponents is discussed in 
the following section.

\subsubsection{Benettin's algorithm}
\label{sec:benettin}

The largest Lyapunov exponents can be evaluated efficiently following
the idea of Benettin \textit{et al} \cite{benettin1980lyapunov}, which is
widely used and illustrated, e.\,g.\ in \cite{skokos2010lyapunov}. Here we 
discuss two different aspects of Benettin's algorithm.

\paragraph{Iterated finite-time method}

Instead of measuring the divergence of one pair of trajectories for
large times, one can rely on iterated measurements of the rate of
divergence for shorter time intervals. Within this approach one computes 
a reference trajectory $\mathbf{x}^{(0)}$ starting at a random initial 
condition $\mathbf{x}^{(0)}(0)$ in the vicinity of the attractor.
The initial condition $\mathbf{x}^{(1)}(0)$ for the auxiliary trajectory
$\mathbf{x}^{(1)}$ is chosen such that the initial distance is
$d^{(1)}(0)=\delta\ll\sigma$ is small compared to the extent $\sigma$
(or the variance $\sigma^2$) of the attractor under investigation.
The divergence of both trajectories is measured by their distance $d^{(1)}(k)$
after $k$ integration steps, viz after an integration time $k\varDelta t$.
The finite-time Lyapunov exponent,
\begin{align}
\lambda_1^{(\text{ft})}=\frac{1}{k\varDelta t}
\log\frac{d^{(1)}(k)}{\delta}~,
\label{eq:lyaploc}
\end{align}
then provides a local estimate of (\ref{eq:lyapdef}).
Next one rescales the auxiliary vector $\mathbf{x}^{(1)}(k)$ to
$\tilde{\mathbf{x}}^{(1)}(k)$, such that the distance 
to the reference trajectory is reset to
\begin{align}
\lVert\tilde{\mathbf{x}}^{(1)}(k)-\mathbf{x}^{(0)}(k)\rVert
=\sqrt{N}\delta\,.
\label{eq:refrescale}
\end{align}
%
\begin{figure*}[t]\centering
\includegraphics[width=0.98\textwidth]{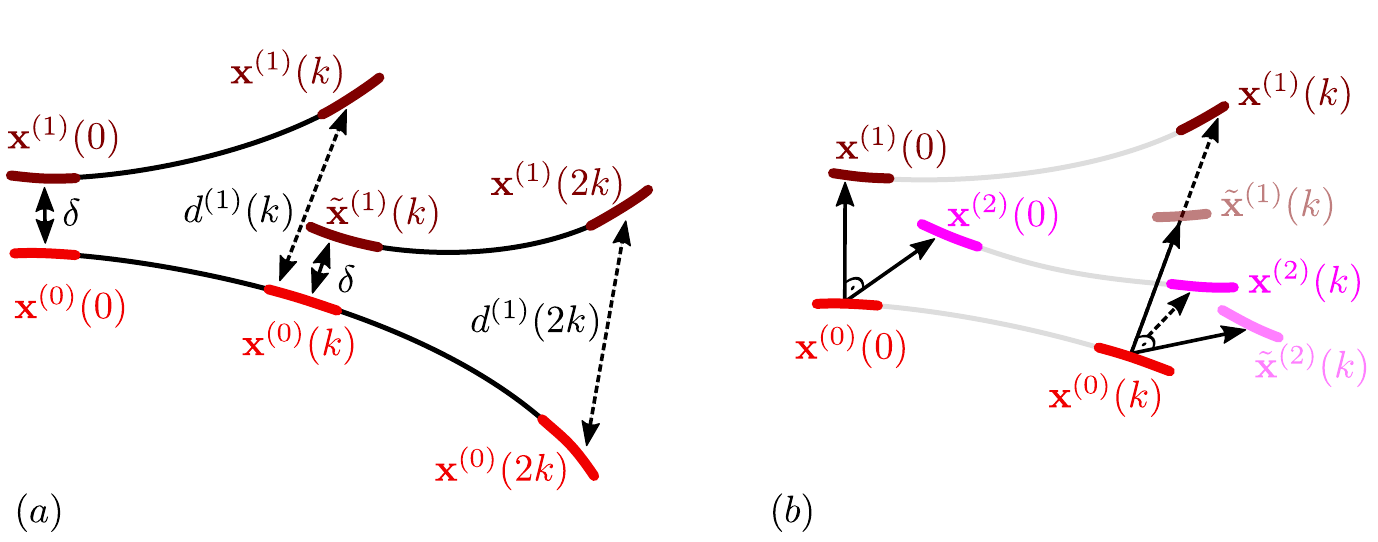}
\caption{\label{fig:lyapbenettin}
Estimating Lyapunov exponents using Benettin's method.  ($a$) Iteratively
measuring the distance $d^{(1)}$ between the reference trajectory
$\mathbf{x}^{(0)}$ and the auxiliary trajectory~$\mathbf{x}^{(1)}$. After each
iteration the auxiliary state is rescaled, such that the initial distance
$d^{(1)}=\delta$ is restored.
($b$) Computing the $L=2$ largest Lyapunov exponents using a reference
trajectory $\mathbf{x}^{(0)}$ and two auxiliary trajectories
$\mathbf{x}^{(1)}$ and $\mathbf{x}^{(2)}$, in the same way as shown 
in ($a$). The set of difference vectors are initially selected to be
orthogonal, and re-orthogonalized and rescaled after every step 
of the iteration. The thick segments of length $\tau$ indicate 
the respective state histories.}
\end{figure*}

The procedure described above is repeated with 
$\mathbf{x}^{(0)}(k)$ and $\tilde{\mathbf{x}}^{(1)}(k)$
being the new pair of starting state histories, 
cf.~Fig.~\ref{fig:lyapbenettin}. After $M$ iterations 
the average of the finite time Lyapunov exponents
\begin{align}
\langle\lambda_1^{(\text{ft})}\rangle&=
\frac{1}{M}\sum\limits_{j=1}^M
\frac{1}{k\varDelta t}\log\left(\frac{d^{(1)}(j\cdot k)}{\delta}\right)
\approx \lambda_1
\quad\qquad (M\to\infty)
\label{eq:lyapaverage}
\end{align}
then converges to the largest Lyapunov exponent~$\lambda_1$. 
Of interest in this context is that the distribution of 
local finite-time Lyapunov exponents can have a variance that 
is large compared to the average value \cite{wernecke2017test}. 

Keeping the direction of the difference vector $\mathbf{d}^{(1)}(k)$ 
in every iteration ensures that $\mathbf{d}^{(1)}(k)$ aligns with the
direction of the largest divergence in phase space. Note that this is not a 
fixed direction, but a direction that depends on the location on the 
attractor. For a given system the integration step size $\varDelta t$, 
the number of steps $k$ during divergence, the initial distance $\delta$ 
and the number of iterations $M$ have to be optimized.

\paragraph{Computing arbitrary many exponents}

The iterated method of diverging trajectories allows also
to compute, when suitably generalized, the $L\leq N$ largest
exponents. In this context the trajectory $\mathbf{x}^{(0)}$ 
serves as reference for a set of $L$ auxiliary trajectories 
of state histories,
$\left\{ \mathbf{x}^{(1)},\ldots, \mathbf{x}^{(L)} \right\}$.
The initial conditions $\mathbf{x}^{(i)}(0)$ of the auxiliary 
trajectories are selected such that their distance to the reference 
state history $\mathbf{x}^{(0)}(0)$ is small, $d^{(i)}(0)=\delta$, 
with the difference vectors with respect to the reference point 
being mutually orthogonal:
\begin{align}
 \mathbf{d}^{(i)}(0)\cdot\mathbf{d}^{(j)}(0)&=\left\{\begin{array}{cl}
N\delta^2&\text{ if }\; i=j\\ 0 & \text{ else} \end{array}\right.\,.
\end{align}
From the distances $d^{(i)}(k)$ after $k$ integration steps 
the local Lyapunov exponents $\lambda_i^{(\text{ft})}$ are 
estimated following (\ref{eq:lyaploc}). In order to prepare 
the next iteration, the first reference state is rescaled 
$\mathbf{x}_1(k)\to\tilde{\mathbf{x}}_1(k)$ according to 
(\ref{eq:refrescale}). The remaining reference states are 
then modified, such that all $L$ difference vectors form 
an orthogonal set with each vector having an average length
$d^{(i)}(k)=\delta$. This procedure is sketched in
Fig.~\ref{fig:lyapbenettin} for $L=2$. Performing the 
iteration $M$ times allows then to estimate the average 
Lyapunov exponents $\langle\lambda_i^{(\text{ft})}\rangle$ via
Eq.~(\ref{eq:lyapaverage}).

Using in every time step an iterative algorithm for the 
orthogonalization, such as the Gram-Schmidt procedure,
ensures that the $i$th difference vector aligns with 
the $i$th direction of divergence and that the average Lyapunov exponents
$\langle\lambda_\text{max}^{(\text{ft})}\rangle=\langle\lambda_1^{(\text{ft})}\rangle
\geq\langle\lambda_2^{(\text{ft})}\rangle\geq\ldots\geq\langle\lambda_L^{(\text{ft})}\rangle$
are ordered \cite{trefethen1997numerical,ruhe1983numerical}.

Benettin's method is widely used for the evaluation of the 
Lyapunov spectrum of a dynamical system. Its accuracy is
limited however in particular by the numerical restrictions 
arising from the Gram-Schmidt orthogonalization procedure. 
Alternative concepts for addressing the Lyapunov exponents 
and the corresponding directional vectors are consequently of
interest \cite{pazo2010characteristic}.


\subsubsection{Extracting Lyapunov exponents from the Euler map}
\label{sec:eulermapjacobian}

Lyapunov exponents may be extracted, as mentioned already in
Sect.~\ref{sec_Lyapunov_maps}, also from the discretized 
system (\ref{eq:eulermap}), viz from the Euler map 
$\mathbf{x}(k+N)=\mathbf{G}(\mathbf{x}(k))$, where
$G$ is a $N\times N$ matrix.
The respective Jacobian 
matrix of derivatives $J(k)=\left\{ J_{lm}(k) \right\}$
evaluated for a state $\mathbf{x}(k)$ of the map~(\ref{eq:eulermap}),
\begin{align}
J_{lm}(k)&=\frac{\partial G_l\big(\mathbf{x}(k)\big)}{\partial x_m(k)},
\qquad\quad
0\leq l,m < N~,
\label{eq:eulerjac}
\end{align}
has $N$ complex eigenvalues 
$\sigma_j(k)\equiv\sigma_j=\sigma_j^\prime+\imath\sigma_j^{\prime\prime}$, 
with real and imaginary parts $\sigma_j^\prime$ and 
$\sigma_j^{\prime\prime}$, that describe the dynamics 
of the Euler mapping~(\ref{eq:eulermap}) in tangent space
\cite{pikovsky2016lyapunov, sandor2018world}.
We now consider with
\begin{equation}
\mathbf{d}^{(1)}(k)=
\mathbf{x}^{(1)}(k)- \mathbf{x}^{(0)}(k)
=\delta\, \mathbf{e}_j(k)
\label{eq:dpe}
\end{equation}
two state vectors $\mathbf{x}^{(0)}(k)$ and $\mathbf{x}^{(1)}(k)$
for which the distance vector $\mathbf{d}^{(1)}(k)$ is aligned
to the $j$th (normalized) eigenvector $\mathbf{e}_j(k)$ of the
Jacobian $J$ of the Euler map.
The Jacobian maps this distance vector to
$J\mathbf{d}^{(1)}(k)=\delta\sigma_j\mathbf{e}_j(k)$, 
viz to a vector having the norm
\begin{equation}
\lVert J(k)\,\mathbf{d}^{(1)}(k)\rVert
 = \lVert\delta\,\sigma_j(k)\, \mathbf{e}_j(k)\rVert
 = \delta\,\lVert\sigma_j(k)\rVert\,.
\label{eq:dpe_dsquared}
\end{equation}
It is known that the local Lyapunov exponents of a map equal the
logarithm of the eigenvalues of the corresponding Jacobian matrix
\cite{gros2015complex}.
For the Euler map the logarithm of the Jacobian's eigenvalues $\sigma_j$
converges for $N\to\infty$ to the local Lyapunov exponents $\Lambda_j$
of the approximated DDE:
\begin{equation}
 \frac{1}{\tau}\log\sigma_j\to\Lambda_j\quad
 \text{ for }N\to\infty\,,\qquad
 \sigma_j=\lVert\sigma_j\rVert\; \exp(\imath\arg\sigma_j)\,,
 \label{eq:eigenvaluesmap}
\end{equation}
where $\arg\sigma_j$ denotes the argument of a complex number,
which corresponds to its phase angle in polar representation.
The normalization factor $1/\tau$ in Eq.~(\ref{eq:eigenvaluesmap})
stems from the fact that the Euler map evolves
by a time difference of $\tau$ in every iteration.

One may thus use the modulus $\lVert\sigma_j(k)\rVert$ of the $j$th
eigenvalue of the Jacobian of the Euler map to approximate
the real part $\Lambda_j^\prime$ of the local Lyapunov exponent
at a point in phase space, which corresponds to the state $\mathbf{x}(k)$
of the Euler map~(\ref{eq:eulermap}):
\begin{align}
\Lambda^\prime_j&=\lim\limits_{N\to\infty}\frac{1}{\tau}\log \lVert\sigma_j(k)\rVert\,,\qquad
\lVert \sigma_j\rVert^2 = (\sigma_j^{\prime})^2 + (\sigma_j^{\prime\prime})^2\,,
\label{eq:lyapfromeig} 
\end{align}
compare Eq.~(\ref{eulerMap_e_lambda}).

\section{Conclusions}

Dynamical systems with retarded interactions constitute
an active and rapidly developing research field with
increasing relevance for real-world applications. An 
example is the proposal \cite{gros2017entrenched}, that 
the time delays resulting from entrenched election cycles 
may contribute to destabilizing modern democracies, 
in particular if the presumption holds that the technological 
progress induces a continuously accelerating opinion dynamics. 

A defining feature of time delay system is the enlarged phase
space, which becomes formally infinite-dimensional when 
retarded feedback is introduced into a finite dimensional
dynamical system. The existence of an infinitely large phase 
space, the space of state histories, raises a series of
interesting questions, as pointed out in the introduction,
Sect.~\ref{sec:intro}, such as: How to treat an infinite 
dimensional system and it's diverging spectrum of Lyapunov exponents? 
How does the phase space compactify in the limit of vanishing
time delays?

Time delays come in large varieties, as detailed out
systematically in Sect.~\ref{sec:types}, one of
the fascinating aspects of the field. The possibilities
range here from time delays that are characterized by
their dependence on time, on the state, or by a statistical
distribution. Time delays may be classified furthermore by 
alternative criteria, such as being conservative or dissipative.

The distinct types of time delays lead to a corresponding
large range of dynamical behaviors, as discussed in 
Sect.~\ref{sec:tests}, in particular for chaotic states,
the central theme of this review. Chaos may be classical
or partially predictable, weak or strong, intermittent or
laminar. Of particular importance in this respect are
binary tests for chaos, which we also included in
Sect.~\ref{sec:tests}. There exists furthermore a range of 
complementing proposals for the dimension of a chaotic 
attractor, for which we discussed the respective implementations 
for time delay systems.

Numerical simulations of time delay systems is generally demanding,
as explained in Sect.~\ref{sec:numerics},
as a consequence of the formally diverging dimension of
the phase space of state histories. A central algorithm
for the evaluation of the spectrum of global Lyapunov 
exponents is here Benettin's method, but it is also of
interest, as we point out in Sect.~\ref{sec:numerics},
to cross-check with the results obtained from the Euler map.

Overall we hope that this review serves its purpose as a 
concise compendium of the state of the field that provides
in addition tools for a comprehensive classification of time 
delay systems and of the respective induced types of chaotic 
dynamics. We interseeded the discussion with several educational
examples aiming to provide a self-contained presentation of 
the material. Our intention
is that this comprehensive review may also serve as an entry points 
for both practitioners and newcomers to the field.

\section{Competing interests}
The authors declare that they have no competing interests.

\section{Funding}
This research was funded by the German research foundation (DFG).
H.\,W.\ acknowledges the financial support from Stiftung Polytechnische Gesellschaft Frankfurt am Main.

\section{Authors' contributions}
The paper was mostly written by H.\,W.\ and C.\,G., B.\,S.\ adding some paragraphs and contributing to the analytical derivations and numerical methods part.
H.\,W.\ did all simulations and prepared all figures.
All authors reviewed the manuscript.

\section*{Acknowledgements}
\addcontentsline{toc}{section}{Acknowledgements}
The authors acknowledge the financial support from the German research foundation~(DFG).
H.\,W.\ acknowledges support from Stiftung Polytechnische Gesellschaft Frankfurt~am~Main.
Further, H.\,W.\ thanks the organizers of the 675th Heraeus seminar on `Delayed complex systems',
G{\"u}nter Radons, Andreas Otto, and Wolfram Just for an inspiring conference that facilitated
the richness of topics in this article.
The authors wish to thank
Sue Ann Campbell, 
Georg Gottwald, 
Thomas J{\"u}ngling, 
Cristina Masoller, 
Andreas Otto, and
Eckehard Sch{\"o}ll
for their comments on the manuscript and useful hints.



\end{document}